\documentstyle[12pt,epsf]{article}

\def\TKk1{\Theta_{Kk_1}}
\def\TKk2{\Theta_{Kk_2}}

\def\TPk{\Theta_{Pk}}
\def\TPk1{\Theta_{Pk_1}}
\def\TPk2{\Theta_{Pk_2}}
\def\t12{\theta_{12}}
\def\tij{\theta_{ij}}
\def\theta{\vartheta}
\def\th{\vartheta}

\def\al{$ \alpha_s $\ }
\def\col{c_a}
\def\k1n{(k_1,..,k_n)}
\def\xxem{\ln\left(K\theta_1^m/\Lambda\right)}
\def\xxiL{\ln\left(K\theta_{1i}/\Lambda\right)}
\def\xxeL{\ln\left(K\theta_1/\Lambda\right)}
\def\upo{^{(1)}}
\def\upt{^{(2)}}
\def\r2{r^{(2)}}
\def\la{\lambda}
\def\remark#1{}
\def\rem1#1{}
\def\alim{$a-$limit\ }
 \newcommand{\labl}[1]{\label{#1}}
\newcommand{\as}{\alpha_s}
\newcommand{\pa}{\partial}
\newcommand{\ve}{\varepsilon}
\newcommand{\om}{\omega}
\begin{document}
       \rem1{powtit4}
\noindent
\hspace*{104mm} MPI-PhT/94-94\\
\hspace*{104mm} TPJU-26/94 \\
\hspace*{104mm} December 15, 1994\\
\hspace*{104mm} hep-ph/9412384\\
\vfill
%\vspace{10mm}
\begin{center}
{\Large\bf
 ANGULAR STRUCTURE OF QCD JETS }\\
%%\mbox{ }\\
\vfill
Wolfgang Ochs \\
\mbox{ }\\
Max Planck Institut f\"ur Physik \\
F\"ohringer Ring 6, D-80805 Munich, Germany \\
\mbox{ }\\
and\\
\mbox{ }\\
Jacek Wosiek\footnote{Work supported in part by KBN grant No.
      PB0488/P3/93.}\\
Jagellonian University, Reymonta 4,
PL 30-059 Cracow, Poland

\end{center}
%%\vspace{20mm}
\vfill

\begin{abstract}
We derive the angular
correlation functions between an arbitary
number of partons inside a quark or gluon jet emerging
from a high energy hard collision. As an application
results for the correlation in the relative
angle between two partons as well as the multiplicity
moments of any order for partons in a sidewise angular region are calculated.
At asymptotic energies the distributions of rescaled angular variables
approach a scaling limit of a new type with two redundant quantities.
All observables reveal the power behaviour characteristic to
the fully developed selfsimilar cascade in appropriate regions of phase space.
We present analytical results from the double logarithmic approximation
as well as Monte Carlo results.
\end{abstract}

 \vfill
%\noindent
%{\small
%\end{document}
       \newpage
\section{Introduction}
    \rem1{powintv5}
\rem1{powintv5}
Originally Perturbative QCD has been applied to the calculation of
quantities with the hadronic final states
largely averaged over, either total cross sections or jet production
in hard collision processes.    Such calculations
turned out to be very successful in comparison with experiment. On the
other hand there is continued effort to apply PQCD also to the more
detailed description of the final states \cite{DDT,KUV}.
Such studies have been performed     in Double Log
Approximation (DLA)  summing the infrared and
collinearly divergent terms and taking also into account the soft
gluon interference effects by ``angular ordering''\cite{BAS,QCD}.
A prominent example is the prediction of the energy dependence
of total multiplicity in a jet \cite{MULT}.
Another noteworthy
result is the derivation of the ``drag'' or ``string''
effect \cite{DRAG} suggested earlier on phenomenological grounds \cite{LUND}.
Also the one-particle momentum
spectrum is successfully predicted indicating that the QCD evolution
can be meaningfully continued to low momentum scales of the order of hadron
 masses
%$O(m_\pi)$
 which led to the notion of ``Local Parton Hadron Duality''
(LPHD) \cite{LPHD}. Recently the two-particle momentum correlations
have been studied
\cite{FW} and the shape although not the normalization had been
experimentally verified \cite{ACTON}.
The aim of such studies is not primarly
a further test of PQCD at a fundamental
level, rather one wants to find out the limiting scale for the application
of PQCD, and thereby to learn about the onset of
nonperturbative confinement forces.

This paper summarizes our study of further details
of the final states in QCD
%inner structure of jets in QCD
(for selected results see also \cite{OW1}-\cite{WOSIEKC}).
We give a systematic derivation of a variety
of multiparton correlations starting
from the master equation for the generating functional in DLA
\cite{DFK}.
This approach allows to
attack for the first time the computation of observables for
multiparticle final states and to find some simple asymptotic formulae.
All the observables  considered
approach a universal scaling limit at high energies described by a single
function of a scaling variable.
In this way quite different observables become closely
related.
The validity of our analytic formulae is checked by Monte Carlo calculations
(using the program
HERWIG \cite{HERWIG}).
The experimental confirmation of such relations would support
the idea that the underlying structure of multiparticle production
is the parton cascade with little
disturbance by confinement effects.

Our results also allow to directly adress the question of the fractal
nature of the QCD cascade  \cite{VENFR,GUST}.
Indeed we find the kinematic regime where
our observables depend like a power on the resolution scale.
This  behaviour is found sufficiently far away from the cutoff scale.
The simplest example is the well known result for the total multiplicity
$\bar n\approx \exp (\beta\sqrt{P\Theta/\Lambda})
=(P\Theta/\Lambda)^{a(P\Theta)}$  in the cone with angle $\Theta$
of a jet with momentum $P$. The energy dependence is determined
by the anomalous dimension $a\equiv\gamma_0=\sqrt{6\alpha_s/\pi}$. For the
higher
order correlations we find power laws of similar type in the
appropriate phase space region.
The power behaviour of multiplicity moments
suggested as consequence of an underlying scale invariant
dynamics (``Intermittency'' \cite{BP}) has been looked for in
many collision processes \cite{INTERM}. We specify the variables and
the kinematic regime
appropriate for such studies and calculate explicitly the intermittency
exponents for the QCD cascade.

The problem of angular correlations with application
to azimuthal correlations has also been treated in detail in
\cite{DMO} where the same scaling property at high
energy is found as in case of our variables. Results on
multiplicity moments
have been obtained in parallel works by two
other groups \cite{DD,BMP} applying the
method of \cite{DMO} which is based on the same principles
but differs in details.
Whereas \cite{DD} also gives some corrections due to the MLLA
and \cite{BMP} derives further
phenomenological applications, we first construct the fully differential
$n$-particle angular correlations and then find the universal
properties of various specific projections.

The paper is organized as follows. In the next section we derive the integral
equations for multiparticle distributions of arbitrary order
and discuss our approximations.
In Section 3 we give results on
single particle observables partly known already. Section 4 explains our
methods to solve the integral equations for multiparticle angular correlations.
These are applied in Section 5 to the derivation of the
 two parton angular correlation
function, the $n$-particle cumulant correlation functions and
in Section 6 to the $n$-particle
multiplicity moments in different angular regions.
In Section 7 we compare our analytic results with Monte Carlo
calculations. Conclusions are presented in the last Section.
The Appendices contain various cross-checks and
details of the calculations.
\section{Integral equations for multiparton correlations}
         \rem1{pow2v5}
 \rem1{pow2v5}
 \subsection{Generating function}

A compact description of the
multiparticle distributions is provided by
the generating functional
\cite{MULG}
\begin{equation}
  Z\{u\}=\sum_n \int dk_1^3\ldots\int dk_n^3
  u(k_1)\ldots u(k_n) p(k_1,\ldots,k_n)
  \labl{zdef}
\end{equation}
where $p(k_1,\ldots,k_n)$ is the probability density for exclusive
production of particles with 3-momenta $k_1\ldots k_n$
and $u(k_i)$ can be thought of as profile or acceptance functions.
$Z\{u\}$ contains the complete information about the
multiparticle production.
For example
$ {1 \over n!} \delta^n / \delta u(k_1)$ $\dots
\delta u(k_n) Z\{u\}|_{u=0} = p(k_1,\dots ,
k_n)$ reproduces the
probability densities, while Taylor
expansion around $u(k)=1$ generates the inclusive densities
\begin{equation}
\rho^{(n)} (k_1,...,k_n)= \delta^n Z\{u\}/\delta u(k_1)...
\delta u(k_n)\mid_{u=1} .     \labl{den}
\end{equation}
Setting $u(k)=v$ for all $k$ gives
the generating function of the distribution of the total multiplicity.
Let us also introduce the connected correlation functions
(also refered to as cumulants)
%and their
%generatung function $W\{u\}$. If we define $W$ by
%\begin{equation}
%  Z_P\{u\} = e^{W\{u\}},  \labl{zw}
%\end{equation}
%Than the connected correlations are given by
which are obtained from
\begin{equation}
\Gamma^{(n)} (k_1,\ldots,k_n)= \delta^n \ln Z\{u\}/\delta u(k_1)\ldots
\delta u(k_n)\mid_{u=1} .     \labl{con}
\end{equation}
They are related to the inclusive densities by
\begin{equation}
\rho^{(n)} = \Gamma^{(n)} + d^{(n)}_{prod}. \labl{roga}
\end{equation}
where $d^{(n)}_{prod}$ is build from products of correlations
of lower order than $n$, for example
\begin{eqnarray}
 d_{prod}^{(2)} (1,2) & = &
\rho^{(1)} (1) \rho^{(1)} (2) \labl{prod1}\\
d_{prod}^{(3)} (1,2,3) &=&
 \prod_{i=1}^{3} \rho^{(1)}(i) +
\rho^{(1)}(1) \Gamma^{(2)} (2,3)
+ cycl.
\labl{dprod}
\end{eqnarray}

The generating functional approach to
QCD jets has been first developped in Refs.\cite{KUV,HZ}.
 In the double log approximation (DLA) the
 master equation for the generating functional of multiparton
 densities for an initial parton a (either a gluon g or a quark q) reads
 \cite{DFK}

\begin{equation}
   Z_{P,a}\{u\}=
\exp \left( \int_{\Gamma_P(K)} {\cal M}_{P,a}(K) [u(K) Z_{K,g}\{u\} -1]
d^3 K \right),
\labl{mez}
\end{equation}
The subscript $P$ denotes collectively the momentum vector of the parent
parton and
the half opening angle ($P=\{\vec P,\Theta\}$) of a jet it generates;
$\Gamma_P(K)$ stands for the phase space of the intermediate parent $\vec{K}$
($\Gamma_P(K)=\{K : K<P, \Theta_{KP}<\Theta,K\Theta_{KP}>Q_0\}$
where $Q_0$ is a transverse momentum cutoff
parameter).
${\cal M}_{P,a}(K)$ is the probability for bremsstrahlung of a single gluon
off a primary parton $a$
which reads for small angle $\Theta_{PK}$
\begin{equation}
 {\cal M}_{P,a}(K)d^3 K=\col a^2(K\Theta_{PK})\frac{dK}{K}
\frac{d\Theta_{PK}}{\Theta_{PK}}\frac{d\Phi_{PK}}{2\pi},
\labl{adef}
\end{equation}
where $c_g=1,~c_q=C_F/N_c=4/9$
and $a^2\equiv\gamma_0^2=6\alpha_s/\pi$ is the QCD anomalous dimension
controlling
the multiplicity evolution. We treat $\alpha_s$ either as constant or as
running with $a^2(p_T)=\beta^2/
\ln(p_T/\Lambda)=\beta^2/\lbrack\ln(p_T/Q_0)+\lambda\rbrack$
with the QCD scale $\Lambda$, $\lambda=\ln(Q_0/\Lambda)$
and $\beta^2=12(\frac{11}{3}N_c-\frac{2}{3}N_f)^{-1}$
(we use $N_f=5$, so $\beta^2=\frac{36}{23}$).
In this approximation the recoil of the radiating parton is neglected,
 i.e. its momentum remains unaltered. Also in (\ref{mez})
the pair creation of quarks inside a jet is neglected (the production
of gluons dominates because of an infrared singularity and the larger
color factor).

Densities defined by Eqs.(\ref{den},\ref{mez}) describe only the
radiated gluons excluding the primary parton.
For example, the total multiplicity
$\overline{n}=\int \rho^{(1)}d^3K$ does not contain the beam (leading)
parton which also contributes to the final state. To account for the
leading particle one can define another generating function
\cite{QCD}
\begin{equation}
\tilde Z_{P,a}\{u\} = u(P) Z_{P,a}\{u\}.\labl{gen2}
\end{equation}
The master equation for $\tilde Z_{P,a}$ follows readily from Eq.(\ref{mez})
\begin{equation}
 \tilde Z_{P,a}\{u\}= u(P)
\exp \left( \int_{\Gamma_P(K)} {\cal M}_{P,a}(K) \left[
\tilde Z_{K,g}\{u\} -1\right]
d^3 K \right),
\labl{mew}
\end{equation}
For many problems, like production of particles into sidewise
detectors this distinction is irrelevant. For fully inclusive observables
the leading particle should be included; in cases studied by us
with an angular average its effect
however decreases with increasing energy.
We shall therefore mainly use the first formulation, pointing when necessary,
to the differences between the two.

\subsection{Inclusive densities}
By  differentiations (\ref{den},\ref{con}) of the generating functional
(\ref{mez}) we get the integral equations
\begin{equation}
\Gamma_P^{(n)}(k_1,..,k_n)=d_{nest}^{(n)}(k_1,..,k_n)+\int
      d^3K{\cal M}_{P,a}(K) \rho_{g,K}^{(n)}(k_1,..,k_n)
\labl{FF3}
\end{equation}
where the ``nested'' term
\begin{equation}
d^{(n)}_{P,nest} \k1n={\cal M}_{P,a}(k_1)
\rho_{g,k1}^{(n-1)}(k_2,..,k_n)+cycl. perm.
\labl{nest}
\end{equation}
refers to the direct emission of one parton followed by the ``decay''
into the remaining $n-1$ ones.
Here and in the following we drop the index $a$ of correlation functions
if the equation holds for both cases $a=q$ and $a=g$.
Using (\ref{roga}) we get integral
equations for densities, explicitly for $n=1,2$
\begin{eqnarray}
\rho^{(1)}_P (k) & = & {\cal M}_{P,a} (k)
              +  \int_{\Gamma} {\cal M}_{P,a} (K) \rho^{(1)}_{g,K}(k) d^3 K ,
              \labl{ro1} \\
\rho^{(2)}_P (k_1,k_2) & = & \rho^{(1)}_P (k_1) \rho^{(1)}_P (k_2)
 +  {\cal M}_{P,a} (k_1) \rho^{(1)}_{g,k_1} (k_2)
  +  {\cal M}_{P,a} (k_2) \rho_{g,k_2}^{(1)} (k_1)  \nonumber\\
  & + &
\int_{\Gamma} {\cal M}_{P,a} (K) \rho^{(2)}_{g,K} (k_1,k_2) d^3 K ,
\labl{mero}
\end{eqnarray}
and for arbitrary order (see Fig.1)
\begin{equation}
\rho^{(n)}_P (k_1,\dots,k_n) = d^{(n)}_P (k_1,\dots,k_n) +
 \int_{\Gamma}
 {\cal M}_{P,a} (K) \rho^{(n)}_{g,K} (k_1,\dots,k_n) d^3 K. \labl{men}
\end{equation}
%The inhomogenous terms $d_P$ are constructed from various products of
%correlation functions and Born cross sections ${\cal M}_P$
%as in Eq.(\ref{mero})
%for $n=2$.
%It follows from Eqs.(\ref{mez},\ref{men}) that for general $n$
%$d^{(n)}$ can be split into a sum
with
%\begin{equation}
$
d^{(n)}=d^{(n)}_{prod} + d^{(n)}_{nest}.
$
%  \labl{prones}
%\end{equation}
%where the "product" terms are constructed from the products
%of various correclation functions, and the "nested" contributions
%contain products of densities and Born terms.
%It follows from
%Eqs.(\ref{mez},\ref{zw}) that the product terms are in fact
%the difference between the connected and disconnected correlation
%functions.

%The parton densities $\rho^{(n)}$ rise exponentially
%with energy. Therefore
 Contributions containing
Born terms ${\cal M}_P$
are non-leading in the high energy limit so they can be
neglected at asymptotically high energies.
For example, for $n=2$ the ``nested'' term
${{\cal M}}_P(k_1) \rho_{k1}^{(1)} (k_2)$ , which
corresponds to chain emission $P\to k1\to k2$, is nonleading at
large $P$ as compared to the product term
$ \rho^{(1)}_P (k_1) \rho^{(1)}_P (k_2)  $.
Here, and in the following, the subscript $\Gamma$ denotes generically
all boundaries of the phase space integration.

Cumulant correlations containing
leading particles are easily derived from Eqs.(\ref{con},\ref{gen2})
\begin{equation}
\tilde{\Gamma}_P^{(n)} \k1n = (-1)^{n-1} (n-1)!\prod_{i=1}^n
   \delta (P-k_i)+ \Gamma_P^{(n)} \k1n
  \labl{gamtil}
\end{equation}
%\begin{eqnarray}
%\tilde{\rho}^{(1)}_P (k) & = & \delta(P-k)
% +  \int_{\Gamma} {\cal M}_P (K) \tilde{\rho}^{(1)}_K (k) d^3 K ,
%              \labl{rob1} \\
% \tilde {\rho}^{(2)}_P (k_1,k_2) & = & \rho^{(1)}_P (k_1) \rho^{(1)}_P (k_2)
% - \delta (P-k_1)\delta (P-k_2)
%  \nonumber\\
% & + &
%int_{\Gamma} {\cal M}_P (K) \tilde {\rho}^{(2)}_K (k_1,k_2) d^3 K .
%label{merobar}
%end{eqnarray}
The corresponding
densities are found from Eq. (\ref{roga}) to be linearely related
to the densities without leading particles, for example
\begin{eqnarray}
\tilde{\rho}^{(1)}_P (k) & = & \delta(P-k)
 +  \rho^{(1)}_P (k) ,
              \labl{robb1} \\
\tilde{\rho}^{(2)}_P (k_1,k_2) & = &
     \delta(P-k_1) \rho^{(1)}_P (k_2)
  +   \delta(P-k_2) \rho^{(1)}_P (k_1)
 + \rho^{(2)}_P (k_1,k_2)
\labl{robb2}
\end{eqnarray}
These quantities obey similar integral equations as
in (\ref{ro1},\ref{mero}).

The above equations all hold for
either quark or gluon jets. It should be noted, however,
that
in the jet we assumed production of gluons only. Therefore in
the integral equations the correlation functions under the
integral always refer to gluon initiated intermediate jets
whereas the quark degree of freedom only enters the
emission from the primary parton described by ${\cal{M}}_{P,a}$
in our formulae.

The difference between quark and gluon jets is seen most easily
for the cumulant correlations which correspond
to one primary gluon emission and are therefore proportional to
$c_a$. From Eqs.(\ref{FF3},\ref{nest}) one finds
\begin{equation}
\Gamma^{(n)}_{P,a} = c_a \Gamma^{(n)}_{P,g} \label{quarkgam}
\labl{gama}
\end{equation}
{}From this result the more complicated equations for
$\rho^{(n)}_{P,a}$ or the correlations with leading
particles can be derived.

Finally we introduce our notation for inclusive densities.
Unless stated otherwise a distribution $\rho(x)$ is
differential in its argument $x$, i.e. $\rho(x)\equiv dn/dx$.
For example, the distributions in spherical and polar angles
are related by  $\rho(\Omega)=\rho(\theta)/(2\pi\theta)$.

\subsection{Connected correlation functions}
%{\em Connected correlations.}
In the solution of (\ref{men}) a careful treatment of the boundaries
$\Gamma$ is required. These depend on the multiparticle
kinematics
and therefore also on the structure of the direct terms $d_P^{(n)}$.
In particular,
if one solves Eq.(\ref{men})
by iteration one finds in general different bounds for the
first iteration (integral $\int d^3 K d^{(n)}_K$)  and for
all higher iterations. For illustration we consider
in more detail the bounds
for the angular integrals in the left column (``precise bounds'')
of Fig.2.
The simplest case is Eq.(\ref{ro1}) for $\rho^{(1)}_P
(\Omega_1)$. In the lowest order there is the chain $P
\rightarrow K \rightarrow 1$. The limits of integration come
from the lower $p_T$ cutoff $(k_1 \Theta_{Kk_1} > Q_0$, the
small circle around particle 1 in Fig.2a) and the
requirement of angular ordering, namely, the intermediate
parton $K$ can emit parton 1 only in a cone around $K$
limited by $P$, i.e. $\Theta_{Kk_1} < \Theta_{PK}$, which
leads to the outer boundary in the figure. For two particles
the integral bounds $\int d^3K d^{(2)}_{K,prod}$ involving
the product term have, in the first iteration of
Eq.(\ref{mero}), to obey the same constraints as above twice
which yields the contour in Fig. 2b (bound $\Gamma''$).
If we iterate
Eq.(\ref{mero}) once more according to the chain emission $P
\rightarrow K_1 \rightarrow K_2 \rightarrow (1,2)$ the
angular ordering requires the same outer bound for $K_1$ as
for $K_2$ before but now the lower limit in the $K_1$
integral is given by the minimal circle enclosing the final
particles, because of angular ordering $\Theta_{K_2k_1},
\Theta_{K_2k_2} < \Theta _{K_1K_2}$ (see Fig.2.c for the
$K_1$-integration domain). Similarly, bounds in all higher
iterations are sensitive only to the minimal virtuality of the
last parent (bound $\Gamma'$).

This distinction applies to an
arbitrary number of partons in the final state:
whereas the bound $\Gamma''$ depends on the
coordinates of all individual momenta, the bound $\Gamma'$
depends only on those of the circle.
It is then advantageous to work with the connected correlation
functions. They satisfy the
following equation, c.f. (\ref{roga},\ref{FF3})
%%\begin{equation}
%$
%\Gamma_P^{(n)}=\rho_P^{(n)}-d_P^{(n)},
%$
%%\labl{F2a}
%%\end{equation}
\begin{eqnarray}
\Gamma_{P,a}^{(n)}(k_1,..,k_n)&=&d_{P,nest}\k1n +\Delta_{P,a}^{(n)}(k_1,..,k_n)
   \nonumber\\
&+&\int_{\Gamma'}
      d^3K{\cal M}_{P,a}(K) \Gamma_{g,K}^{(n)}(k_1,..,k_n)
\labl{F3}
\end{eqnarray}
which has the same boundary $\Gamma'$ for all iterations.
On the other hand the inhomogeneous term is defined in terms of the
different boundary~$\Gamma''$
\begin{equation}
\Delta_{P,a}^{(n)}(k_1,..,k_n)=\int_{\Gamma''}
      d^3K{\cal M}_{P,a}(K) d_{g,K,prod}^{(n)}(k_1,..,k_n).
\labl{F4}
\end{equation}
The contribution of the nested term can be neglected in the high energy
limit.

It follows from Eqs.(\ref{F3},\ref{F4}) that the connected correlations
contain one more iteration of the Born kernel than the corresponding
densities. This has two consequences: a) the perturbative expansion of the
cumulants starts at one power of $a^2$ more, and b) the
connected and disconnected correlations have different
factorisation properties. This last feature will be
exploited later in detail.
\subsection{Pole approximation}
The double logarithmic approximation consists of identifying
regions of the phase space which give dominant contributions
to the integrals involved.
In these domains only simple poles from inner emissions dominate -
all other slowly varying functions are replaced by their central values
at singular points.
This ``pole dominance'' approximation is an important part of the DLA
and was used by many authors  for calculations of the total multiplicities
and momentum distributions \cite{DDT}-\cite{QCD}.
\newline
\subsubsection{Single parton density}
%{\em Sinlgle parton inclusive density.}
As a first example we solve Eq.(\ref{ro1})
for the single parton inclusive distribution in $\vec{k}$.
Writing the momentum integration explicitly for $dn/d\Omega dk$ with $k=|k|$
(see Fig.3),
\begin{equation}
\rho_{P}^{(1)}(\Omega,k)  = k^2 {\cal M}_P (\vec{k})
  + {1 \over 2 \pi} \int_{\Gamma} {dK\over K} {d\Omega_K\over
  \Theta_{PK}^2} a^2(K\Theta_{PK})\rho_K^{(1)}(\Omega_{Kk},k).
  \labl{omega}
\end{equation}
The dominating
singularities are those in $\Theta_{Kk}$, $\Theta_{Kk}\sim 0$
since $\Theta_{PK} > \Theta_{Kk}$ due to the angular ordering.
At this point $\Theta_{PK}\sim
\Theta_{Pk}\equiv \vartheta$. Azimuthal integration
around the
 $\vec{k}$ direction is trivial and we get
\begin{equation}
\rho_{P}^{(1)}(\Omega,k)  = k^2 {\cal M}_P (\vec{k})
  + {1 \over 2\pi \vartheta^2}
  \int_{k}^P {dK\over K} \int_{Q_0\over k}^{\vartheta}
  {d\Theta_{Kk}\over\Theta_{Kk}} a^2(K\theta)
  \left[\Theta_{Kk}^2\rho_K^{(1)}(\Omega_{Kk},k)
  \right],
  \labl{tkk}
\end{equation}
where the factor in the square brackets does not have the Born pole
$\Theta_{Kk}^{-2}$.
The integration domain is shown in Fig. 2a. The upper bound on
$\Theta_{Kk}$ follows from the angular ordering.
A further simplification results if we introduce logarithmic
variables $x=\ln{(P/k)}$, $z=\ln{(K/k)}$,
$\zeta=\ln{(\vartheta k/Q_0)}$,  $\xi=\ln{(\Theta_{Kk}k/Q_0)}$,
 $\lambda=\ln{(Q_0/\Lambda)}$, and the corresponding
density $\rho^{(1)}(x,\zeta)\equiv d^2 n / dx d\zeta = 2\pi k^3 \vartheta^2
\rho^{(1)}_P(\vec{k})$. Then, with $a^2(x)=\beta^2/x$
\begin{equation}
\rho^{(1)}(x,\zeta)  = a^2(x+\zeta+\lambda)
  +  \int_{0}^x dz a^2(z+\zeta+\lambda) \int_{0}^\zeta
  d\xi \rho^{(1)}(z,\xi).
  \labl{xzt}
\end{equation}
This equation has a simple solution for constant \al \cite{QCD}.
\begin{equation}
\rho^{(1)}_P(\vec{k})={\cal M}_P(\vec{k}) I_0(2a\sqrt{x \zeta}).
\labl{I0}
\end{equation}
The case of running \al will be discussed in Section 3.

This completes our first application of the pole approximation.
The single particle density (\ref{I0}) plays also an important role as
the Green function or the resolvent for other equations, see
Sect. 3.1.
% Note the analogy between the lower angular cutoff $Q_0/k$
%for the emission of the elementary parton, Eq.(\ref{tkk}), and
%the opening angle $\sigma$ for the virtual jet, Eq.(\ref{res2}).
%Indeed the inclusive density of ``elementary'' partons
%$\rho_{P}^{(1)}(k)$,
%Eq.(\ref{I0}), coincides with the resolvent, Eq.(\ref{bes}),
%if the angular cutoff for emission of an elementary parton $Q_0/k$
%is replaced by $\sigma$ for a virtual jet.
\subsubsection{Simplified equation for connected correlation functions}
In the second example we use the pole approximation to reveal the
simple structure and natural variables for the connected angular
two body
correlation functions.
Momentum integrated connected correlations satisfy equations
analogous to Eqs. (\ref{F3}) and (\ref{F4}).
Due to the different singularity structure of the product term
$d_{prod}(\Omega_1,\Omega_2)$ and
$\Gamma^{(2)}(\Omega_1,\Omega_2)$ itself, the simplified integrals
over the parent momentum $\vec{K}$ will be different in both equations.
In  (\ref{F4}) the angular integral $d\Omega_K$ is dominated by two
regions $\vec{K}\parallel \vec{k_1}$ and $\vec{K}\parallel \vec{k_2}$.
Choosing appropriate polar variables in both cases we get
neglecting the nested contribution,
\begin{eqnarray}
\lefteqn{
\Delta^{(2)}_P(\Omega_1,\Omega_2)=} \labl{dirter2} \\ & &
\frac{1}{\theta_{1}^2 }
\int_{Q_0/\theta_{12}}^P \frac{dK}{K} a^2(K\vartheta_1)
\rho^{(1)}_K(\Omega_{12})
\int_{\kappa_K}^{\theta_{12}}\frac{d\Theta_{Kk_1}}{\Theta_{Kk_1}}
\left[\Theta_{Kk_1}^2\rho^{(1)}_K(\Omega_{Kk_1})\right]
%\nonumber \\
+ ( 1 \rightarrow 2).
\nonumber
\end{eqnarray}
The singular part around each pole was integrated down to
the elementary cutoff
 $\kappa_K\equiv Q_0/K$ while the
 slowly varying function
 $\Theta_{Kk_2}(\Theta_{Kk_1},\theta_{12},\phi)$
 is approximated by its central value $\theta_{12}$ in all nonsingular
 expressions. Similarly $\Theta_{PK} \rightarrow \vartheta_i, i=1,2$.
 The upper bound for $\Theta_{Kk_1}$ is chosen as $\theta_{12}$ since
 the contribution from beyond this scale is asymptotically negligible.

It is readily seen from Eq.(\ref{dirter2}) that,
 although $d^{(2)}$ contains a product of two poles (or more
general, power singularities),
 the structure of $\Delta^{(2)}$ is simpler.
The pole dominance in $d^3 K$ integration splits $\Delta^{(2)}$ into
a sum of two terms where the parent is almost parallel to {\em each }
of the final partons and
consequently $\Delta^{(2)}$ is a {\em sum} of two contributions.
It is also easy to see that $\Delta^{(2)}_i$ defined by the
decomposition (\ref{dirter2}) depends only on the two variables
$\vartheta_i$ and $\vartheta_{12}, i=1,2$. Hence they may be
regarded as the natural variables for this process.
The integration domains in this case are shown in Fig.2e,g.

This structure of the inhomogenous term $\Delta $
implies simplification of
the parent integration in the integral equation for $\Gamma^{(2)}$
itself , Eq.(\ref{F3}). It turns out
that $\Gamma^{(2)}$ also decouples into a sum of two terms
$\Gamma_{i}^{(2)}\;\;\;(i=1,2) $ and each of them
satisfies the following equation
\begin{eqnarray}
\lefteqn{\Gamma^{(2)}_{P,i} (\Omega_i,\Omega_{12}) =
\Delta^{(2)}_i(\Omega_i,\Omega_{12},P)} \nonumber \\& &
 + \frac{1}{\theta_{i}^2} \int_{Q_0/\theta_{12}}^{P}\frac{dK}{K}
\int_{\theta_{12}}^{\theta_i}\frac{d\Theta_{Kk_i}}{\Theta_{Kk_i}}
a^2(K\vartheta_i)
\left[ \Theta_{Kk_i}^2\Gamma^{(2)}_{K,i}(\Omega_{Kk_i},\Omega_{12})
\right] ,
\theta_i > \theta_{12}.
\labl{Gamsim}
\end{eqnarray}
Again the pole dominance was used. The upper limit on $\Theta_{Kk_i}$
follows from the angular ordering as in the single particle distribution
Eq.(\ref{tkk}), while the lower one is set by the minimal virtuality
required for the subsequent decay into $k_1,k_2$ pair (see Fig.2f,h).
 This discussion provides a quantitative example of the general statement
made at the beginning of Sect.2.3.
Similar equations hold for higher order correlation functions
(see Sect.5).
They all imply a remarkably simple singularity
structure of  the connected
correlation functions.

\subsection{Generic equation}
At this point it is is worthwhile to emphasize
 a large degree of the universality
in the description of seemingly different observables
characterizing the QCD cascade. It turns out that many physically
different processes are described by  mathematically the same
equation requiring only
an appropriate renaming of variables and a proper choice of the inhomogenous
term. For this reason we introduce a generic integral equation
\begin{equation}
h(\delta,\theta,P)=d(\delta,\theta,P)+\int_{Q_0/\delta}^{P} \frac{dK}{K}
\int_{\delta}^{\theta}
\frac{d \Psi}{\Psi} a^2 (p_T) h(\delta,\Psi,K).  \labl{gen}
\end{equation}
where the scale of the running $\alpha_s$ depends on the kinematics of
the specific problem. In principle one should distinguish two classes:
a) $p_T=K\theta (P\Psi) $, and
b) $p_T=K\Psi$.
%With the appropriate interpretation of variable,
%this equation describes most of the situations.
For example, class
a) contains connected angular correlations,
cumulant moments
and doubly differential single parton density,  and class
b)
                                         correlations in the relative
angle $\theta_{12}$ and single parton energy distribution.
In fact
equations describing both classes are related by simple
integration, therefore it suffices to consider only
one class. The inhomogenous term $d$ depends usually only
on two out of the three indicated variables. It is essentially different
for single parton spectra and for higher densities. Typically
for angular correlations it has
the following form for n-parton observables in the running \al
case
\begin{equation}
 d \sim \exp \left(2n\beta\sqrt{\ln(P\delta/\Lambda)}\right).
\labl{gendir}
\end{equation}
Solutions of this equation are discussed in the Sect.4.
\section{Multiplicity and single parton densities}
        \rem1{pow3v8}

This chapter summarizes various techniques of calculating
different single particle densities which we needed in the derivation
of various multiparton correlations.
Most of the results discussed here are known
\cite{QCD,MULT}. The motivation here is to introduce our notation and
the methods, which are used in Sect.4 in the derivation of the
angular observables. Furthermore we want to expose
the emergence of the power behaviour
in the variety of observables. This aspect of the QCD cascade, i.e.
the explicit proof of the selfsimilarity/intermittency in the time-like
partonic showers, has attracted considerable interest only recently
\cite{INTERM}. Some results concerning the integral representations
of various densities were not derived before.
As we shall see, not all observables
show selfsimilarity even for the constant \al . It is one of our
goals to identify the conditions necessary for occurence of the
power behaviour.
\newline
{\em Total multiplicity.} Multiplicity of partons $n(P,\Theta)$
in a jet with
momentum $P$ and half opening angle $\Theta$ is the simplest quantity
which was studied in perturbative QCD \cite{MULT}. The equation
which determines total multiplicity
\begin{equation}
\overline{n}(P,\Theta)=\int_{Q_0/\Theta}^P {dk\over k}
\int_{Q_0/k}^{\Theta}
{d\theta \over \theta} \rho_P^{(1)}(\ln{k},\ln{\vartheta}),
\labl{ntot}
\end{equation}
follows readily from  Eq.(\ref{tkk}) upon integration over the parton
phase space (\ref{ntot}). Due to the logarithmic singularities
of Born diagrams, it is convenient to use the logarithmic variables.
Moreover, the structure of the inhomogenous Born term, together
with the resulting equation, implies that total multiplicity
depends only on the "virtuality" of a parent jet $P\Theta$.
One obtains,
\begin{equation}
 \overline{n}(X_{\la})=b(X_{\la}) + \int_0^X dv \int_0^v du a^2(u+\lambda)
\overline{n}(u+\la)
\end{equation}
\begin{equation}
X=\ln{(P\Theta/Q_0)},u=\ln{(K\Psi/Q_0)}, X_{\la}=X+\la,\;\;
\la=\ln{(Q_0/\Lambda)}.
\labl{ntoteq}
\end{equation}
with the Born term $
b(X_{\la}) = \int_0^X dv \int_0^v du a^2(u+\lambda)$.
For constant $\alpha_s$ $(a(x) \rightarrow a)$ one finds the simple
solution
\begin{equation}
\overline{n}(X)=\cosh{(a X)}-1.
\labl{coshhh}
\end{equation}
Hence in
 the high energy limit $(P\rightarrow \infty)$ multiplicity
grows like a power of the virtuality
\begin{equation}
\overline{n}(P\Theta)\simeq {1\over 2}
\left ( {P\Theta\over Q_0} \right )^a
\labl{ntotpower}
\end{equation}
with the anomalous dimension $\gamma_0\equiv a$.

For running $\alpha_s$ one solves the differential equation
equivalent to Eq.(\ref{ntoteq}), which
can be
reduced to the Bessel equation
by appropriate
change of variables.
The solution satisfying initial
conditions implicit in (\ref{ntoteq}) reads (\cite{QCD})
\begin{equation}
\overline{n}(X_{\la})= 2\beta\sqrt{X_{\la}}
[ K_0(w_0) I_1(2\beta\sqrt{X_{\la}}) +
I_0(w_0) K_1(2\beta\sqrt{X_{\la}}) ] -1,
\labl{nrunn}
\end{equation}
where $ w_0=2\beta \sqrt{\lambda} $.
At high energy using $I_n(z)\simeq e^z/\sqrt{2\pi z},
K_n(z)\simeq e^{-z}\sqrt{\pi/2z}$
one recovers the well known expression
\begin{equation}
\overline{n}(P\Theta)\simeq
 {f \over 2\sqrt{a}}  \left (
{ P \Theta \over \Lambda }
\right )^{2 a(P\Theta)},
{}~~~~f={2\beta  K_0(2\beta\sqrt{\lambda}) \over \sqrt{\pi}}.
\labl{ffflll}
\end{equation}

The results for constant \al can be found in the limit
$\beta,\lambda \to \infty$ with $\beta/\sqrt{\lambda}=a$ kept fixed
which will be referred as the "a-limit". It also implies
\begin{equation}
f\to \sqrt{a} e^{-2\beta\sqrt{\lambda}},~ a(P\Theta)\to a,~
f \left (  { P \Theta \over \Lambda } \right )^{2 a(P\Theta)}
\to \sqrt{a}
 \left (  { P \Theta \over Q_0 } \right )^a
\labl{alimit}
\end{equation}
So in this limit Eqs.(\ref{nrunn},\ref{ffflll})
reduce to
(\ref{coshhh},\ref{ntotpower}) respectively.
It is seen that the power growth of the total multiplicity
occurs in the constant \al  case. As is well known
running \al  violates this simple scaling and multiplicity grows
slower than any power of the energy.
%These effects  are
%usually regarded as the corrections to the simple parton model
%scaling predictions.

%It is instructive to realize that our general resolvent solution
%implies  the existence of simple integral representations for various
%partonic densities.
%They are particularly usefull if the
%direct solutions are not available.

%In the present case applying procedure (\ref{ite})
%- (\ref{res2}) to Eq.(\ref{ntoteq}) with constant \al one obtains
%\begin{equation}
%\overline{n}(X)=b(X)+a\int_0^S \sinh{(a(X-u))} b(u), \;\;\;
%b(x)={1\over 2} a^2 x^2 .  \labl{nsinh}
%\end{equation}
%This result also follows from the direct integraton of the resolvent
%solution, Eq.(\ref{enfres}), over the phase space of a final parton.
%Analogous representations exist also for the running \al  and will
%be later discussed for other distributions.

{\em Angular distributions.}
The angular distribution of partons in a $(P,\Theta)$ jet is defined as
\begin{equation}
\rho^{(1)}(\xi,Y_{\la})=\int_{-\xi}^{Y} \rho^{(1)}(y,\xi,Y_{\la}) d y,
  \labl{ang}
\end{equation}
where $Y_{\la}=Y+\la, Y=\ln{(P\theta/Q_0)},
y=\ln{(k/Q_0)}$ and $\xi=\ln{(\vartheta)}$.
Integrating Eq.(\ref{tkk}) over parton momentum $k$ gives the following
equation
for the angular density (\ref{ang})
\begin{equation}
\rho^{(1)}(\xi,Y_{\la})=b_1(Y_{\la}+\xi)+\int_{-\xi}^{Y} d Z a^2(Z_{\la}+\xi)
\int_{-z}^{\xi} d\zeta \rho^{(1)}(\zeta,Z_{\la}), \labl{rang}
\end{equation}
where $Z=\ln{(K/Q_0)}, \zeta=\ln({\psi})$ measure the momentum
and the emission angle of the intermediate parent parton, $b_1(u)=\beta^2
\ln{(u/\lambda)}$.

Eq.(\ref{rang}) implies that the angular density is independent
of the jet opening angle $\Theta$. This follows from the angular
ordering, and the pole dominance. Namely all emissions from the
intermediate parents
(relative to the
final parton $\vec{k}$ ) are
limited by the final angle $\vartheta=\Theta_{Pk}$ between $\vec{P}$
and $\vec{k}$. This applies to other distributions
 provided {\em none} of the angular variables is integrated.
This simple property has rather important consequences, for example
\begin{equation}
\rho^{(1)}(\xi,Y_{\la})=
{d \overline{n} (Y_{\la}+\Xi) \over d \Xi } \mid_{\Xi=\xi}
= \rho^{(1)}(Y_{\la}+\xi), \;\;\;
\Xi=\ln{\Theta}.
\labl{deri}
\end{equation}
Therefore all results derived previously for the jet multiplicity
can be directly translated for the angular density as well.
In particular, the
complete solution of Eq.(\ref{rang}) for  running \al reads
\begin{equation}
%\rho(X)=2\beta^2
\rho\upo_P(\theta)={2\beta^2 \over \theta}
[ K_0(w_0) I_0(2\beta\sqrt{X_{\la}}) -
I_0(w_0) K_0(2\beta\sqrt{X_{\la}}) ].
\labl{rhoang}
\end{equation}
One may also derive this result solving the
differential equation equivalent to Eq.(\ref{rang}).
In the high energy limit we get
\begin{equation}
\rho\upo_P(\theta)\simeq {f\over 2\theta}
%n(P\vartheta)\simeq {\beta \over \sqrt{\pi}} K_0(2\beta\sqrt{\lambda})
\sqrt{a(P\vartheta)}
\left ( {P\vartheta\over\Lambda} \right )^{2 a(P\vartheta)}
\labl{asyan1}
\end{equation}

The corresponding exact and high energy results for constant \al are
\begin{equation}
\rho\upo_P(\theta)= {a\over \theta} \sinh (a\ln(\theta/\kappa)),~~~
\rho\upo_P(\theta)
        \simeq {a\over 2\theta} \left ( {P\vartheta\over Q_0} \right )^a
\labl{asyan2}
\end{equation}
As for the total multiplicity, the power law emerges
only at very high energies, and is violated by running \al .
Note that the angular density depends on $P\vartheta$, i.e.
on the transverse momentum of a whole jet with respect to the
parton direction. Result (\ref{asyan1},\ref{asyan2}) is also relevant
for the intermittency study. It shows that even the first
factorial moment may have nontrivial intermittency index.

Finally we quote the integral representation for running \al
which follows
%from the resolvent solution, Eq.(\ref{enfres})
%or can be derived
from the direct iteration of Eq.(\ref{rang}), cf. also Eq.(\ref{enfres}) .
\begin{equation}
\rho^{(1)}(Y_{\la})=b_1(Y_{\la})+\int_0^Y d v \int_0^v du R(Y-v,v-u,u+\la)
b_1(u+\la).
\end{equation}
Where $R(x,t,\lambda)=\partial_t\overline{R}(x,t,\lambda)$
 can be calculated from the momentum distribution,
cf. Eqs.(\ref{rder},\ref{emom},\ref{invlap}).

{\em Momentum distribution.}
Contrary to the angular density, the momentum distribution
$\rho^{(1)}(x,t,\lambda)\equiv dn/dx$
depends essentially on two variables, for which we choose
$x=\ln{(P/k)},t=\ln{(\Theta k/Q_0)}$.
As a result the corresponding integral equation,
\begin{equation}
\rho^{(1)}(x,t,\lambda)=\beta^2\ln{t+\lambda\over\lambda} +
\int_0^x dz
\int_0^t d\tau {\beta^2\over z+\tau+\lambda}\rho^{(1)}(z,\tau,\lambda).
\labl{rhoeq}
\end{equation}
has more complex structure, and its analytic solution is not known.
It turns out, however,
 that the moments of the energy distribution can be calculated
analytically. Consequently the distribution itself can be
reconstructed. To this end define the dimensionless moments of the
energy distribution
\begin{equation}
\mu_n(X,\lambda)=P^{-n}\int_{Q_0/\Theta}^P k^n
\rho^{(1)}(k,P,\Theta,\lambda) dk
=\int_0^X e^{-n x} \rho^{(1)}(x,X-x,\lambda) dx, \labl{emom}
\end{equation}
\rem1{(III.15.2, s4)}
with $X=\ln{(P\Theta/Q_0)}$ and $t=X-x$.
%where $x=\ln{(P/k)},X=\ln{(P\Theta/Q_0)},t=\ln{(k\Theta/Q_0)},
%\lambda=\ln(Q_0/\Lambda)$
It follows from the master equation (\ref{rhoeq}) that
$\mu_n$ satisfy
the following differential equation \remark{(III.I.15.2-15.6)}
\begin{equation}
{d^2\over d X^2} \mu_n(X)+n{d\over d X}\mu_n(X)={\beta^2\over X+\lambda}
(\mu_n(X)+1),
\labl{hyper}
\end{equation}
with the boundary conditions
\begin{equation}
\mu_n(0)=0,\;\;\; {d\over d X}\mu_n(0)=0,\labl{bound}
\end{equation}
Eq.(\ref{hyper}) is easily transformed into the confluent hypergeometric
equation. The solution satisfying the boundary conditions (\ref{bound}) reads.
\begin{eqnarray}
\lefteqn{\mu_n(X,\lambda)+1 = } \nonumber  \\
 & & \Gamma(\alpha)Z e^{-Z}\left[
\Phi(\alpha-1,1,n\lambda) F(\alpha,2,Z) +
F(\alpha-1,1,n\lambda) \Phi(\alpha,2,Z) \right]
\labl{momsol}
\end{eqnarray}
\rem1{(pink notes 30,31)}
where $Z=n X_{\la}$, $\alpha=1+\beta^2/n$, and $F$ and $\Phi$ denote the
regular and irregular confluent hypergeometric functions respectively.
Inverse Laplace transform of (\ref{momsol})
\begin{equation}
\rho^{(1)}(x,X-x,\lambda)[\Theta(X-x)-\Theta(x)]=\int_{\gamma} {ds\over 2\pi}
e^{x s} \mu(s,X,\lambda), \labl{invlap}
\end{equation}
can now be used to derive the asymptotic
behaviour of the momentum spectrum at high energies. To this end we
perform the $s$ integration in Eq.(\ref{invlap}) by the saddle point method.
The saddle point $s^*(X,x)\rightarrow 0$ for
$X\rightarrow\infty, x/X=const.$, therefore it is sufficient to use the
appropriate asymptotic
forms of the confluent hypergeometric functions (see Appendix A). Using
Eqs.(\ref{agfas},\ref{uas}) we find the following condition
\begin{eqnarray}
 x+{X\over 2}({s\over \sqrt{A}} -1)&+&
{\lambda\over 2s}( s^2+4a_0^2 )({1\over\sqrt{A}}-{1\over\sqrt{B}})+ \nonumber
\\
{1\over 2\sqrt{B}}(1+{4a_0^2\over s(s+\sqrt{B})}) &=&
{2 \beta^2\over s^2} \ln{\left(\sqrt{X+\lambda\over\lambda}
{s+\sqrt{A}\over s+\sqrt{B}} \right) }, \labl{sad} \\
A=s^2+\beta^2/(X+\lambda) &,& B=s^2+4a_0^2-2s/\lambda
\end{eqnarray}
with $a_0^2=\beta^2/\lambda$.
This determines the saddle point $s=s^*(X,x)$. The integral (\ref{invlap})
is dominated by the neighbourhood of this point with the following leading
result
\begin{eqnarray}
\rho^{(1)}(x,X-x,\lambda) & \sim & \exp{\left( 2 s^* x +
{X(s^* -\sqrt{A})^2\over 2\sqrt{A}}\right) }  \\  \nonumber
\exp{\left( {\lambda\over 2} (\sqrt{A}-\sqrt{B})
(1-{{s^*}^2+4a_0^2\over\sqrt{A B}})\right)} & &
 \exp{\left({1\over 2\sqrt{B}}(s^*+{4a_0^2\over s^*+\sqrt{B}})
\right) }. \labl{hump}
\end{eqnarray}
This is the double logarithmic expression for the famous hump-back plateau
which was confirmed experimentally and is considered as an important
test of the perturbative QCD \cite{DFK}.

We conclude this subsection by pointing out
some interesting limiting cases of the above
expressions.\newline
1. {\em Total multiplicity.} The zeroth moment of (\ref{emom}) gives the
total multiplicity of a jet. Indeed using the asymptotic form of the confluent
functions \cite{ABR} $lim_{a\rightarrow \infty,b=const} F(a,b,x/a)=
\Gamma(b) x^{(1-b)/2} I_{b-1}(2\sqrt{x})$ and $ lim_{a\rightarrow \infty,
b=const}[\Gamma(1+a-b) \Phi(a,b,x/a)] = 2 x^{(1-b)/2} K_{b-1}(2\sqrt{x})$,
 one readily obtains  the already quoted
result (\ref{nrunn}) for $\overline{n}$. \rem1{(notes-III.I.16)} \newline
2. {\em Constant} \al {\em limit}.
This limit, the $a-$limit cf. Eq.(\ref{alimit}),
is realized formally with
$\lambda,\beta
\rightarrow\infty, \beta^2/\lambda=a^2=const$. After some algebra
one obtains from (\ref{momsol}) much simpler expression
(see Appendix B).
\begin{eqnarray}
\mu_n(X)+1 & = & \nonumber \\
{1\over 2\sqrt{n^2+4a^2}  }&  ( &
      (\sqrt{n^2+4 a^2}+n)
      \exp{[{X\over 2}(\sqrt{n^2+4a^2}-n)]} \nonumber    \\
  &+ & (\sqrt{n^2+4a^2}-n)
      \exp{[-{X\over 2}(\sqrt{n^2+4a^2}+n)]}\;\;\;   )  \labl{muca}
\end{eqnarray}
\remark{(III.I.16-21,..17.2)}
which can be also derived from the Laplace transform of the analytic solution
(\ref{rkex}) or by solving Eq.(\ref{hyper}) for the constant \al .

Finally we consider the \alim  of the hump-back formula (\ref{hump}).
Equation (\ref{sad}) simplifies considerably and one readily finds
\begin{equation}
s^*={a(X-2x)\over\sqrt{x(X-x)}},
\end{equation}
and the asymptotic form of the density (\ref{hump}) reads
\begin{equation}
\rho^{(1)}(x,X-x)\sim \exp{(2a\sqrt{x(X-x)})}.  \labl{kaas}
\end{equation}
This agrees with the asymptotic form of the exact solution for the constant
\al which follows for example from Eq.(\ref{I0})
\begin{equation}
\rho(x,X-x)=a\sqrt{{X\over x}-1} I_1(2a\sqrt{x(X-x)}).  \labl{rkex}
\end{equation}

We conclude this Section with two observations. \newline
1. The momentum distribution, Eq.(\ref{kaas})
{\em does not} show the power behaviour in any of the two variables.
Therefore not all observables (and processes) are revealing the fractal nature
of the QCD cascade. The selfsimilarity is destroyed by the second (in addition
 to P) independent scale
$k$ which exists in this process. This observation suggests, that only
processes containing a single scale are suitable for searching for the
intermittency/selfsimilarity. \newline
2. Note also that if the second scale is {\em fixed} to be proportional
to the first one $x=\epsilon X$ say, than we recover the power behaviour.
In our example Eq.(\ref{kaas}) becomes
\begin{equation}
 \rho^{(1)}(x,X-x) \sim
 \left({P\Theta\over Q_0} \right)^{2a\sqrt{\epsilon(1-\epsilon)}},
 \labl{rosim}
\end{equation}
and the intermittency indices depend on the ratio $\epsilon=x/X$.
We shall find this phenomenon also in other more complicated situations
with many scales.
\section{Methods of the solution for higher densities}
        \rem1{pow4v8}
\subsection{Solutions in terms of the resolvents}
\subsubsection{General resolvent representation}

There exists a rather general recursive
scheme for solving Eq. (\ref{men})
for a density of arbitrary order.
First, observe that the inhomogenous term $d^{(n)}_P$,
which corresponds to the direct emission from
the parent parton, is built from various products of the
correlation functions of {\em lower} order. Only if all n derivatives
act on the $Z$ function under the $d^3 K$ integral in the exponent
of Eq.(\ref{mez}) can we recover the n-th order density.
This gives the
last term in Eq.(\ref{men}). In all other cases the $n$
derivatives are
split among various factors resulting from the differentiation
of the
exponent giving various products of the densities of lower order.
Integral
of any particular density, of lower than $n$-th order, can be always
replaced
by the density itself and corresponding inhomogenous term of yet
lower order,
by using appropriate integral equation for the lower order.

Second, the unknown function $\rho^{(n)}$
appears only in the last term in Eq.(\ref{men}).
The simplicity of this structure allows us
immediately to solve equation (\ref{men}) by iterations
\begin{eqnarray}
\lefteqn{\rho^{(n)}_P (k_1,..,k_n) =} \nonumber\\
& & \sum_{r=0}^{\infty} \int_{\Gamma} d^3 K_1 ..
d^3 K_r {\cal M}_{P} (K_1) {\cal M}_{K_1} (K_2) ..
{\cal M}_{K_{r-1}} (K_r)
d^{(n)}_{K_r} (k_1,..,k_n),  \labl{ite}
\end{eqnarray}
or symbolically
\begin{equation}
\rho^{(n)}= (\frac{1}{1-{\hat{\cal M}}}) \circ d^{(n)}.
\end{equation}
Further, we note
that {\em all but one} $K_m$ integrations ($m=1,\dots,r-1$) can
actually be done, since
the complete $K_m$ dependences are given by the Born cross sections
and variable
boundaries of the inner integrations. Hence changing the orders of
integrations we get the following integral representation
of the solution
\begin{equation}
\rho^{(n)}_P (k_1,..,k_n)=d_P^{(n)} (k_1,..,k_n) +
\int_{\Gamma} R_P (\vec{K},\sigma) d^{(n)}_K (k_1,..,k_n) d^3 K,
\labl{res}
\end{equation}
where the resolvent $R_P (\vec{K},\sigma)$ is given by
\begin{equation}
R_P (\vec{K},\sigma) = \sum_{r=1}^{\infty}
\int_{\Gamma,\Theta_{K_{r-1}K}>\sigma}
 d^3 K_{r-1} \dots d^3 K_{1}
{\cal M}_P (K_1) \dots {\cal M}_{K_{r-1}} (K) . \labl{res2}
\end{equation}
For constant $\alpha_s, R_P (\vec{K},\sigma)$ can be calculated explicitly
(see Sect. 3 for more details)
but the results (\ref{res},\ref{res2}) are valid for running
$\alpha_{s}$ as well.
In these formulae $\sigma$ is the minimal opening angle of a jet
$K$ and
has to be determined for each case separately; in general it depends
on all momenta
$\sigma=\sigma(k_1,\dots,k_n,K)$, for example,
$\sigma=\Theta_{k_1K}$  for $n=1$.

The role of $\sigma$ can be better understood if we consider in detail
the contribution from the second iteration, see also Fig.4,
\begin{equation}
\rho^{(n,2)}_P({\cal K})=\int_{\Gamma_1} d^3 K_1 {\cal M}_P(K_1)
\int_{\Gamma_2} d^3 K_2 {\cal M}_{K_1}(K_2) d^{(n)}_{K_2}({\cal K}).
\labl{ron2}
\end{equation}
$\Gamma_1$ and $\Gamma_2$ denote appropriate
boundaries resulting from all
three constraints $\Gamma_P(K_1)$, $\Gamma_{K_1}(K_2)$
and $\Gamma_{K_2}({\cal K})$,
where ${\cal K}$ denotes collectively all momenta of the final partons
${\cal K}=\{k_1,\dots,k_n\}$. After changing
the orders of the $K_1$ and
$K_2$ integration we get
\begin{equation}
\rho^{(n,2)}_P({\cal K})=\int_{\overline{\Gamma}_2} d^3 K_2
\left[
\int_{\overline{\Gamma}_1,\Theta_{K_1 K_2}>\sigma} d^3 K_1
{\cal M}_P(K_1) {\cal M}_{K_1}(K_2)
\right]  d^{(n)}_{K_2}({\cal K}), \labl{exch}
\end{equation}
which allows to identify the term in the square brackets with the
second order contribution to the resolvent, Eq. (\ref{res2}). However,
because the $K_1$ integration is now done {\em at fixed} $K_2$, its
boundaries
are influenced by the constraints implied on $K_2$ by the configuration
of the
final partons ${\cal K}$. In particular, angular ordering requires that
$\Theta_{K_1 K_2} > \sigma({\cal K},K_2)$ with $\sigma$ defined as
above.
Similar considerations lead to the general formula,
Eqs. (\ref{res2},\ref{bes}),
with $\sigma$ being the lower cutoff of the emission angle
$\Theta_{PK}$
and, at the same time, the opening angle or a measure of
the ``virtuality''
of the jet $K$.

Eq. (\ref{res}) has a simple interpretation in terms of the
cascading process
(see Fig.1). In fact, the resolvent $R_P (K,\sigma)$ is nothing but
the
inclusive distribution of a jet $(K,\sigma)$
in a jet  $(P,\Theta)$. Note the dual interpretation of the angles
$\sigma$ and $\Theta$, they denote opening angle of the corresponding
jets, but also they are simply related to the viruality of the
parent partons.

Eqs.(\ref{roga},\ref{res}) give
 the recursive prescription for explicit calculation of   general
 (fully or
partly differential) multiparton densities of arbitrary order.

\subsubsection{Resolvents for fixed $\alpha_s$}

The explicit form of the resolvent can be readily derived by considering
the integral equation for $R$ itself. It can be written in many equivalent
forms. We shall use the forward master equation for the density in the
logarithmic variables   $x=\ln{(P/K)},t=\ln{(\vartheta/\sigma)}$.
\begin{equation}
R\left(x,t\right)={dn_{jet}\over
dxdt}={R_P(\vec{K},\sigma)\over  { \cal M}_{P}(\vec{K}) }
%R\left(\ln({P\over K}),\ln({\vartheta\over\sigma})\right)={dn_{jet}\over
%d\ln{K}\ln{\vartheta}}=2\pi K^3\vartheta^2R_P(K,\sigma).
\labl{Rdef}
\end{equation}
{}From Eq.(\ref{res2}) we have (symbolically)
\begin{equation}
R={\cal M} + {\cal M} \circ R,
\end{equation}
or explicitly
\begin{equation}
R(x,t)=a^2+a^2\int_0^{x} dz \int_0^{t} d\tau R(z,\tau)., \labl{Rmst}
\end{equation}
 The pole approximation was
%where $x=\ln{(P/K)},t=\ln{(\vartheta/\sigma)}$. The pole approximation was
used similarly to Sect. 2.4.1 to simplify the three-dimensional integrals.
$K\sigma$ is the virtuality of the $(K,\sigma)$ jet emitted at the polar angle
$\vartheta$. Equation (\ref{Rmst}) is identical with Eq.(\ref{xzt}) for
constant \al with the replacement $Q_0\rightarrow K\sigma$. Hence the
solution (\ref{bes}) follows.
\begin{equation}
R_P (\vec{K},\sigma) = { \cal M}_{P}(\vec{K})
I_0 (2a\sqrt{\ln(P/K)\ln(\Theta_{KP}/\sigma)}),   \labl{bes}
\end{equation}
 Solution (\ref{res}) together with the formula (\ref{bes}) provides
useful integral expressions of many observables for constant \al .
We quote here several results leaving other more involved applications
for subsequent sections \remark{( .. 12.1.4)}
($Y=\ln(P/Q_0),y=\ln(k/Q_0),\Xi=\ln\Theta,\xi=\ln\theta$).
\newline
1. {\em Total jet multiplicity and total factorial moments.}
\begin{equation}
\overline{n}(X)=b(X)+a\int_0^S \sinh{(a(X-u))} b(u), \;\;\;
b(x)={1\over 2} a^2 x^2 ,  \labl{nsinh}
\end{equation}
\begin{equation}
f_n(Y+\Xi)=b_n(Y+\Xi)+\int_{-\Xi}^Y dZ \int_{-Z}^{\Xi} d\zeta
I_0(2a\sqrt{(Y-Z)(\Xi-\zeta)})
b_n(Z+\zeta), \labl{fnres}
\end{equation}
where $f_1(X)\equiv
\overline{n}(X), b_1(X)=a^2 X^2/2, b_2=\exp(2aX)$, and higher
are obtained recursively, (see Appendix D). \newline
2. {\em Energy density} $\rho^{(1)}(y,Y,\Xi)\equiv dn/dy$.
\begin{equation}
\rho^{(1)}(y,Y,\Xi)=b(\Xi+y)+\int_{y}^Y dZ \int_{-y}^{\Xi} d\zeta
I_0(\sqrt{(Y-Z)(\xi-\zeta)})
b(y+\zeta),  \labl{efen}
\end{equation}
with $b(X)=a^2 X$.\newline
3. {\em Angular distribution} $\rho^{(1)}(\xi,Y)=dn/d\xi$.
\begin{equation}
\rho^{(1)}(\xi,Y)=b(Y+\xi)+\int_{-\xi}^Y dZ \int_{-Z}^{\xi} d\zeta
I_0(\sqrt{(Y-Z)(\xi-\zeta)})
b(Z+\zeta),
\end{equation}
where $b(X)=a^2 X$.\newline
4. {\em Inclusive density in both variables} $\rho^{(1)}(y,\xi,Y)\equiv
d n/dy d\xi$.
\begin{equation}
\rho^{(1)}(y,\xi,Y)=a^2+\int_{y}^Y dZ\int_{-y}^{\xi} d\zeta
I_0(\sqrt{(Y-Z)(\xi-\zeta)}) a^2.
\labl{iro1}
\end{equation}
5. {\em Energy-energy correlations} $\rho^{(2)}(y_1,y_2,Y,\Xi)\equiv
d^2 n/dy_1 dy_2$.
\begin{equation}
\rho^{(2)}(y_1,y_2,Y,\Xi)=b(y_1,y_2,Y,\Xi)+\int_{y_{>}}^Y
dZ\int_{-y_{<}}^{\Xi}
d\zeta I_0(2a\sqrt{(Y-Z)(\Xi-\zeta)}) b(y_1,y_2,Z,\zeta),
\end{equation}
where
%\begin{eqnarray}
%\lefteqn{b=\rho^{(1)}(y_1,Y,\Xi)\rho^{(1)}(y_2,Y,\Xi) + } \\
% & &{a^2(\Xi+y_{<})\over |y_1-y_2|}
%   I_2(2a\sqrt{|y_1-y_2|(\Xi+y_{<})}),\\
%\rho^{(1)}(y,Y,\Xi)&=&a\sqrt{(\Xi+y)\over(Y-y)}I_1(2a\sqrt{(Y-y)(\Xi+y)})
%\end{eqnarray}
\begin{equation}
b=\rho^{(1)}(y_1,Y,\Xi)\rho^{(1)}(y_2,Y,\Xi) +
{a^2(\Xi+y_{<})\over |y_1-y_2|}
   I_2(2a\sqrt{|y_1-y_2|(\Xi+y_{<})}),
\end{equation}
\begin{equation}
\rho^{(1)}(y,Y,\Xi) = a\sqrt{(\Xi+y)\over(Y-y)}I_1(2a\sqrt{(Y-y)(\Xi+y)})
\end{equation}
and $y_{>(<)}$ denotes the greater (smaller) one
of the two "rapidities" $y_1,y_2$.
\remark{(notes-II.12.5)}
Some of these expressions can be further simplified, for example
the explicit forms of the total multiplicity and of the
single parton densities
quoted in points ( 2 - 4 )  are well known \cite{QCD}.

Equations (\ref{efen} - \ref{iro1} ) can be derived,
either as the resolvent solutions of the
appropriate
integral equations, or as the pole approximation of Eq.(\ref{res}).
For example, solution (\ref{res}) of Eq.(\ref{ro1}) for the single parton
density reads
\begin{equation}
\rho^{(1)}_P(\vec{k})={\cal M}_P(\vec{k})+\int d^3 K R_P(\vec{K},\Theta_{Kk})
 {\cal M}_K(\vec{k}).  \labl{enfres}
\end{equation}
In this case the minimal opening angle of a $(K,\sigma)$ jet
 $\sigma=\Theta_{Kk}$.
The integral is saturated by the $\Theta_{Kk}\sim 0$ pole of the inner
bremsstrahlung. Approximating $\Theta_{PK|\Theta_{Kk}=0}\simeq\Theta_{Pk}$
at the pole, and changing variables gives Eq.(\ref{iro1}).

\subsubsection{Resolvents for running $\alpha_s$}

The forward master equation for the resolvent reads in this case
\begin{equation}
R\left({P\over K},{\vartheta\over\sigma},{K\sigma\over\Lambda}\right)=
a^2(K\vartheta)+\int_K^P{dQ\over Q}\int_{\sigma}^{\vartheta}{d\Psi\over\Psi}
a^2\left({Q\vartheta\over\Lambda}\right) R\left({Q\over K},{\Psi\over\sigma},
{K\sigma\over\Lambda}\right). \labl{req1}
\end{equation}
There are two novel features: the coupling constant $a^2$ depends on the
appropriate transverse momentum, and the resolvent
depends on the {\em three}
dimensionless
ratios as opposed to the constant \al case. In the logarithmic variables
 $x'=\ln{(P/K)},t'=\ln{(\vartheta/\sigma)},\lambda^{\prime}
=\ln{(K\sigma/\Lambda)}$
Eq.(\ref{req1}) takes the form
\begin{equation}
R(x',t',\lambda^{\prime})=a^2(t'+\lambda^{\prime})+\int_0^{x'} dz \int_0^{t'}
d\tau a^2(z+t'+\lambda^{\prime})
R(z,\tau,\lambda^{\prime}). \labl{req2}
\end{equation}
It will prove useful to introduce the integral over the polar angle
$\overline{R}(x',t',\lambda^{\prime})=
\int_0^{t'} d\tau R(x',\tau,\lambda^{\prime})$
which satisfies
the equation
\begin{equation}
\overline{R}(x',t',\lambda^{\prime})=\beta^2\ln{t'+
\lambda^{\prime}\over\lambda^{\prime}}
+ \int_0^{x'} dz
\int_0^{t'} d\tau' {\beta^2\over z+\tau'+\lambda'}\overline{R}(z,\tau',
\lambda^{\prime}).
\labl{rqe3}
\end{equation}
This is equivalent with the integral equation for the inclusive momentum
distribution of partons (\ref{rhoeq}).
Therefore we conclude that
\begin{equation}
\overline{R}(x',t',\lambda^{\prime})= \rho(x',t',\lambda^{\prime}).
 \labl{rder}
\end{equation}
\remark{(notes - III.I.13)}
It is worth to emphasize the different physical meaning of the parameters
$\lambda$  and $\lambda^{\prime}$.
While in $\rho$:  $\lambda=
\ln{(Q_0/\Lambda)}$, on the LHS:
$\lambda^{\prime}=\ln{(K\sigma/\Lambda)}$. This is
a natural and important generalization from the elementary partons to the
virtual jets.
While for partons $Q_0$ is the fixed cut-off, for jets its role
is taken by the lower scale (virtuality) from which we are following
their perturbative evolution.

To summarize, we find the resolvent $R(x,t,\lambda)$ by solving
Eq.(\ref{rhoeq})
for the inclusive momentum distribution of partons with the appropriate
reinterpretation of variables. This will be further discussed in the next
section.

Once the resolvent $R(x',t',\lambda^{\prime})$ is known, any observable can be
directly calculated from the representations analogous to Eqs.(\ref{efen}-
\ref{iro1}) also in the running \al case.

\subsubsection{Saddle point solution of the generic equation}

We are now ready to solve the generic equation (\ref{gen}).
Only class b) be will be considered. First, observe that the equation
\begin{equation}
h_n(\delta,\theta,P)=d_n(\delta,P)+\int_{Q_0/\delta}^{P} \frac{dK}{K}
\int_{\delta}^{\theta}
\frac{d \Psi}{\Psi} a^2 (K\Psi) h_n(\delta,\Psi,K),  \labl{genn2}
\end{equation}
is identical with the the integral equation for the energy distribution
\begin{equation}
\rho(k,P,\Theta)=b(k,\Theta)+\int_{k}^{P} \frac{dK}{K}
\int_{Q_0/k}^{\Theta}
\frac{d \Psi}{\Psi} a^2 (K\Psi) \rho(k,K,\Psi),  \labl{rhoen}
\end{equation}
if we replace in (\ref{rhoen})
\begin{equation}
k\rightarrow\delta, P\rightarrow\vartheta, \Theta\rightarrow P,
\Psi \rightarrow K, K\rightarrow \Psi,b\rightarrow d_n. \labl{subst}
\end{equation}
\remark{(III.II.R1(1) )}
Together with Eq.(\ref{enfres}) this gives the following resolvent
solution of Eq.(\ref{genn2})
\begin{equation}
h_n(\delta,\theta,P)=d_n(\delta,P)+\int_{Q_0/\delta}^{P} \frac{dK}{K}
\int_{\delta}^{\theta}
\frac{d \Psi}{\Psi} \overline{R}\left({\vartheta\over\Psi}
,{P\over K}, {K\Psi\over\Lambda} \right)
{ \partial\over \partial ln{K} } d_n(\delta,K).
 \labl{hnres}
\end{equation}
\remark{(III.II.R1(4) )}
The leading contribution is yet simpler (see Appendix C for the details)
\begin{equation}
h_n(\delta,\theta,P)=d_n\left({P\delta\over\Lambda}\right)+
\int_{Q_0/\delta}^{P} \frac{dK}{K}
\overline{R}\left({\vartheta\over\delta}
,{P\over K}, {K\delta\over\Lambda} \right)
d_n\left({K\delta\over\Lambda}\right).
 \labl{hnrsim}
\end{equation}
Using (\ref{momsol}) for the energy moments we get from the
inverse Laplace transform, Eq.(\ref{invlap}) and from Eq.(\ref{rder})
\begin{eqnarray}
\lefteqn{h_n(\delta,\theta,P)=} \nonumber  \\  & &
{1\over 2\pi i} \int_{\gamma} ds e^{s\ln{\vartheta/\delta}}
f(s,\ln{P\vartheta/\Lambda})
\int_{Q_0/\delta}^{P} \frac{dK}{K} {\cal A}(s,\ln{K\delta/\Lambda})
d_n\left(\ln{(K\delta/\Lambda)}\right),
 \labl{hnrhyp}
\end{eqnarray}
with ${\cal A}(s,\lambda)=\Gamma(\alpha)\Phi(\alpha-1,1,s\lambda)$,
$f(s,X)=sX e^{-sX} F(\alpha,2,sX)$  and $\alpha$ as
in Eq. (\ref{momsol}).
\footnote{The contribution from the second term in Eq.(\ref{momsol}) is
negligible for large $X$.}.
Performing both integrals in the
saddle point approximation
%and neglecting all preexponential factors
gives our final result (see Appendix A) \remark{(III.II.F4-F.7)}
\begin{equation}
h_n(\delta,\vartheta,P) = {\cal P}_n {\cal D}_n
\exp{\left(2\beta\sqrt{\ln{(P\vartheta/\Lambda)}}\omega(\epsilon,n)\right)},
\labl{finsol}
\end{equation}
where
\begin{equation}
\epsilon={\ln{(\vartheta/\delta)}\over \ln{(P\vartheta/\Lambda)}
}. \labl{epsi}
\end{equation}
The universal function
$\omega(\epsilon,n)$ is given by
\begin{eqnarray}
\omega(\epsilon,n)&=&\gamma(z(\epsilon,n))+\epsilon
z(\epsilon,n),\labl{ome}\\ \gamma(z)&=&{1\over 2} (\sqrt{z^2+4}-
z),\labl{gamaofz}
\end{eqnarray}
and $z(\epsilon,n)$ is the
solution of the following algebraic equation
\begin{equation}
\gamma^2(z)-
2\ln{\gamma(z)} - \epsilon z^2 = n^2 - 2\ln{n}. \labl{alg}
\end{equation}
%Coefficients $f$ and $F_n$ are given by Eqs.(\ref{ffflll}) and (\ref{fqas})
%respectively.
Equations (\ref{ome}-\ref{alg}) completely
determine the scaling function $\omega$, in particular its
behaviour around the end points of the  kinematic range
 can be readily obtained
\begin{eqnarray}
\omega(\epsilon,n)\simeq & n-{n^2-1\over 2 n} \epsilon, & \epsilon \sim 0,
\labl{olin} \\
\omega(\epsilon,n)\simeq & \sqrt{1-\epsilon}
\sqrt{\ln{1\over 1-\epsilon} +n^2-2-2\ln{n}}, & \epsilon \sim 1.
\end{eqnarray}
In Fig.5 we show the results for $\omega(\epsilon,n)/n$
with $n=2,3,4$. They approach a finite limit for large n, $\epsilon$ fixed,
see Eq.(\ref{omn}) below.
 The normalization of (\ref{finsol}) can be conveniently split
into two factors. The first one is independent of the inhomogenous
term
\begin{equation}
{\cal P}_n  =
%{a(P\vartheta) F_n \over 2\pi} &\sqrt{z\over 1-n^2}
%\left({f\over 2\sqrt{a(P\vartheta)}} \right)^n &
% \left(1+{4\over z^{2}}\right)^{-{1\over 4}}
%\left( {\gamma^2(z)-1 \over 2\gamma(z)+z}
%+\epsilon z \right)^{-{1\over 2}}
\sqrt{n^2\over n^2-1}
\sqrt{ z^2 \over \gamma(z)^2-1 + \epsilon z (2\gamma(z)+z) },  \labl{presol}
\end{equation}
while the second one ${\cal D}_n$ represents the normalization of the
inhomogenous term $d_n$
 ,
hence it depends on the considered process.
It can be readily calculated for the cases of interest
(cf. Chapter 6 and Appendix A).
At fixed $\epsilon$ both factors provide a nonleading (of the relative
order $1/\sqrt{\ln{(P\theta/\Lambda)}}$)  corrections to the
asymptotic scaling  function $\omega(\epsilon,n)$.
Since ${\cal P}_n(0)=1$ and it is regular at $\epsilon=0$ it
remains close to one for small $\epsilon$.
${\cal P}_2$ rises up to 1.6 for $\epsilon=0.5$ and
${\cal P}_n < {\cal P}_{n-1}$.

%can be also calculated once the
%normalization of the ingomogenous term $d_n$ is known.

\subsection{WKB solution of the corresponding \newline
differential equation}
\subsubsection{The case of running $\as$}

An alternative method to solve the generic equation
(\ref{gen}) consists in solving the corresponding differential
equation (for $p_T=K\Psi$)
\begin{equation}
\frac{\pa h}{\pa u\pa v} - a^2 h =0
\labl{gendif}
\end{equation}
in the logarithmic variables $u=\ln (P\delta/\Lambda)$
and $v=\ln
(\theta/\delta)$ with boundary conditions along $u=\ln(Q_0/\Lambda)$
and $v=0$.
Whereas we obtained from
 the resolvent method the exact asymptotic
solution,  the WKB solution yields easily  power expansions
and nonasymptotic corrections.

As we are interested in the exponential (power) behaviour and corrections
to it, we write a differential equation for the exponent and study
its expansion in appropriate variables (generalized WKB-method)
\begin{equation}
S(u,v)=\ln h(u,v)
\labl{sdef}
\end{equation}
\begin{equation}
\frac{\delta^2 S}{\delta u \delta v}+\frac{\delta S}{\delta u}
\frac{\delta S}{\delta v} -a^2(u,v)=0
\labl{wkb}
\end{equation}
with $a^2=\beta^2 /(u+v)$.
The boundary conditions are obtained from the inhomogenous term
(\ref{gendir}) as
\begin{equation}
S(u,0)=2n\beta\sqrt{u}
\labl{boundr}
\end{equation}
In our high energy approximation  the second boundary conditons
for $u=\ln(Q_0/\Lambda)$ will be satisfied only approximately.

In order to find convenient
expansion variables for the solution of
(\ref{wkb}) with (\ref{boundr}) we perform
the first few iteration steps of the generic equation
(\ref{gen}) with (\ref{gendir}). In the transformed variables
$X=2n\beta\sqrt{u} =2n\beta\ln^{1/2} (P\delta/\Lambda)$
and $v=\ln(\theta/\delta)$ and after expansion of
$a^2(K\Psi)$ this can be written in the form
\begin{equation}
h(X,v)=e^X+2\beta^2\int\limits^X_{X_0}\frac{dx}{x}
\int\limits^v_0\sum^\infty_{m=0} c_m\left(\frac{v'}{x^2}\right)^m
h(x,v') dv'
\labl{genx}
\end{equation}
Starting with $h_0 (X,v)=\exp X$ one finds for the first iteration
after successive partial integrations in $X$
\begin{eqnarray}
h_1(X,v) &=&e^X\frac{v}{X}(c_{00}+c_{10}\frac{v}{X^2} +
       c_{20}\frac{v^2}{X^4}+\ldots) \nonumber\\
         &+&e^X\frac{v}{X^2} (c_{01} + c_{11}\frac{v}{X^2}+
c_{21}\frac{v^2}{X^4}+\ldots)+\ldots\labl{h1}
\end{eqnarray}

Here we neglected the contribution from the lower limit
$X_0$ of the integral, i.e.\ we require $u$ (or the
momentum $P$) sufficiently large.
In every higher iteration step for $h_n$ the factor
$v/X =X\cdot (v/X^2)$ multiplies $h_{n-1}$.
The expressions for $h_n$ like (\ref{h1})
suggest a double expansion of $h$  in $1/\beta\sqrt{u}$
and $v/u$. Instead of calculating the coefficients directly
from the iteration we expand the eikonal $S$ in these variables
and determine the coefficients from the differential equation (\ref{wkb}).
However  $v/u$ is not a good expansion variable as it can vary between
0 and $\infty$. Therefore it
is more convenient, but equivalent,
to replace $u\to u+v $  and to choose as
expansion parameters
\begin{equation}
\varepsilon =\frac{v}{u+v}\equiv
\frac{\ln(\theta/\delta)}{\ln (P\theta/\Lambda)} \qquad
\zeta =\frac{1}{\beta\sqrt{u+v}}\equiv
\frac{1}{\beta\sqrt{\ln (P\theta/\Lambda)}}\/,
\labl{epszet}
\end{equation}
with $0\leq\varepsilon\leq 1$. This choice also leads to a full
cancellation of the $a^2(u,v)$ term in Eq.(\ref{wkb})
in leading order of
$\zeta^2\sim (u+v)^{-1}$ and $\ve$
and to a rapid convergence of the expansion
in the region $\ve < 1/2$. Therefore we write
\begin{eqnarray}
S(\ve,\zeta) &=&\frac{1}{\zeta}\sum^\infty_{j=0}\sum^\infty_{k=0}
c_{kj} \epsilon^k\zeta^j     \labl{sexpa}\\
 &\equiv & \frac{2}{\zeta}\sum^\infty_{j=0} \om_j(\epsilon)
    \zeta^j    \labl{somex}
\end{eqnarray}
The coefficients $c_{kj}$ can be obtained successively from (\ref{wkb});
the boundary condition (\ref{boundr}) implies $\om_0(0)=n$.
Whereas the expansion
of $h$ in (\ref{h1}) would involve also negative
powers of $\zeta$ the limitation $j\geq 0$
in (\ref{somex}) is required by the boundary condition for small $u$.
As $u$ itself should be sufficiently large we consider the limit
$u/v \ll 1$ or $\epsilon\approx 1$. Consistency with (\ref{boundr})
can only be obtained for nonnegative $j$. In the high energy limit
$\zeta \to 0$ with $\epsilon$ fixed we obtain
\begin{equation}
S(u,v)\to 2\om_0(\ve)/\zeta  \labl{slim}
\end{equation}
Note that this limiting form agrees with the result from the resolvent
method (\ref{finsol}). Inserting the expansion (\ref{somex})
into (\ref{wkb}) we can also obtain differential
equations for $\om_j(\ve)$.

In leading order $O(\zeta^2)$ the double differential
first term in (\ref{wkb})
drops and one finds for $\om(\ve)\equiv\om_0(\ve)$
the WKB equation
\begin{equation}
(\om -2\ve\om')(\om +2(1-\ve)\om')=1   \labl{omdif}
\end{equation}
which easily yields the Maclaurin-expansion  for $\omega$
\begin{equation}
\omega(\epsilon,n) = n - \frac{1}{2} \frac{n^2-1}{n}\epsilon
{}~\bigl(1 + \frac{n^2+1}{4n^2} \epsilon + \frac{1}{8n^4}
(n^4 + \frac{4}{3} n^2 + \frac{5}{3}) \epsilon^2 + \cdots\bigr).
\labl{ommcl}
\end{equation}
The differential equation, Eq.(\ref{omdif}) together with the boundary
condition, corresponds to our algebraic result Eq.(\ref{ome}-\ref{alg}).
As a consistency check we have reproduced the first four terms of the expansion
(\ref{ommcl}) starting from the definition (\ref{ome}-\ref{alg}).

For the terms nonleading in $\zeta$ we obtain in lowest $\epsilon$ orders
\begin{eqnarray}
\omega_1(\epsilon,n)&=&\frac{1}{4n^2}\epsilon -
\frac{n^2-3}{8n^4}\epsilon^2+\cdots,
\nonumber\\
\omega_2(\epsilon,n)&=&\frac{1}{4n^3}\epsilon+\cdots,\qquad
\omega_3(\epsilon,n)=\frac{3}{8n^4}\epsilon+\cdots.
\labl{omk}
\end{eqnarray}
Alternatively we may solve Eq.(\ref{omdif}) in an expansion
in $1/n$; note that Eq.(\ref{ommcl}) suggests $\om(\ve, n)
=nf(\ve)+g(\ve)/n$. The leading term $nf(\epsilon)$ is
found as $n\sqrt{1-\ve}$
and solves Eq.(\ref{omdif}) if the r.h.s. is neglected.
Then one obtains for $g(\ve)$
\begin{equation}
g'(\ve)+g(\ve)/(2-2\ve)-1/(2\sqrt{1-\ve})=0  \labl{geq}
\end{equation}
with $g(0)=0$. Finally one finds
\begin{equation}
\om(\ve, n)=n\sqrt{1-\ve}~(1-\frac{1}{2n^2}\ln (1-\ve))+\cdots.
\labl{omn}
\end{equation}
Remarkably, this approximation has an accuracy of better
than 1\% for $\ve <0.95$ already for $n=2$ and is therefore
well suited for numerical analysis.

The same type of asymptotic behaviour (\ref{slim}) was found for
the correlations studied
in \cite{DMO}, their effective power $\nu_{eff}$ is related to our $\omega$
by $\nu_{eff}=(n-\omega(\epsilon,n))/(1-\sqrt{1-\epsilon})$.

So far we have analysed the generic equation (\ref{gen})
with $p_T=K\Psi$ which applies exactly in this way to the 2-particle
correlation $\rho^{(2)}(\theta_{12})$. For the cumulant
correlation $\Gamma^{(n)}(\{\Omega_i\})$ the same equation holds
but with $p_T=K\theta$;
also the inhomogeneous term obtains a variable prefactor
\begin{equation}
\Delta (u,v)\sim (u+v)^{-1} u^\alpha exp(2n\beta\sqrt{u})
      \labl{delmod}
\end{equation}
Taking into account the mixed scale $K\theta$ and writing $\Gamma =
\exp S$, $\Delta =\exp S'$ we arrive at the modified differential
equation
\begin{eqnarray}
\frac{\pa S}{\pa u}\frac{\pa S}{\pa v}&-&\frac{\beta^2}{u+v}+
\frac{\pa S}{\pa u\pa v}+\frac{1}{u+v}\frac{\pa S}{\pa u}\nonumber \\
&=&e^{S-S'}\left(\frac{\pa S'}{\pa u\pa v}+\frac{\pa S'}{\pa u}
     \frac{\pa S'}{\pa v}+\frac{1}{u+v}\frac{\pa\Delta}{\pa u}\right)
\labl{sspr}
\end{eqnarray}
The form (\ref{delmod}) for $\Delta$ and (\ref{slim},\ref{omn})
for $S$ implies $S-S'\sim -b^2(\ve)\sqrt{u+v}$ for large
$u+v$, therefore an exponential decrease of the r.h.s. of
(\ref{sspr}). In leading order of $\zeta^2=(u+v)^{-1}$
only the first two terms of the left-hand-side contribute
as before, therefore one obtains the same asymptotic solution (\ref{slim}),
and differences appear only at the nonleading level.

\subsubsection{Results for fixed $\as$}
We start from the generic equation (\ref{gen}) with constant $a$ but
with inhomogeneous term
\begin{equation}
d\sim \exp \left(na\ln(P\delta/Q_0)\right)
 \labl{gendirf}
\end{equation}
The approriate expansion variables are now $u=\ln(P\delta/Q_0)$
and $v=\ln(\theta/\delta)$ as before, also $\epsilon=v/(u+v)$.
As above we arrive at the same differential equation (\ref{wkb})
but now with boundary condition
\begin{equation}
S(u,0)=nau.
\labl{boundf}
\end{equation}
Similar reasoning as above leads to the ansatz
$S=a(u+v)\om(\ve,n)$ where $\om(\ve,n)$
satisfies a differential equation as in (\ref{omdif}) but without the
factors 2. The solution is found as $\om(\ve,n)=n-(n-\frac{1}{n})\ve$,
so finally
\begin{eqnarray}
S &=& an(u+v)\left(1-(1-n^{-2})\ve\right) \labl{sfal}\\
  &=& a\left(nu + n^{-1} v\right) \labl{sfal1}
\end{eqnarray}
As is easily seen this linear function in $\epsilon$ is actually
the exact solution of
Eq.(\ref{wkb}) with (\ref{boundf}), i.e.\ of the generic
integral equation (\ref{gen}) where the contribution of the lower
limit from the $K$ integral is neglected. So the running of
the coupling is most directly seen in the nonlinear behaviour
of the $\omega(\epsilon,n)$ function.
\section{From integrated to differential correlations}
        \rem1{pow5v10}
\subsection{Multiplicity correlator $F_q$}
\rem1{pow5v10, 15.XII.94}

The results in this subsection are needed later for consistency
checks of our differential results; they are therefore rederived
using our methods although they are largely known already (see \cite{QCD},
for example).

The second factorial moment, or the
total number of pairs in a jet,
provides the normalization of the two parton densities
\begin{equation}
f_2 = < n(n-1) > = \int d^3 k_1 d^3 k_2
\rho_{P}^{(2)} (\vec{k_1},\vec{k_2})  \labl{f2def}
\end{equation}
Its integral equation follows from Eq.(\ref{mero}) for the two parton
differential density, $X=\ln{(P\Theta/Q_0 )},
X_{\la}=X+\la=\ln{(P\Theta/\Lambda)} $
\begin{equation}
 f_2(X_{\la})=g_2(X_{\la}) + \int_0^X d v \int_0^v d u a^2(u+\lambda)
 f_2(u+\la).  \labl{f2eq}
\end{equation}
This differs from Eq.(\ref{ntoteq}) for the total multiplicity only
by the form of the
inhomogenous term
%\begin{equation}
$
g_2 =
 \int d^3 k_1 d^3 k_2
d_{P}^{(2)} (\vec{k_1},\vec{k_2})
%\labl{g2def}
$
%\end{equation}
with $d=d_{prod}+d_{nest}$ given by (\ref{prod1},\ref{nest}).
For constant \al
one gets explicitly
%\begin{equation}
$g_2(X) = \overline{n}^2 + 2 \overline{n} - a X^2 $
%\end{equation}
and the solution of Eq.(\ref{f2eq}) reads
\begin{equation}
f_2(X)={4\over 3} ( {\rm cosh} (a X) -1 )^2 \simeq {1\over 3}
\left ( {P\Theta\over Q_0} \right )^{2a}  \labl{f2sol}.
\end{equation}
The normalized correlator $F_2=f_2/\bar n^2$ is then found from
(\ref{coshhh}) as $F_2={4\over 3}$ \cite{QCD}.

Using the resolvent representation Eq.(\ref{fnres}) one can
easily show that for constant \al
\begin{equation}
f_q(X)=F_q <n(X)>^q \;\;\;X\rightarrow\infty,  \labl{fqas}
%f_n(X)=F_n {e^{naX}\over 2^n}, \;\;\;X\rightarrow\infty,  \labl{fqas}
\end{equation}
with $<n>=\exp(aX)/2$ and the coefficients
$F_q$ are given by the recursive relation
%and the coefficients $F_n$ are given by the recursive relation
(see Appendix D for the details)
\begin{equation}
F_q={q^2\over q^2-1}\sum_{k=1}^{q-1} {q-1\choose k }
 {F_k F_{q-k}\over (q-k)^2},
\;\;\; q>1,\;\;\;F_1=1.  \labl{recur}
\end{equation}
 Equation (\ref{fqas}) generalizes to the case of running \al.
%For running \al Eq.(\ref{fqas}) is generalized to
%\begin{equation}
%\lim_{X\rightarrow\infty} {f_n(X)\over <n>^n} = F_n, \labl{fqrun}
%\end{equation}
%with the pure algebraic coefficients $F_n$ independent of the coupling.
%Also the recursive relation identical to Eq.(\ref{recur}) was derived
%\cite{QCD}.
%It is worth pointing out that
In fact, the {\em independence} of
the coefficients $F_q$ in (\ref{fqas}) of the running coupling
(or more precisely of $\la$ and $\beta$) is sufficient
to assert that the recursion (\ref{recur}), derived for constant \al ,
must also hold for the running \al. Otherwise the usual correspondence
between the constant and running coupling results would be violated.

%For running \al we have the following integral representation of
%the solution
%\begin{equation}
%asymptotics for running alpha
%\end{equation}
%We shall frequently refer to these results as the cross check of the
%formulae for more differential distributions.

\subsection{Correlations in the relative angle}
\subsubsection{Definitions and integral equation}
The distribution of the relative
angle $\t12$
between two partons
in a jet with momentum $P$  inside a cone of
half opening angle
$\Theta$ is defined as follows
(see Fig.6)
\begin{equation}
\rho^{(2)}(\theta_{12},P,\Theta)
=\int \rho_{P}^{(2)}(k_1,k_2) \delta(\Theta_{k_1k_2}-\theta_{12}
)d^3 k_1 d^3 k_2.
\labl{the12}
\end{equation}
This quantity shares the simplicity (small number of variables)
of global observables and, at the same time, provides the first nontrivial
information about the differential structure of the correlations.
We also consider two normalized densities. Firstly,
\begin{equation}
r(\theta_{12})={\rho^{(2)} (\theta_{12}, P, \Theta)  \over
          d_{prod}^{(2)} (\theta_{12}, P, \Theta) }
\labl{r12}
\end{equation}
where the normalizing quantity in the denominator
is defined as in (\ref{the12}) but
with $\rho^{(2)}_P(k_1,k_2)$ under the integral replaced by
$\rho^{(1)}_P(k_1)\rho^{1)}_P(k_2)$; then, for an uncorrelated distribution
$\rho\upt$, the normalised correlation is $r(\t12)=1$
\footnote{This type of normalisation is also suggested in \cite{CORRI}
("correlation integrals"). Experimentally this normalisation is
obtained conveniently by "event mixing" as is often done for the determination
of Bose-Einstein correlations.}.

Secondly we define
\begin{equation}
\hat r(\theta_{12})={\rho^{(2)} (\theta_{12}, P, \Theta)  \over
          \bar n^2(P, \Theta) }
\labl{r12hat}
\end{equation}
with the full multiplicity in the forward cone as normalizing
factor. The importance of these different normalizations will
become clear in Sect. 7 where $\hat r$ is found to exhibit better
scaling properties.

The correlation function is defined in (\ref{the12})
as an integral over the
full forward cone, therefore it matters in general whether
the leading particle is included or not. We will derive first
the correlation without leading particle $\rho(\t12)$, the one
with leading particle $\tilde\rho(\t12)$ is then obtained from
Eqs. (\ref{robb2},\ref{the12}) as
\begin{equation}
\tilde\rho\upt_P(\t12)=\rho\upt_P(\t12)+2\rho\upo_P(\t12).
\labl{rho12new}
\end{equation}
Whereas these two cases are clearly separated in the theoretical analysis
it is not so straightforward to realise experimentally the case with the
leading particle subtracted. One could think of a flavour tag, for example
subtracting the charmed particle in a charm quark jet.

For clarity we consider first the gluon jet and come back
to the quark jet at the end of this subsection.

The integral equation for $\rho\upt_P(\t12)$ is obtained by
 integrating Eq.(\ref{mero}) over the phase space of final partons
 at fixed
 $\theta_{12}$
\begin{equation}
\rho^{(2)}(\theta_{12},P,\Theta)=d\upt(\theta_{12},P,\Theta)
     + \int_{Q_0/\theta_{12}}^{P}
\frac{dK}{K} \int_{\theta_{12}}^{\Theta} \frac{d\Psi}{\Psi}
  a^2(K\Psi) \rho\upt(\theta_{12},K,\Psi).
\labl{ang2}
\end{equation}
The derivation of this equation is similar to that of
Eq.(\ref{tkk}). Now however the polar angle $\Psi=\Theta_{P K}$
and
as a consequence the relative angle $\theta_{12}$ determines the lower
bounds
for $\Psi$ and $K$ integrations in accordance with general rules
discussed in Sect. 2.4.1.
 Namely, the leading logarithms always emerge from
$\vec{K}\parallel \vec{k}_1 (\vec{k}_2)$. In this configuration the minimal
virtuality of a jet $K$, which subsequently emits a pair $(k_1,k_2)$
with
given angular separation, is controlled by the relative angle
$\theta_{12}$.
Note that the individual directions $\hat{n}_{k_1}, \hat{n}_{k_2}$
are
already integrated over, therefore the ``most narrow'' singularities
$\Theta_{Kk_1}, \Theta_{Kk_2} \sim Q_0/K$ are included in $\rho^{(2)}
(\theta_{12},\Psi,K)$. One can say that the relative angle $\theta_{12}$
is the
only remaining scale controling the singularities of the integrand in
Eq.(\ref{ang2}). On the contrary, for more differential distributions
the boundaries depend in general on other different scales entering the
problem.

The direct term in Eq.(\ref{ang2}) is defined in analogy to (\ref{the12})
\begin{equation}
d\upt(\theta_{12},P,\Theta)=\int d_{P}^{(2)}(k_1,k_2)
\delta(\Theta_{k_1k_2}-\theta_{12}
)d^3 k_1 d^3 k_2.
\labl{dirterm}
\end{equation}
where $d^{(2)}_{P}(k_1,k_2)$ is given explicitly by the
inhomogenous part of
Eq.(\ref{mero})
and can be written as
$d^{(2)}=d_{prod}\upt+d_{nest}\upt$
as discussed in Sect. 2.2.
%After the the momentum integrals in Eq.(\ref{dirterm}) being carried out
%the remaining angular integrals yield
%the direct term
%can be represented in terms of single particle angular
%distributions $\rho^{(1)}_P(\theta_i)$.
%\begin{equation}
%\rho^{(1)}_P(\theta_i)={a\over \theta_i}
%     \rm{sinh}(a\ln(\theta_i/\kappa)) ~(\kappa=Q_0/P,~
%\theta_i \equiv \Theta_{Pk_{i}}).
%\labl{single}
%\end{equation}
They can be derived from the angular distribution $\rho\upo_P(\th)$
\begin{eqnarray}
   d\upt_{prod}(\theta_{12},P,\Theta)&\simeq&
   \rho\upo_P(\t12)\int_{\kappa}^{\t12} d\theta_1\rho\upo_P(\theta_1)
   +(1\leftrightarrow 2),   \labl{dprodpa}\\
   d\upt_{nest}(\theta_{12},P,\Theta)&=&
      \int_{Q_0/\t12}^P {dk_1\over k_1} \rho_{k_1}\upo(\t12)
      \int_{\t12}^{\Theta}{d\theta_1 \over \theta_1} a^2(k_1\theta_1),
    \labl{dnestgen}
\end{eqnarray}
where $\kappa=Q_0/P$.
The integration of
           the product term over the angles has been
done in the pole approximation as explained in Sect. 2.4.2,
where the first term corresponds to the pole for $\theta_1=0$, $\theta_2
\approx \t12$.

\subsubsection{Exact results for constant $\alpha_s$}
For constant \al the solution of Eq.(\ref{ang2}) can be written as
\begin{equation}
   \rho^{(2)}(\theta_{12},P,\Theta)=d\upt(\theta_{12},P,\Theta)+
a^2\int_{Q_0/\theta_{12}}^{P} \frac{dK}{K} \int_{\theta_{12}}^{\Theta}
 \frac{d\Psi}{\Psi} I(\frac{P}{K},\frac{\Theta}{\Psi})
 d\upt(\theta_{12},K,\Psi),
\labl{sol12}
\end{equation}
with the resolvent
\begin{equation}
 I(x,y)= I_0(2a\sqrt{ln(x)ln(y)}). \labl{resI}
\end{equation}
If the leading particle is not included the direct term
%$d^{(2)}=d_{prod}\upt+d_{nest}\upt$
is easily calculated from Eqs. (\ref{dprodpa},\ref{dnestgen})
%The nested term can be integrated analytically
with $\rho\upo_P(\th)$ from (\ref{asyan2})
\begin{eqnarray}
   d\upt_{prod}(\theta_{12},P,\Theta)&\simeq&\frac{2a}{\theta_{12}}
   \rm{sinh}(a\ln
       \frac{\theta_{12}}{\kappa})
       \{\rm{cosh}(a\ln\frac{\theta_{12}}{\kappa})-1\}.
\labl{rhoprod}\\
%\end{equation}
%\begin{equation}
  d\upt_{nest}(\theta_{12},P,\Theta)&=&\frac{2a^2}{\theta_{12}}
       \{\rm{cosh}(a\ln\frac{\theta_{12}}{\kappa})-1\}
       \ln\frac{\Theta}{\theta_{12}}.
\labl{rhonest}
\end{eqnarray}
Note that integrating the direct term $d\upt_{prod}$ over the
whole range of
the relative angles gives the square of the total multiplicity in a cone
\begin{equation}
\int_{\kappa}^\Theta d \theta_{12} d\upt_{prod}(\theta_{12},
P,\Theta) = \left(\rm{cosh}(a\ln{(P\Theta/Q_0)})-1 \right)^2 =
\overline{n}(P\Theta)^2,
\labl{nbsq}
\end{equation}
as it should. This shows that the pole approximation
 correctly identifies all
leading logarithms also in this case.

With the direct term given by Eqs.(\ref{rhoprod}) and (\ref{rhonest})
the integral in Eq.(\ref{sol12}) can be done analytically (see Appendix E)
and our final result without leading particle reads
\begin{equation}
\rho^{(2)} (\theta_{12}, P, \Theta)=
 \frac{a}{\theta_{12}} \sum^\infty_{m=0} y^{2m+1} I_{2m+1}(z)
 -\frac{2 a}{\theta_{12}}
\sinh{ (a\ln  \frac{\t12}{\kappa} )},
%\sinh{ (      \frac{1}{4} z y  )},
\labl{rhobess}
\end {equation}
where $y=2\sqrt{\ln{(\theta_{12}/\kappa)}/\ln{(\Theta/\theta_{12})}}$,
$z=2a\sqrt{\ln{(\theta_{12}/\kappa)} \ln{(\Theta/\theta_{12})}}$,
and $I_l(z)$ is the modified Bessel function of the order $l$.
The same result (\ref{rhobess})
can be obtained also by solving the partial
 differential equation corresponding to Eq.(\ref{ang2})
 which is separable in the variables $y$ and $z$
together with the appropriate boundary conditions.

If the leading particle is included one has to drop the (-1)
in Eqs. (\ref{rhoprod})-(\ref{nbsq}) and also the second term in
Eq.(\ref{rhobess})
as is found from relation (\ref{rho12new}).

It is easy to check the backward compatibility of these results with the
second factorial moment $f_2$ discussed in the previous Section.
Integrating Eq.(\ref{ang2}) over the relative angle one readily recovers
the integral equation (\ref{f2eq}) for $f_2$.
Also the direct integration of our solution (\ref{rhobess}) gives back
the expression (\ref{f2sol}) for the total number of pairs
(see Appendix F \rem1{"II.16.4"}).

In the high energy limit, $\kappa=Q_0/P \to 0$
with $\t12$ kept fixed, the r.h.s  of
Eq.(\ref{rhobess})  exhibits   a power behaviour
irrespective of the leading particle contribution,
(see Appendix G for the details)
\begin{equation}
\rho^{(2)}(\theta_{12} ,P,\Theta)
\simeq
\frac{a}{2\theta_{12}}\left(
 \frac{\theta_{12}}{\kappa} \right)^{2a}
  \left(\frac{\Theta}{\theta_{12}}\right)^{{a\over 2}},
\labl{selfsim}
\end{equation}
which proves a selfsimilarity property of the QCD cascade for
fixed $\alpha_s$.
Note that this simple behaviour emerges only for the well developed
cascade, i.e. for sufficiently large angles $\t12 \gg\kappa$.
%i.e. for $\theta_{12}$ much bigger than the elementary cut-off  $Q_0/P$.
Eq.(\ref{selfsim}) represents the two components
of Eq.(\ref{sol12}): the first two factors correspond to the
direct term
$(d\sim d_{prod})$, the  last factor to the
enhancement from the emissions of the intermediate parent jets.
The second term in our solution
(\ref{rhobess}) which distinguishes between including or not including
the  leading particle
is nonleading at high energies
$\sim(\t12/\kappa)^a$.
The asymptotic result (\ref{selfsim}) can be obtained also easily
by solving the integral equation (\ref{ang2}), for example by iteration,
with the asymptotic form for the direct term $d\upt \simeq d\upt_{prod} \simeq
(a/2\t12)(\t12/\kappa)^{2a}$, neglecting the nonleading
nested term $\sim (\t12/\kappa)^a$.

For the normalised density $r(\t12)$ in Eq.(\ref{r12}) the leading
power from $d\upt_{prod}$ cancels and one obtains
%The exponent in the first term $2a=4\sqrt{3\alpha_{s}/2\pi}$
% agrees with the asymptotic
%result of Gustafson and Nilsson derived with a different method and for
%different observables \cite{GN}.
%However the last factor changes this result
%by $25\%$.
%In fact the leading behaviour often cancells in the measured observables.

\begin{equation}
 r(\t12) \simeq  \left({\Theta\over\theta_{12}}\right)^{{a\over 2}}
% \;\;\;\;\; \theta_{12}/\kappa  \gg  1,
  \labl{r12as}
\end{equation}
which is sensitive only to the nonleading exponent $a/2$.
The leading exponent is not canceled for the alternative normalization
Eq.(\ref{r12hat}) in which case we find
\begin{equation}
\hat r (\t12) \simeq  {2 a \over \theta_{12}}
\left({\Theta\over\theta_{12}}\right)^{-{3a\over 2}}.
  \labl{rh12as}
\end{equation}
It is noteworthy that these results
are infrared safe as they don't
%and free of narrow divergences. It does not
depend on the cutoff $Q_0$.
%nor on the scale parameter $\Lambda$.
%   check tha int(.)=f2 (16.4.II) check that int(Eq(ro12))=Eq(f2) (ibid)
%   \newline RUNNING     \newline (III.F )

\subsubsection{High energy behaviour for running \al}
For running \al we derive the asymptotic behaviour of the density
(\ref{the12}) using methods discussed in Sect.4. The integral equation
(\ref{ang2}) is indeed one of the special cases of the generic equation
(\ref{gen}) with the substitutions
\begin{equation}
%substitutions from the generic equation, \labl{subst12}
\delta \to \t12,\qquad \theta \to \Theta, \qquad h \to \t12 \rho\upt_P
\labl{subsr12}
\end{equation}
The inhomogeneous term is calculated from the one particle
angular distribution in high energy
approximation Eq.(\ref{asyan1}) using pole dominance Eq.(\ref{dprodpa}) and
neglecting the nonleading nested term as before. We obtain
\begin{equation}
%d^{(2)}(\t12)&\simeq &{f^2\over 2\t12}
d^{(2)}(\t12) \simeq  {f^2\over 2\t12}
\exp \left(4\beta\sqrt{\ln(P\t12/\Lambda)}\right)
\labl{d2t12a}
%\\& =&{f^2 \over 2\t12} \left(P\t12/\Lambda\right)^{4a(P\t12)}
%\labl{d2t12}
\end{equation}
which has the form (\ref{gendir}) considered for the generic equation.
Hence, applying the results of Sect.4,
the high energy asymptotic behaviour is
governed by the function $\omega(\epsilon,2)$ with
the scaling variable
\begin{equation}
\epsilon= {\ln{(\Theta/\vartheta_{12})} \over
\ln{(P\Theta/\Lambda)} }
\labl{scal12}
\end{equation}
%For normalized density, Eq.(\ref{r12}), we obtain
%\begin{equation}
%r_{12}(\vartheta_{12})=e^{2\beta
%                      \sqrt{ \log{(P\Theta/\Lambda)}}
%                          \omega_{12}(\epsilon_{12})
%                         }
%\end{equation}
%with $\omega_{12}(z)=\omega(z,2)-2 \sqrt{1-z}$ and $\omega(z,n)$ given by
%Eq.(\ref{omofz}).
%In the constant \al limit
%$\epsilon_{12}\simeq {1\over \lambda}
%\log{(\Theta/\vartheta_{12})} $
%and, using the
%Maclaurin expansion
%of $\omega(\epsilon,n) \simeq n-{n^2-1 \over 2n}\epsilon$.
%one readily obtains Eq.(\ref{r12as}). Higher (up to 4) terms of
%he expansion were also checked against the results of the WKB method.
%n fact we have prooven that $\omega(\epsilon)$, as defined by the
%addle point conditions Eqs.(\ref{} and \ref{}),  satisfies the
%ifferential equation implied by (\ref{}) and (\ref{}).
%
%ndeed the RHS of Eq.(\ref{omega}) can be rewritten as
%
%\left({P\theta \over \Lambda}
%right)^{2a(P\theta) \omega(\epsilon,n)}
%$ which agrees with Eq.(\ref{F16}) to the first two orders in
%$\epsilon$
%after appropriate normalization is taken into account.
and we obtain for
the unnormalised and normalised correlation
functions
\begin{eqnarray}
\rho^{(2)}(\t12)&=&{f^2\over 2 \t12}\exp \left(2\beta\omega(\epsilon,2)
 \sqrt{\ln(P\Theta/\Lambda)}\right)
\labl{rho12ra}\\
  r(\t12)&=&\exp \left(2\beta
 \sqrt{\ln(P\Theta/\Lambda)} \left(\omega(\epsilon,2)-2\sqrt{1-\epsilon}\right)
    \right)
\labl{r12ra}
\end{eqnarray}
In  the normalized correlation $r(\t12)$
the leading term $2\sqrt{1-\epsilon}$ in the exponent
cancels in the approximation (\ref{omn})
and we find in this case
\begin{equation}
  r(\t12)\simeq \exp \left(-{\beta \over 2}
 \sqrt{\ln(P\Theta/\Lambda)} \sqrt{1-\epsilon}\ln (1-\epsilon)
    \right).
\labl{r12raa}
\end{equation}

As to the correlation $\hat r$ defined in Eq.(\ref{r12hat})
we note that it behaves like $\hat r(\t12) \sim 1/\t12$,
therefore it is more convenient to consider the distribution
differential in the logarithmic variable $\epsilon$, i.e.
$\hat r(\epsilon)=\t12\rho^{(2)}(\t12)\ln(P\Theta/\Lambda)$.
Using Eq.(\ref{ffflll}) for $\bar n$ we find

 \begin{equation}
 \hat r(\epsilon)=2\beta\sqrt{\ln(P\Theta/\Lambda)} \exp \left(2\beta
 \sqrt{\ln(P\Theta/\Lambda)} \left(\omega(\epsilon,2)-2\right)
    \right).
\labl{rh12ra}
 \end{equation}
The prefactors in Eqs.(\ref{rho12ra},\ref{r12ra}) and (\ref{rh12ra})
provide the proper normalization at $\epsilon=0$ consistent with
(\ref{d2t12a}). We stress that they are nonleading and therefore less
reliable. Nonleading terms beyond DLA are not included in
the present approach.

For sufficiently large angles $\theta_{12}$ (small
$\epsilon$) we apply the linear approximation $\omega\simeq 2-3\epsilon/4$
and obtain
\begin{equation}
\rho^{(2)}(\t12) \simeq
    {f^2 \over 2\t12} \left({P\Theta\over\Lambda}\right)^{4a(P\Theta)}
    \left({\Theta\over \t12}\right)^{-{3\over 2}a(P\Theta)};
\labl{rho2pra}
% r(\t12)&\simeq &
%   \left({\Theta\over \t12}\right)^{{1\over 2}a(P\Theta)}
%\labl{r2pra}
\end{equation}
for the normalized correlations $r$ and $\hat r$ we
recover the   power behaviour in the relative angle
as in the case of fixed $\alpha_s$, Eqs.(\ref{r12as},\ref{rh12as}), but
with exponents which depend on energy and angle through
the anomalous dimension $a(P\Theta) =\sqrt{6\alpha_s(P\Theta)/\pi}$.
The formulae  (\ref{selfsim}-\ref{rh12as}) for
constant $\alpha_s$ are recovered in the a-limit
 $\lambda, \beta\to\infty$ ($\epsilon \to 0$), see Eq.(\ref{alimit}).
%\beta\to\infty, \beta/\sqrt{\lambda}\to a$ which yields $f^2 \times
%(P\Theta/\Lambda)^{2a(P\Theta)}\to a\times (P\Theta/Q_o)^a$ and
%$a(P\Theta)\to a$.

It is worth pointing out the different scaling behaviour for the
cases of fixed and running $\alpha_s$. In the latter case a scaling
limit is approached in $\epsilon$ for the quantity
\begin{equation}
{\ln  r(\t12) \over
 \sqrt{\ln(P\Theta/\Lambda)}} = 2\beta
 \left(\omega(\epsilon,2)-2\sqrt{1-\epsilon}\right).
\labl{rascal}
\end{equation}
For fixed \al we define
\begin{equation}
\tilde{\epsilon}      = {\ln{(\Theta/\vartheta_{12})} \over
\ln{(P\Theta/Q_0)} }
\labl{epstil}
\end{equation}
i.e. replace $\Lambda$ by $Q_0$ in (\ref{scal12}); then in the limit
$\tilde \epsilon$ fixed, $P\to\infty$ we find again the result (\ref{selfsim})
or
\begin{equation}
{\ln  r(\t12) \over
 \ln(P\Theta/ Q_0)} = {1\over 2}a \tilde\epsilon.
\labl{fascal}
\end{equation}
The comparison of both cases shows (neglecting the difference between
$\epsilon$ and $\tilde\epsilon$) that first, the scaling behaviour occurs
for differently normalized quantities,
and second, the limiting functions are different:
the linear behaviour in (\ref{rascal}) (namely ${1\over 2}\beta\epsilon$)
is approached only for small $\epsilon$. Therefore the study of the quantity
$r(\t12)$ allows for some novel tests of the running of \al in multiparticle
final states and the same is true for the correlation $\hat r(\t12)$ in
complete analogy.

\subsubsection{The behaviour for small relative angles}

If the relative angle $\t12$ approaches the lower cutoff $\kappa=Q_0/P$
the DLA becomes more and more unreliable. Here we consider
the limit of small angles $  \t12\to \kappa,\quad \t12 \ll \Theta$ but
with $\t12$ still
sufficiently above the cutoff $\kappa$. In this limit the leading particle
plays an important role, in particular for the correlation $r(\t12)$
which is considered first.

For constant \al this limit in Eqs.(\ref{rhobess},\ref{rhoprod}) corresponds
 to the lowest order in the coupling $a$ , i.e. to the Born term.
Without leading particle one obtains
\begin{eqnarray}
\rho\upt_P(\t12) &\simeq& d\upt_{nest}(\t12)\simeq
   (a^4/\t12)\ln^2(\t12/\kappa)\ln(\Theta/\t12)
    \labl{nestsma}\\
    d\upt_{prod}(\t12)&\simeq& (a^4/\t12)\ln^3(\t12/\kappa).
    \labl{prodsma}
\end{eqnarray}
Both distributions vanish for $\t12\to\kappa$ but
the nested term dominates by one power in $\ln(\t12/\kappa)$.
Then the normalised correlation becomes singular in this limit
$r(\t12) \to \ln(\Theta/\kappa)/\ln(\t12/\kappa)$ or
\begin{equation}
r(\t12) \to {\kappa\ln(\Theta/\kappa) \over \t12-\kappa}
\qquad  \hbox{(without leading particle)}.
\labl{r12bo}
\end{equation}

On the other hand, if the leading particle is included
the additional term
 $2\rho\upo_P(\t12)\simeq (2a^2/\t12)\ln(\t12/\kappa)$
in (\ref{rho12new})  dominates both $\rho\upt_P(\t12)$ and
$d\upt_{prod}(\t12)$ in Eqs.(\ref{nestsma}) and (\ref{prodsma}), so
%\begin{equation}
%\rho\upt_P(\t12) \simeq
%  (2a^2/\t12)\ln(\t12/\kappa), \qquad
%   d\upt_{prod}\simeq (2a^2/\t12)\ln(\t12/\kappa)
%labl{rhosmlp}
%\end{equation}
consequently for $\t12 \to \kappa$
\begin{equation}
r(\t12) \to 1
\qquad  \hbox{(with leading particle)}.
\labl{r12bolp}
\end{equation}
i.e. the strong correlation has turned into a vanishing correlation.

In case of running \al
one finds the small angle behaviour again from  the Born approximation.
With $\rho\upo_P(\theta)\simeq
       (\beta^2/\theta)\ln(\ln(P\theta/\Lambda)/\lambda)$
the direct terms
without leading particle are given by

\begin{eqnarray}
d^{(2)}_{prod}(\t12,P,\Theta)&\simeq&
{2\beta^4\lambda\over\th}\ln z(z\ln z-z+1)\/,
\quad z={1\over\lambda}\ln {P\t12\over\Lambda} \labl{dprodsmr}\\
d^{(2)}_{nest}(\t12,P,\Theta)&\simeq&
{\ln z'\over \ln z}d^{(2)}_{prod}(\t12,P,\Theta)\/,
\quad z'={1\over\lambda}\ln{P\Theta\over\Lambda} \labl{dnestsmr}
\end{eqnarray}
This yields for the normalized correlation
\begin{equation}
r(\t12) \to {\ln z'\over \ln z}\simeq{\lambda\kappa
\ln\ln{\Theta\over\kappa}\over \t12 -\kappa}.
\labl{r12br}
\end{equation}
One observes that the Born term becomes less important at
high energies in comparison to the fixed \al case.
After inclusion of the leading particle one finds again
the dominance of the additional contribution with the result
(\ref{r12bolp}).

The correlation $\hat r(\t12) = \rho^{(2)}(\t12)/\bar n^2$
vanishes for $\t12 \to \kappa$ in any case either in first or second
order of $\ln (\t12/\kappa)$.

 \subsubsection{Comparison of various approximations for $r(\t12)$.}

In Fig.7 we compare various analytic and numerical results
on the (rescaled)  normalized correlation
$r(\theta_{12})$. First there is the exact numerical result
 with and without leading particle
at the finite energy   for $Y=\ln
P/Q_0=5$ appropriate for LEP (full curves).
It shows a linear regime for small $\epsilon\leq 0.3$.
The full result with leading particle vanishes for $\t12 \to \kappa=Q_0/P$,
i.e. for $\epsilon\to \epsilon_0=\ln(P\Theta/Q_0)/\ln(P\Theta/\Lambda)<1$,
on the other hand without leading particle the correlation rises strongly
in this kinematic limit.
These results are obtained by solving the integral equation (\ref{ang2})
numerically\footnote{
Applying Simpson's rule to the integral in (\ref{ang2}) on a grid of
values for the logarithmic variables $y_i=\ln K_i,~ \chi_k=\ln\Psi_k$
in the range  $(Q_0/\t12<K<P$;$\t12<\Psi<\Theta)$ we can express
$\rho\upt$ at each point $(y_i,\chi_k)$ by the respective
values at the points $(y_{i-1}\chi_k)$, $(y_i,\chi_{k-1})$
and $(y_{i-1},\chi_{k-1})$. Starting with $\rho\upt=d\upt$
at the lower limit of the integrals one can successively calculate $\rho\upt$
for all points on the grid, finally for $P,\Theta$.}
and with (\ref{rho12new}).
The inhomogeneous term is calculated using Eqs.(\ref{dprodpa},\ref{dnestgen})
with the exact $\rho\upo(\t12)$ from (\ref{rhoang})
where the $k_1$-integral in (\ref{dnestgen}) is derived also numerically.
With increasing energy the cutoff $\epsilon_0\to 1$ and the effect
of the leading particle vanishes for $Y\to\infty$.

For these high energies the rescaled correlation
$\ln r(\t12)/\sqrt{\ln(P\Theta/\Lambda)}$ approaches
a limit for fixed $\epsilon$
as given by Eq.(\ref{rascal}).
The linear approximation ${1\over 2}\beta\epsilon$ of this quantity
(dashed line)
represents the asymptotic result well for $\epsilon\leq 0.6$ but does not fully
match the result for $Y=5$.

The nonleading terms in $\zeta=1/(\beta \sqrt{\ln(P\Theta/\Lambda)}$
given in (\ref{omk}) provide a finite energy correction to $r(\t12)$.
In leading order of $\epsilon$ one obtains from (\ref{r12ra},\ref{omk})
the expansion
\begin{equation}
\ln r(\t12)/\sqrt{\ln(P\Theta/\Lambda)}={1\over 2}\beta \epsilon
(1+{1\over 4}\zeta + {1\over 8}\zeta^2 +{3\over 32}\zeta^3 \dots)
\labl{rzeta}
\end{equation}
The first term corresponds to the leading power $a/2$ in (\ref{r12as}),
the terms of higher order in $\zeta$ which amount to about 10\%
match better the exact results at small $\epsilon$
(upper straight line in figure).
 We conclude that
the expansion of $\ln \rho^{(2)}$ in $\zeta$
is rapidly converging for $\epsilon \leq {1 \over 2}$
and already the lowest
order term which corresponds to a running $\alpha_s$ with
scale $Q=P\Theta/\Lambda$ gives a fair approximation
for large opening angles.

\subsubsection{Correlation functions for quark jets}

In the study of $e^+e^-\to$ hadrons one is
interested in
comparison of data with predictions for quark jets. The
correlation function $r_a(\theta_{12})$ for arbitrary
initial parton $a$ can be derived from our results on gluons
using relation  Eq.(\ref{gama}).
Including the leading particle according
to Eq.(\ref{rho12new}) we obtain
\begin{eqnarray}
\tilde{\rho}^{(2)}_a(\t12) &=& c_a\rho^{(2)}(\t12)
-(c_a-c_a^2) d^{(2)}_{prod}(\t12) +2c_a\rho^{(1)}(\t12)
\labl{rhoat}\\
\tilde{d}^{(2)}_{prod,a}(\t12) &=& c_a^2d^{(2)}_{prod}(\t12)
   +2c_a \rho^{(1)}(\t12)
\labl{dat}
\end{eqnarray}
and finally
\begin{equation}
\tilde{r}_a(\theta_{12}) = 1 +
\frac{\rho^{(2)}(\theta_{12}) - d^{(2)}_{prod}(\theta_{12})}
{c_a d^{(2)}_{prod}(\theta_{12})+2 \rho^{(1)}(\theta_{12})}
\labl{r12at}
\end{equation}
where the distributions on the r.h.s. refer to initial gluons.
For sufficiently high energies
$\rho^{(1)}(\theta_{12}) \ll d_{prod}^{(2)}(\theta_{12})$,
 then the leading particle effect becomes negligable and we obtain
\begin{equation}
\tilde{r}_a(\theta_{12}) = 1 + \frac{1}{c_a}
(\tilde{r}_g(\theta_{12}) - 1). \label{ralpqhe}
\end{equation}
This equation means that the normalized cumulant
correlation $(r-1)$ is rescaled by $1/c_a$ (because
$\Gamma^{(2)}$ is multiplied by $c_a$ and $d^{(2)}_{prod}$
by $c^2_a$ according to Eq.(\ref{gama})). Ultimately in exponential accuracy
$r_q(\t12)\to r_g(\t12)$ for infinite energies ($a(P\Theta)\to 0$).

In Fig.8a we display as dashed curves the predictions
on $\tilde{r}_a(\t12)$ for quark and gluon jets from Eq.(\ref{r12at})
where we insert the high energy approximations for gluon jets
from Eqs.(\ref{rho12ra},\ref{d2t12a},\ref{asyan1}).
The full curve represents the asymptotic limit for both types of jets.

Next we consider the correlation function $\hat{r}$ with the
alternative normalization $\hat{r}_a=\tilde{\rho}^{(2)}_a/\bar n ^2_a$.
This quantity is calculated from Eq.(\ref{rhoat}) using the
approximate formulae
Eqs.(\ref{rho12ra},\ref{asyan1},\ref{ffflll}) and
$\bar n_a\approx c_a\bar n_g$ as
\begin{eqnarray}
\hat{r}_a(\epsilon)&=&c_a^{-1}y\exp (-y(2-\omega(\epsilon,2)) \nonumber \\
 & - & (c_a^{-1}-1)y\exp(-2y(1-\sqrt{1-\epsilon})) \nonumber  \\
 & + & 2\beta\sqrt{2y}/(c_af(1-\epsilon)^{{1\over 4}})
        \exp (-y(2-\sqrt{1-\epsilon}))
\labl{rhqg}
\end{eqnarray}
where $y=2\beta\sqrt{\ln{(P\Theta/\Lambda})}$.
In Fig.8b we show the predictions for this quantity
at finite energy for quark and gluon jets ($c_a=4/9$ or $c_a=1$ resp.)
as well as the asymptotic limit Eq.(\ref{rh12ra}).
The finite energy effect at large $\epsilon$ is smaller than for
$r(\t12)$ where the leading term in the exponent
is canceled.
The shape of the
asymptotic curve in the figure
$\sim 2\beta(1-\omega(\epsilon,2)/2$) is reached faster
with increasing energy than its normalization.

%In fig.~7.1 we show the results on $r_a(\theta_{12})$ for
%quark and gluon jets, first the exact numerical solution of
%the integral equation Eq.(\ref{gen}) and second the
%predictions using the high energy approximation
%Eq.(\ref{lnr12}).
%
%The comparison between our analytic calculation and the MC
%result shows that the overall shapes are similar: there is
%a certain range of linear increase and drop towards the small
%ANGLE LIMIT $\kappa =Q_0/P$. However the absolute
%normalization is off considerably (the $g$-jet calculation
%would actrually fit better contrary to expectation.)

\subsection{Fully differential connected angular correlations}
\subsubsection{2-parton correlations for fixed \al}
This calculation shares common features of both cases considered
earlier,
namely the angular ordering, as in the single parton density,
Eq.(\ref{tkk}),
and integration of a product of the singularities as
in Eq.(\ref{dirter2}). We consider here the case of constant \al .
Results for running \al will be discussed in the next section
together with the ones for general order $n$.
Integrating our resolvent representation, Eq.(\ref{res})
over parton momenta gives
\begin{equation}
   \rho^{(2)}_P(\Omega_1,\Omega_2)=g^{(2)}_P(\Omega_1,\Omega_2)+
%  \rho^{(2)}_P(\Omega_1,\Omega_2)=g^{(2)}_P(\theta_1,\theta_2)+
   {a^2\over 2\pi} \int \frac{dK}{K}
   \frac{d\Omega_K}{\Theta_{PK}^2} I(\frac{P}{K},
   \frac{\Theta_{PK}}{\sigma})
   g^{(2)}_K(\Omega_{Kk_1},\Omega_{Kk_2}), \labl{o1o2}
%  g^{(2)}_K(\Theta_{Kk_1},\Theta_{Kk_2}), \labl{o1o2}
\end{equation}
with the resolvent $I(x,y)$ defined in Eq.(\ref{resI})
and the inhomogenous
term given by the product of the single particle densities
$\rho^{(1)}(\Omega)\equiv\rho^{(1)}(\theta)/(2\pi\theta)$ in (\ref{asyan2}).
Similarly to the previous cases, nested terms do not contribute to the
leading behaviour in the high energy limit and will be neglected.
If needed, they can be included without any difficulty.
Saturating the angular
integration by two poles $\Theta_{Kk_1} (\Theta_{Kk_2}) \sim 0$ we get
for the connected correlation function,
%$\Gamma^{(2)}\equiv\rho^{(2)}-\rho^{(1)}\rho^{(1)}$,
\begin{eqnarray}
\lefteqn{\Gamma^{(2)}_P(\Omega_1,\Omega_2) =
 \frac{a^3}{2\pi} \int_{Q_0/\theta_{12}}^P \frac{dK}{K}
I(\frac{P}{K},\frac{\theta_1}{\theta_{12}})}
\nonumber \\& &
 \left( \frac{1}{\theta_1^2} \rho^{(1)}_K(\Omega_{12})
%\left( \frac{1}{\theta_1^2} \rho^{(1)}_K(\theta_{12})
        \int_{\kappa_K}^{\theta_{12}}
\frac{d\Theta_{Kk_1}}{\Theta_{Kk_1}}
\left( \Theta_{Kk_1}\rho_{K}^{(1)}(\Theta_{Kk_1}) \right)
%\rm{sinh}(a\ln{(
%            \frac{\Theta_{Kk_1}}{\kappa_K} )} )
 + (1\rightarrow 2) \right) .    \labl{cora1}
\end{eqnarray}
As before $\hat{n}_K$ was replaced by $\hat{n}_{k_1}
(\hat{n}_{k_2}$ in the $1\leftrightarrow 2$ term) in all nonsingular
expressions. In particular, $\Theta_{PK}\to \theta_1$ and
$\Theta_{Kk_2} \to \theta_{12}$. Note also that
$\sigma(k_1,k_2,K)=\theta_{12}$ in this case since the minimal
virtuality
(emission angle $\Theta_{PK}$) required for the parent $K$ to emit
$k_2$
is indeed controlled by $\theta_{12}$ if $\vec{K}\|\vec{k}_1$. The
upper
limit of the $\Theta_{Kk_1}$ integration is given by $\theta_{12}$ and
 not by $\Theta_{Pk_1}$ as one might have guessed from
the angular ordering. This is a consequence of the arguments following
Eq.(\ref{dirter2})
 \footnote{However the connected correlation $\Gamma$
in Eq.(\ref{cora1}) does not vanish only if
$\Theta_{Pk_1}(\Theta_{Pk_2}) > \theta_{12}$
hence the angular ordering along the cascade is preserved.}.
Similarly
 the lower bound of the $K$ integration is
controlled by $\theta_{12}$ and not by $\Theta_{Pk_1}$. Integration
similar to one performed explicitly in Appendix E gives
\begin{eqnarray}
\lefteqn{\Gamma^{(2)}_P(\Omega_1,\Omega_2) =}
\nonumber \\& &
  = \frac{a^2}{(2\pi)^2 \theta_1^2 \theta_{12}^2 }
\sum_{m=0}^\infty (2^{2m}-1) \left( \frac{l}{L_1} \right)^{m+1}
I_{2m+2}(2a\sqrt{l L_1}) +(1\leftrightarrow 2),\labl{cora2}
\end{eqnarray}
where $l=\ln{(\theta_{12}/\kappa)},
L_1=\ln{(\theta_1/\theta_{12})}$
and $\theta_1 (\theta_2) > \theta_{12}$ respectively.
Eq.(\ref{cora2}) is the new result for the full angular dependence of
the two parton correlation function in DLA.
Similarly to the density
in the relative angle, Eq.(\ref{selfsim}), we find a characteristic
power behaviour at high energies (see Appendix G)
\begin{equation}
\Gamma^{(2)}_P(\Omega_1,\Omega_2)
\simeq \frac{a^2}{2(4\pi)^2} \frac{1}{\theta_{12}^2}
\left( \frac{\theta_{12}}{\kappa} \right)^{2a}
\left ( \frac{1}{\theta_{1}^2}
 \left (\frac{\theta_1}{\theta_{12}}\right )^{a/2}
+ \frac{1}{\theta_{2}^2}
\left (\frac{\theta_2}{\theta_{12}}\right )^{a/2} \right),
\labl{Gas}
\end{equation}
which shows again the fractal (or intermittent) nature of the cascade.

Again one can prove "backward compatibility" with earlier results.
For example we show in Appendix H that integrating Eq.(\ref{cora1})
over $\Omega_1, \Omega_2$ at fixed relative angle reproduces the
second term of the resolvent representation for $\rho^{(2)}(\vartheta_{12})$,
Eq.(\ref{sol12}) as it should. Similarly one can integrate directly the
asymptotic form, Eq.(\ref{Gas}), to obtain Eq.(\ref{selfsim}). Note that
in both cases one should include the disconnected term $g^{(2)}_P$
which is not negligible in the high energy limit. This has an interesting
consequence: even though only the connected correlation $\Gamma$ has
a simple power behaviour,
on the more inclusive level
it is the {\em density} $\rho^{(2)}(\vartheta_{12})$
which is power behaved.

% Similar equations hold for arbitrary order $n$, the
%only difference consists of the more complicated
%expression for the inhomogenous term $\Delta^{(n)}$. Note however,
%that although $d^{(n)}$ contains product of many poles (or more
%general, power singularities), which result from varius products
%of lower
%correlation functions, the structure of $\Delta^{(n)}$ is simpler.
%The pole dominance in $d^3 K$ integration splits $\Delta^{(n)}$ into
%a sum of terms where the parent is almost parallel to {\em each one}
%of the final partons. Consequently $\Delta^{(n)}$ has a structure
%\begin{equation}
%\Delta^{(n)}=\int d^3 K {\cal M}_P(K) d^{(n)}_K=
%(\sum_{i=1}^n \Delta_i^{(n)}(\Theta_{Pk_i},\chi),
% \labl{Delta}
%\end{equation}
%where $\chi$ denotes all remaining
%variables. It follows that $\chi$ contains only relative angles
%$\vartheta_{jl}$ between final partons.

%he structure (\ref{Delta}) of the inhomogenous term implies that
%he parent integration in the integral equation for $\Gamma^{(n)}$
%tself , Eq.(\ref{eqgam}), can be also simplified. It turns out
%that $\Gamma^{(n)}$ also decouples into a sum of the type
%(\ref{Delta}) and each of individual terms $\Gamma_{i}^{(n)}$
%satisfies the following equation

\subsubsection{Running \al results for general order n}
Next we compute the cumulant correlation functions
$\Gamma_P^{(n)} (\{\Omega\})$ fully differential in the angles
$\{\Omega\} = (\Omega_1,...,\Omega_n)$ for running $\alpha_s$.
In the DLA we require for the polar angle $\theta_i$ of particle $i$
and the relative angles $\theta_{lm}$
\begin{equation}
\theta_i,\theta_{lm}\gg\Lambda/P.
\labl{thlim}
\end{equation}
We study the correlations again in the high energy limit
with the quantities
\begin{equation}
\epsilon^i_{lm} = \ln(\theta_i/\theta_{lm})/\ln (P\theta_i/\Lambda)
\labl{epsilm}
\end{equation}
kept fixed within the allowed range
$(0 < \epsilon^i_{lm} < 1)$. We
consider here the typical configurations where the various relative
angles $\theta_{lm}$ between partons are of comparable order
but with
\begin{equation}
|\delta^{ij}_{lm}| \ll 1, \qquad
\delta^{ij}_{lm} = \ln(\theta_{ij}/\theta_{lm})/\ln(P\theta_{lm}/\Lambda)
\labl{delijlm}
\end{equation}
Note that this condition can always be satisfied for any $\epsilon^i_{lm}$ for
sufficiently high energies as
$
\delta^{ij}_{lm}=\ln(\theta_{ij}/\theta_{lm})/\left((1-\epsilon^i_{lm})
\ln(P\theta_i
   /\Lambda)\right)$.

We start by deriving the
integral equation  for $\Gamma_P^{(n)}$ from Eqs.
(\ref{F3}),(\ref{F4}). First we carry out the momentum
integral for $\Delta^{(n)}_P$ in (\ref{F4}), thereafter
the angular
integral is obtained  applying the pole approximation:
the integrand becomes singular whenever the direction of the
intermediate jet K becomes parallel to any of the momenta
$k_i$ because of the singularities $1/\theta^2_{Kk_i}$
in $d^{(n)}_{g,K,prod}$. The integral can then be written as
a sum over n terms arising from the leading logarithmic
singularities $\Delta_P^{(n)} = \sum^n_{i=1}
\Delta^{(n)}_{P,i}$ with
\begin{equation}
\Delta_{P,i}^{(n)}(\Omega_i,\{\Omega_{k_i,k_j}\})=\int_{Q_0/\theta_i^m}^P
{dK\over K} \int_{Q_0/K}^{\theta_i^m}
{d\Theta^2_{Kk_i} \over 2\theta_i^2}
a^2(K\theta_i)
 d^{(n)}_{K,prod}(\Omega_{Kk_i},\{\Omega_{k_i k_j}\}).
\labl{deli}
\end{equation}
Each term has the singularity $1/\theta^2_i$ but
depends otherwise only on the $(n-1)$ relative angles
${\Omega_{k_ik_j}}$ between parton i and the others. This
results from replacing the slowly varying functions
$\theta_{Kk_j}$($\Theta_{PK}$) by the value $\theta_{k_ik_j}$($\theta_i$) for
$K\|k_i$. The integral equation for $\Gamma^{(n)}_P$
(\ref{F3}) is then solved
-neglecting the contribution from the nonleading nested term-
by the analogous decomposition
$\Gamma_P^{(n)} = \sum^{n}_{i=1} \Gamma^{(n)}_{P,i}$ with
each term obeying the integral equation
\begin{equation}
\Gamma_{P,i}^{(n)}(\{\Omega\})= \Delta_{P,i}^{(n)}+
 {1\over \theta_i^{2} }  \int_{Q_0/\theta_i^P}^P
    {dK\over K} \int_{\theta_i^M}^{\theta_i}
    {d\Theta_{Kk_i}\over\Theta_{Kk_i}}
     a^2(K\theta_i)
\left[ \Theta_{Kk_i}^2 \Gamma_{K,i}^{(n)}(\{\Omega_K\})\right].
\labl{gami}
\end{equation}
The bounds in the angular integrals of
(\ref{deli}),(\ref{gami}) correspond to the different regions
$\Gamma{''}$ and $\Gamma'$ discussed in section 2:
$\theta_i^m$ is taken as
minimal angle between particle $i$ and the others,
$\theta^M_i$ is the maximal angle. For n=2, $\theta_i^m
= \theta_i^M = \theta_i^P = \theta_{12}$. The lower
limit in the K integration with angle $\theta_i^P$ does
not enter our high energy approximation. Furthermore in
(\ref{gami})
the angular ordering condition
requires for all relative angles
$\theta_{ij}<\theta_i$.

First we calculate $\Delta^{(n)}_P$ for $n=2$ with
$d^{(2)}(1,2) = \rho^{(1)}(1) \rho^{(1)}(2)$ and $\rho^{(1)}
(\Omega)$ from (\ref{asyan1}). In the integral over the
first pole for $\Theta_{Kk_1}\approx 0$ we
replace $\rho^{(1)}(\Omega_2)$ by $\rho^{(1)}(\Omega_{12})$. The
angular and momentum integrals can be integrated by part
whereby terms of relative order $1/\sqrt{\ln
(P\theta_{12}/\Lambda})$ are neglected (see (\ref{thlim})).
Then we obtain
%with $\epsilon_i =
%\ln (\theta_i/\theta_{12})/\ln (P\theta_i/\Lambda)$
\begin{equation}
\Delta^{(2)}_P(\Omega_1,\Omega_2) \simeq
{f^2 a(P\bar\theta_1)
%\sqrt{1-\epsilon^i_{12}}
\over 2(4\pi)^2\theta_{12}^2\theta_1^2}
\biggl({P\theta_{12}\over \Lambda}\biggr)^{4a(P\t12)}
+ (1 \leftrightarrow 2).
\labl{del2s}
\end{equation}
where we have written $a(P\bar\theta_1) =a^2(P\theta_1)/a(P\t12)$.
To derive $\Gamma_P^{(2)}$ we note that the integral
equation (\ref{gami}) is of the generic form (\ref{gen})
with $h\equiv\theta_i^2\Gamma_i^{(2)}$ and $p_T=K\theta_i$.
Our inhomogeneous term
(\ref{del2s}) can be written in the form $\Delta^{(2)}_P\sim
\exp (4\beta\sqrt{\ln(P\theta_{12}/\Lambda)})$ just as in (\ref{gendir}).
So we
can take the solution for $h$ at high energies from Sect.4.
The prefactors
$a(P\bar\theta_1)$
%and $\sqrt{1-\epsilon^i_{12}}$ in
in (\ref{del2s}) as well as the ``mixed scale''
$a(K\theta_i)$ in the integral equation (\ref{gami})
influence the result only at the next to leading order of
$1/\sqrt{\ln P\theta_i/\Lambda}$ (see Sect.4). We obtain
\begin{equation}
\Gamma_P^{(2)}(\Omega_1,\Omega_2)\simeq
   {f^2a(P\bar\theta_1)\over 2(4\pi)^2\t12^2\theta_1^2}
      \exp\left(2\beta\omega
(\epsilon^1_{12},2)\sqrt{\ln (P\theta_1/\Lambda})\right)+(1\leftrightarrow 2).
\labl{gam2as}
\end{equation}
where the prefactors ensure the correct
limit $\Gamma^{(2)}\to\Delta^{(2)}$
for $\theta_1\to\theta_1^M=\theta_{12}$ according to (\ref{gami}).
With
$\omega(\epsilon,2)\approx 2-3\epsilon/4$ for small
$\epsilon$ (sufficiently large $\theta_{12}$) we obtain
\begin{equation}
\Gamma_P^{(2)}(\Omega_1,\Omega_2)\simeq
   {f^2a(P\bar\theta_1)\over 2(4\pi)^2\t12^2\theta_1^2}
   \biggl({\theta_1 \over \t12}\biggr)^{-{3\over 2}a(P\theta_1)}
   \biggl({P\theta_1\over \Lambda}\biggr)^{4a(P\theta_1)}
   +(1\leftrightarrow 2)
\labl{gon2s}
\end{equation}
which has again an approximate power behaviour in the angles.
In the a-limit (\ref{alimit}) of constant $\alpha_s$
one obtains back the previous result
Eq.({\ref{Gas}).

These results can be generalized to an arbitrary
number of particles. The inhomogenous term $d^{(n)}$ is now built up
either from
the product $\prod^n_{i=1} \rho^{(1)} (i)$ or from combinations
of correlation functions $\Gamma^{(m)}$ of lower order $m<n$.
Although for rising $n$ there is an increasing number of terms
corresponding to these various different combinations
the final result can be written in compact form as follows
( see Appendix I for more details)

\begin{equation}
\Gamma_P^{(n)}(\{\Omega\})\simeq
\left({f\over 4\pi}\right)^n {1\over n}
\sum_{i=1}^n {a^{n\over 2}_i(P\bar\theta_i)\over \theta_i^2}
 \exp\left(2\beta\omega(\epsilon^i_{iM},n)\sqrt{\ln(P\theta_i/\Lambda)}\right)
   F_i^n (\{\theta_{kl}\})
\labl{gamnom}
\end{equation}
The intermediate scale $P\bar\theta_i$ is defined by
$a(P\bar\theta_i)^{{n\over 2}}=a^2(P\theta_i) a^{{n\over 2}-2}(P\theta_i^m)$
where $\theta_i^m$ is the minimal or because of (\ref{delijlm}),(\ref{dexpand})
any other relative angle and $\theta^M_i$ is the maximal relative angle.
The functions $F_i^n$ in (\ref{gamnom})
are homogeneous in the relative angles $\theta_{kl}$
of degree $p=-2(n-1)$ and can be derived recursively.
They are built from factors
$[kl]=1/\theta_{kl}^2$ and read for $n=2,3$ explicitly
\begin{eqnarray}
F^2_1&=&F_2^2=[1,2]   \labl{Ffcts2}\\
F^3_1&=&\{3[12][13]+[23]([12]+[13])\}/2;\quad F^3_2,F^3_3\quad \rm{cycl.}
\labl{Ffcts}
\end{eqnarray}

In the linear approximation for $\omega(\epsilon,n)$  we obtain
\begin{equation}
\Gamma_P^{(n)}(\{\Omega\})\simeq
\left({f\over 4\pi}\right)^n {1\over n}
\sum_{i=1}^n {a^{n\over 2}_i(P\bar\theta_i)\over \theta_i^2}
   \biggl({\theta_i \over \theta_i^M}\biggr)^{-(n-{1\over n})a(P\theta_i)}
   \biggl({P\theta_i\over \Lambda}\biggr)^{2na(P\theta_i)}
   F_i^n (\{\theta_{kl}\})
\labl{gamnp}
\end{equation}
with the by now well known power behaviour.
Within our accuracy (\ref{thlim}) this result approaches
$\Delta_P^{(n)}$ for $\theta_i\to \theta_i^M$.

The correlations implied by Eq.(\ref{gamnom}) are dominated by the collinear
singularities from the gluon emissions. In each term $i$ the particle
$i$ is connected with the initial parton direction (factor $\theta_i^{-2}$)
and all other partons $k,l$ are connected among themselves
(factor $\theta_{kl}^{-2}$).
This structure is illustrated for $n=3$ in Fig.9.
The "star" connections in the left column originate from
the first term in (\ref{dprod}) or (\ref{d3ang}) in App. I,
the "snakes" in the other diagrams
from the terms involving the connected parts.
Our results for the QCD cascade correspond then to the phenomenological
model by Van Hove \cite{VHOVE} which builds up multiparticle
cumulant correlations from products of two particle correlations as in
Fig.9,
whereas in the linked pair approximation \cite{LPA} the "stars" are absent
(for a discussion, see \cite{UGL}).
 \section{Local multiplicity moments}
         \rem1{pow6v6}
\subsection{Definition of observables}

In this chapter we discuss moments of multiplicity distributions
in sidewise angular regions: (a) a cone of half opening
angle $\delta$ at angle $\theta$ with respect to the initial
parton and (b) a ring of half opening $\delta$, symmetric
around the initial parton at angle $\theta$. The two cases
correspond to dimensions $D=2$ and $D=1$ of the phase space cells
of  the volume $\sim\delta^D$ (see Fig.10).
The unnormalized factorial and cumulant moments of the
multiplicity distribution are defined as integrals of the
respective correlation functions over the angular region
$\gamma(\theta,\delta)$
\begin{equation}
f^{(n)} (\theta,\delta) = \int_{\gamma(\theta,\delta)}
dk_1 \dots dk_n \rho^{(n)} (k_1\dots k_n)
\labl{fnd}
\end{equation}

\begin{equation}
c^{(n)} (\theta,\delta) = \int_{\gamma(\theta,\delta)} dk_1 \dots dk_n
\Gamma^{(n)} (k_1 \dots k_n)
\labl{cnd}
\end{equation}
These quantities are conventionally normalized by powers of
the average multiplicity in the respective angular region
\begin{equation}
F^{(q)} = f^{(q)} / \bar n^q,\quad C^{(q)} = c^{(q)} / \bar n^q
\labl{momn}
\end{equation}
The factorial moments can then be calculated as the
event average of multiplicities in the region
$\gamma(\theta,\delta)$ by
\begin{equation}
F^{(q)}= <n(n-1) \dots (n-q+1)> / \bar n^q
\labl{fnor}
\end{equation}
The derivation of analytic results for these moments from
our correlation functions $\Gamma^{(n)}$ require some
approximations. They can be avoided for moments which are
differential in one particle momentum. So we also consider
the differential cumulant moments
\begin{equation}
h(k,\delta) = {1\over n} \sum_{i=1}^n \int_{\gamma(k,\delta)}
dk_1\dots dk_n \Gamma^{(n)} (k_1, \dots k_n) \delta (k-k_i)
\labl{hdef}
\end{equation}
where one particle is kept fixed at three-momentum $k$ (or at
angle $\theta $ after integration over $|k|$) and all others
are counted in the region $\gamma$ around three-momentum $k$
(or $\theta$)
\footnote{Such moments  (``star
integrals'') have been also discussed recently by Eggers et al.
\cite{STAR}}. We are particularly interested to investigate
under which conditions the moments behave like a power
\begin{equation}
F^{(n)} \sim (\theta/\delta)^{\phi_n}
\labl{fpow}
\end{equation}
Such behaviour is indicative of a fractal behaviour of the
multiplicity fluctuations, also called intermittency. A
jet with such a property can be assigned a fractal dimension
\cite{BUSCH}, \cite{CELLO}
\begin{equation}
D_n = D(1-{\phi_n\over n-1})
\labl{dfdef}
\end{equation}
For a random distribution of points in $D$ dimensions one
obtains a Poisson distribution in each subdivision, so for
all $\delta$ the factorial moments are $F^{(n)}=1$ and
$\phi_n=0$, so $D_n=D$. On the other hand, if in any
subdivision with cell size $\delta$ there is only one cell
occupied and the others are empty, i.e. the distribution
contracts to a point, then one obtains $\phi_n=n-1$ or
$D_n=0$. The  general formula Eq.(\ref{dfdef}) covers also these
two limiting cases.

\subsection{Integral equation for moments}
\subsubsection{The equation}
One can derive QCD predictions for the moments (\ref{fnd})-(\ref{hdef})
in two ways. In this Section we obtain and solve the corresponding
integral equation. The asymptotic behaviour at larger angles can be also
obtained from the direct integration of our results (\ref{Gas},\ref{gamnp})
for connected correlations.

It follows from the previous Section that the
connected correlation functions depend on a limited subset
of all angles which characterize the
configuration of the final partons. For example
$\Gamma^{(2)}_{P,i}(\Omega_1,\Omega_2)$
 depends only on two variables $\vartheta_i,\vartheta_{12}$.
As a consequence   the integral equation
satisfied by moments themselves can be readily derived.

We begin with $n=2$. In this case integration (\ref{hdef})
over the sideways cone $\gamma(\vartheta,\delta)$ reads explicitly
\begin{equation}
h^{(2)}_P(\Omega,\delta)=4 \pi \int_{\kappa}^{\delta}
 d\vartheta_{12} \vartheta_{12} \Gamma^{(2)}_{P,i}(\Omega,
 \Omega_{12}). \labl{n2cone}
\end{equation}
Where both $i=1,2$ give the same contribution.
Integration of Eq.(\ref{Gamsim}) over the cone,
Eq.(\ref{n2cone}), gives for each $i$ $(i=1,2),\Omega=(\vartheta,\phi),
\Omega_i=(\vartheta_i,\phi_i)$,
\begin{equation}
h_{i P}^{(2)}(\Omega,\delta)=d^{(2)}_i +
   { 2\pi \over\theta^2}
\int_{\kappa}^{\delta} \theta_{12} d\theta_{12}
\int_{Q_0/\theta_{12}}^P
{d K \over K} \int_{\theta_{12}}^{\theta} \theta_i d\theta_i
a^2(K\vartheta) \Gamma_{i K}^{(2)}(\Omega_i,\Omega_{12}).
\end{equation}
Changing orders of the $\theta_{12}$ and $(K,\theta_i)$
integrations produces two terms
\begin{eqnarray}
\lefteqn{h_{i P}^{(2)}(\Omega,\delta)=d^{(2)}_i +}    \nonumber  \\
 & & {1 \over\theta^2}
 \int_{Q_0/\delta}^P
{d K \over K} \int_{Q_0/K}^{\delta} \theta_i d\theta_i
a^2(K\vartheta) \left( 2\pi
\int_{Q_0/K}^{\theta_i} \theta_{12} d\theta_{12}
\Gamma_{i K}^{(2)}(\Omega_i,\Omega_{12})\right) +  \nonumber \\
    &  &  {1\over\theta^2}
 \int_{Q_0/\delta}^P
{d K \over K} \int_{\delta}^{\theta} \theta_i d\theta_i
a^2(K\vartheta) \left( 2\pi
\int_{Q_0/K}^{\delta} \theta_{12} d\theta_{12}
\Gamma_{i K}^{(2)}(\Omega_i,\Omega_{12})  \right). \labl{67}
\end{eqnarray}
Inner integrals are nothing but the moments of the $K$ jet
Eq.(\ref{n2cone}).
Adding the contributions from both $i$ gives
\begin{eqnarray}
h^{(2)}_P (\Omega, \delta)=d^{(2)}_P(\Omega,\delta) +
{1 \over \theta^2}  \int_{Q_0/\delta}^P \frac{dK}{K}
 \int_{Q_0/K}^\delta
  \psi d\psi a^2(K\vartheta) h_{K}^{(2)}(\Omega_\psi,\psi) +
                                            \nonumber \\
 {1\over\theta^2} \int_{Q_0/\delta}^P \frac{dK}{K}
 \int_{\delta}^{\theta}
\psi d\psi a^2(K\vartheta) h^{(2)}_K(\Omega_\psi,\delta) ,
\labl{cumeq2}
\end{eqnarray}
This result remains valid for arbitrary $n$ as well. The
essential property, namely emergence of the two terms in the integral
Eq.(\ref{67}), which can be reinterpreted according to the definition
(\ref{n2cone}) is independent of the order of the moment.
Equation (\ref{cumeq2}) can be brought to the more symmetric form
introducing the
logarithmic density $\overline{h}_P(\Omega,\delta)=
\vartheta^2 h_P(\Omega,\delta)$.
We get for arbitrary $n$
\begin{eqnarray}
\overline{h}^{(n)}_P(\Omega_{\vartheta},\delta)=
\overline{d}^{(n)}_P(\Omega_{\theta},\delta) +
 \int_{Q_0/\delta}^P \frac{dK}{K} \int_{Q_0/K}^\delta
\frac{d\psi}{\psi} a^2(K\vartheta) \overline{h}_{K}^{(n)}(\Omega_{\psi},\psi)
                                            \nonumber \\
 + \int_{Q_0/\delta}^P \frac{dK}{K}\int_{\delta}^{\theta}
\frac{d\psi}{\psi} a^2(K\vartheta) \overline{h}^{(n)}_K(\Omega_{\psi},\delta) .
\labl{cumeq}
\end {eqnarray}
Inhomogeneous parts $\overline{d}_P^{(n)}$ are given by the corresponding
integrals of, once iterated, disconnected contributions
$\Delta^{(n)}_P$
\begin{equation}
\overline{d}^{(n)}_P(\Omega_{\vartheta},\delta) =
{\theta^2\over n} \sum_{i=1}^n \int_{\gamma(\vartheta,\delta)}
dk_1\dots dk_n \Delta^{(n)}_P (k_1, \dots k_n) \delta (\hat{k}-\hat{k}_i)
. \labl{Delta2D}
\end{equation}
In particular
\begin{equation}
\overline{d}^{(2)}_P(\Omega_{\vartheta},\delta)=2\pi \sum_{i=1}^2
 \int_{\kappa}^{\delta}
\left(\theta^2\Delta^{(2)}_i(\Omega,\Omega_{12},P)\right)
\theta_{12} d\theta_{12}, \labl{d2bar}
\end{equation}
with $\Delta_i^{(2)}$ given by the $i-th$ term of
Eq.(\ref{dirter2}).
Even though
equation (\ref{cumeq}) has slightly more complicated structure
due to the
term $\overline{h}(\Omega_\psi,\psi)$, it can be solved
in two steps -
each step involving the type of equation encountered earlier.
To see this
note that at $\delta=\theta$ the last term vanishes and we
get the simpler
equation for the boundary value
$ \hat{h}^{(n)}_P(\psi) \equiv  \overline{h}^{(n)}_P(\Omega_\psi,\psi) $  ,
\begin{equation}
 \hat{h}^{(n)}_P ( \delta)=\hat{d}^{(n)}_P(\delta) +
 \int_{Q_0/\delta}^P \frac{dK}{K} \int_{\kappa_K}^\delta
\frac{d\psi}{\psi} a^2(K\delta) \hat{h}_{K}^{(n)}(\psi).  \labl{boun}
\end{equation}
Equations of this type are satisfied by global quantities like
the
total average multiplicity $\overline{n}(P\Theta)$, Eq.(\ref{ntoteq}),
or factorial moments
in the whole cone $(P,\Theta)$, Eq.(\ref{f2eq}).
For constant \al they can be easily
solved by iteration
\begin{equation}
\hat{h}^{(n)}(x)={d\over dx} \left(\hat{d}^{(n)}(x)+a\int_0^x
\sinh{[a(x-t)]}
\hat{d}^{(n)}(t) dt \right),\;\;\; x=\ln{\left({P\delta
\over Q_0}\right)},
\labl{sinsol}
\end{equation}
while for running \al the asymptotic behaviour of the solution
will be given below.
Now equation (\ref{cumeq}) reads in two variables
\begin{equation}
\overline{h}^{(n)}_P (\Omega, \delta)=
\hat{h}^{(n)}_P(\delta) +
\int_{Q_0/\delta}^P \frac{dK}{K}\int_{\delta}^{\theta}
\frac{d\psi}{\psi} a^2(K\vartheta) \overline{h}^{(n)}_K(\Omega_\psi,\delta) ,
\labl{ctwov}
\end{equation}
with the known boundary term. Hence the
equation for moments, Eq.(\ref{cumeq}), was reduced to,
standard by now,
integral equation with slightly more complicated inhomogenous term.

%\subsection{Results for the full angular region}
\subsubsection{The inhomogenous term}

The two  step procedure leading to Eq.(\ref{ctwov}) has a simple physical
interpretation. The inhomogenous term
$\hat{h}^{(n)}_P(\delta)=\hat{h}^{(n)}_P(\Omega_{\delta},\delta)$
is the cumulant moment in the maximal sideway cone which is allowed by the
angular ordering, see Fig.11. Therefore it should be related to the
total cumulant moment over the full cone. In fact we shall
prove that with $x_{\la}=\ln(P\delta/\Lambda), X_{\la}=\ln(P\vartheta/\Lambda)$
\begin{equation}
\hat{d}^{(n)}(x_{\la})\equiv  \hat{d}^{(n)}_P(\delta)=
{1\over 2\pi }{d\over d X_{\la}}
r^{(n)}(X_{\la})_{|X_{\la}=x_{\la}}, \labl{srul}
\end{equation}
where
\begin{equation}
r^{(n)}(X_{\la})=\int_{\Gamma(P,\vartheta)} \Delta^{(n)}_P(\Omega_1\dots
\Omega_n)
d\Omega_1\dots\Omega_n, \labl{rdef}
\end{equation}
characterizes the full forward cone $(P,\vartheta)$. Since the asymptotic
behaviour of $r$ is known, Eq.(\ref{srul}) gives immediately the
corresponding expression for the inhomogenous term $d$. This determines
$\hat{h}^{(n)}(\delta)$ via Eq.(\ref{boun}).
Relation (\ref{srul}) explains also why $\hat{h}^{(n)}(\delta)$ satisfies
the integral equation which is characteristic of the global observables.
To prove Eq.(\ref{srul}), rewrite the
definition (\ref{Delta2D}) using Eq.(\ref{F4})
\begin{equation}
\hat{d}^{(n)}_P(\delta)={\delta^2\over 2\pi} \int^P_{Q_0/\delta} {d K \over
 K}
a^2(K\Theta_{PK})
{d\Omega_K \over \Theta^2_{PK} } \int d_{prod,K}^{(n)}(\Omega_{\delta},
\Omega_2,\dots ,\Omega_n) d\Omega_2\dots d\Omega_n.
\end{equation}
Due to the angular ordering all internal angles in the cascade are
negligible in comparison with $\Theta_{PK}$, therefore
$\Theta_{PK}\simeq\Theta_{Pk_1}=\delta=\Theta_{Pc}$\footnote{The cone
is defined around the direction of the first parton $k_1$}
, see Fig.11, and integration
over the {\em initial} parent $d\Omega_K$ at fixed $\vec{k}_1$ can
be transformed into the
integration over the {\em final} parton $d\Omega_{k_1}$ at fixed $\vec{K}$,
see Appendix J for the detailed discussion of the $n=2$ and $n=3$ cases.
We obtain
\begin{equation}
\hat{d}^{(n)}_P(\delta)={1\over 2\pi}\int_{Q_0/\delta}^P {d K\over K}
a^2(K\delta) t^{(n)}_K(\delta), \labl{dbyt}
\end{equation}
where
\begin{equation}
t^{(n)}_K(\delta)=\int_{(K,\delta)} d^{(n)}_{prod,K}(\Omega_1,\dots ,
\Omega_n) d\Omega_1\dots \Omega_n, \labl{tdef}
\end{equation}
is integrated over the complete forward cone of an intermediate
parent parton $(K,\delta)$.
On the other hand inserting Eq.(\ref{F4}) into Eq.(\ref{rdef}) and
interchanging orders of $d^3 K$ and angular integrations one obtains
\begin{equation}
r^{(n)}_P(\vartheta)=\int_{Q_0/P}^{\vartheta} {d\Psi\over\Psi}
\int_{Q_0/\Psi}^P {d K\over K} a^2(K\Psi)
t^{(n)}_K(\Psi) . \labl{rbyt}
\end{equation}
Comparing Eqs.(\ref{dbyt}) and (\ref{rbyt}) gives the relation (\ref{srul}).

Once the rule (\ref{srul}) for the inhomogenous term $\hat{d}$ is
established, we immediately see that the {\em whole} solution
$\hat{h}^{(n)}_P(\delta)=\hat{h}_P^{(n)}(\ln(P\delta/Q_0))$
of the boundary equation(\ref{boun}) satisfies an analogous relation. To derive
it, integrate Eq.(\ref{boun}) over $\delta$.
 One obtains
\begin{equation}
g^{(n)}(X_{\la})\equiv\int_{Q_0/P}^{\vartheta} \hat{h}^{(n)}_P(\delta)
{d\delta\over \delta} = r^{(n)}(X_{\la})+\int_{\lambda}^{X_{\la}} dy
\int_{\lambda}^y dz a^2(z) g_n(z).
\labl{ign}
\end{equation}
Since $r^{(n)}$ is given by Eq.(\ref{rbyt}) and $t^{(n)}$ is the source
(the inhomogenous term)
of the {\em global} factorial moment $f_n$, cf. Eqs.(\ref{f2def},\ref{men})
\footnote{Neglecting nested terms at high energy.}, we conclude that
the solution of Eq.(\ref{ign}) is simply
\begin{equation}
g^{(n)}(X_{\la})
 = \int_{\lambda}^{X_{\la}} dy \int_{\lambda}^y dz a^2(z) f_n(z),
\labl{solgn}
\end{equation}
which can be also checked by the direct inspection.
Therefore we have proved that
\begin{equation}
\hat{h}^{(n)}(x_{\la})={1\over 2\pi} \int_{\lambda}^{x_{\la}}
 dz a^2(z)f_n(z).
\labl{frul}
\end{equation}
This relation determines immediately the high energy behaviour of the
inhomogenous term in the two variable master equation for moments,
Eq.(\ref{ctwov}). For, inserting  Eqs.( \ref{fqas} ) and (\ref{ffflll})
into (\ref{frul}) gives the following asymptotic result
 \footnote{Equation
(\ref{fqas}) is also valid for the running \al with the replacement
$X\rightarrow X_{\la}$. }
$(\hat{h}^{(n)}(X_{\la})\equiv \hat{h}_P^{(n)}(\delta))$
\rem1{(III.II.H.1)}
\begin{equation}
\hat{h}^{(n)}(x_{\la})={\cal D}_n \exp{(2n\beta
\sqrt{x_{\la}})}, \labl{finhat}
\end{equation}
and
\begin{equation}
{\cal D}_n={ a(x_{\la}) F_n\over 2\pi n}
\left( {f\over 2\sqrt{a(x_{\la})}}\right)^n ,
\labl{depref}
\end{equation}
with coefficients $f$ and $F_n$  given by Eqs.(\ref{ffflll}) and (\ref{fqas})
 respectively.

\subsection{The  solution}
 \subsubsection{The leading contribution}
 Integral equation (\ref{ctwov}) is identical with the generic equation
Eq.(\ref{gen}) if we substitute
$p_T\rightarrow K\vartheta$.
 %Similarly equation for the
%two parton density
%in the  relative angle, Eq.(\ref{ang2}),  is identical with
%Eq.(\ref{ctwov}) if we substitute
%\begin{equation}
%\Theta \rightarrow \theta ,
%\theta_{12} \rightarrow \delta.  \labl{subst}
%\end{equation}
%The leading, when $P\to\infty$,  behaviours of the inhomogenous
%terms also coincide for $n=2$,
%hence the asymptotic form of the cumulant moment $C^{(2)}$
%for constant \al can be read off directly
%from our asymptotic solution, Eq.(\ref{selfsim}) $fromresolvent$
%upon substitution
%(\ref{subst}) \cite{we}.
Due to this universality, the
asymptotic solution for running \al is readily obtained from
the solution (\ref{finsol}) of Eq.(\ref{gen}).
% upon the appropriate
%interpretation of variables.
In particular the scaling variable is now
\begin{equation}
\epsilon=
\ln{(\vartheta/\delta)}/\ln{(P \vartheta/\Lambda)}
\end{equation}
As was discussed in the previous Section, the asymptotic behaviour of
 the inhomogenous term is directly given by that of the global moments. With
$d $ of Eq.(\ref{gen}) given by Eqs.(\ref{finhat}) and (\ref{depref}),
and using the machinery of Sect.4,
we get the following remarkably simple result for the leading contribution
to the unnormalized cumulant moments
\begin{equation}
\overline{h} ^{(n)}(\delta,\theta,P) \sim
% = {\cal P}_n {\cal D}_n
\exp{\left(2\beta \sqrt{\ln{(P\theta/\Lambda)}}
                  \omega(\epsilon,n)
\right)     }, \labl{hmomas}
\end{equation}
with the universal scaling function $\omega(\epsilon,n)$.
This equation is one of the main results of this paper. It
forms the basis  of the following discussions and further
approximations.
The function $\omega(\epsilon)$ is
determined  algebraically by the saddle point condition,
Eqs.(\ref{ome}-\ref{alg}).
% and the prefactors defined
% in Eq.(\ref{presol}) and (\ref{depref}) respectively.
A power expansion in  $\epsilon$ can be also easily
obtained by the WKB technique proposed in Section 4.2,
cf. Eq.(\ref{ommcl}), as well as the ${1/n}$ expansion,
Eq.(\ref{omn}), which
approximates the full result at the percent level.
%We have
%checked up to the fourth order in $\epsilon$ that the saddle
%point and the WKB method  give the same
%results. WKB provides also in a  rather simple manner  the
%nonleading contributions (in $1/\sqrt{\ln{(P\theta/\Lambda)}}$)
%as power series in $\epsilon$.

The cumulant moments, Eq.(\ref{cnd}), are simply obtained by the
 integration of $h=\overline{h}/\theta^2$ over the cone $(D=2)$
and over the ring $(D=1)$ for the two- and one-dimensional moments
respectively.
\begin{equation}
c^{(n)} (\theta,\delta) \sim
\left({\theta\over\delta}\right)^{-D}
\left({P\theta\over\Lambda}\right)^{2a(P\theta)\omega(\epsilon,n)}
\labl{cmomas}
\end{equation}
 After
normalisation by the n-th power of the multiplicity in
the respective angular region,
\begin{equation}
\bar n(\theta,\delta)={1\over 4} \sqrt{a(P\theta)} f
\left({\delta\over\theta}\right)^D
\left({P\theta\over\Lambda}\right)^{2a(P\theta)},
\labl{nbardr}
\end{equation}
 one obtains
\begin{equation}
C^{(n)}(\theta,\delta) \sim
%= {F_n {\cal P}_n \over   n 2^{(n-1)(3-2D)} }
% \left({1\over a}\right)^{n-1}
  \left({\theta\over\delta}\right)^{D(n-1)}
\left( {P\theta\over\Lambda} \right)^{2a(\omega(\epsilon,n)-n)}.
\labl{cnas}
\end{equation}
 where again $D$ labels the dimensionality of the phase space cell.
\subsubsection{Nonleading terms}

The methods developed in Sect.4 provide also the nonleading
contribution to the asymptotic solution.
In particular the normalisation of Eq.(\ref{hmomas}) is also
available. Since however the double log equations by definition
do not include all next-to-leading effects, the normalization of $\overline{h}$
constitutes only a part of the nonleading terms. However we quote for
the completness the full result given by the saddle point calculation
\begin{equation}
\overline{h} ^{(n)}(\delta,\theta,P)
 = c_a{\cal P}_n {\cal D}_n
\exp{\left(2\beta \sqrt{\ln{(P\theta/\Lambda)}}
                  \omega(\epsilon,n)
\right)     }, \labl{hmomsad}
\end{equation}
with the prefactors ${\cal P}_n$ and ${\cal D}_n$
given by Eqs.(\ref{presol}) and (\ref{depref}) respectively.
The factor $c_a$ distinguishes between quark and gluon jets
according to Eq.(\ref{gama}).
The factor ${\cal D}_n$ generalizes the $n=2$ result, which
can be also derived by the direct integration of the connected
correlation function, see Appendix K.
One obtains for the unnormalised cumulant moments
\begin{equation}
c^{(n)} (\theta,\delta)= 4^{2-D}\pi c_a {\cal D}_n {\cal P}_n
\left({\theta\over\delta}\right)^{-D}
\left({P\theta\over\Lambda}\right)^{2a(P\theta)\omega(\epsilon,n)},
\labl{cmomvo}
\end{equation}
and for moments normalized by the respective multiplicities
\begin{equation}
C^{(n)}(\theta,\delta) = {F_n {\cal P}_n c_a^{1-n} \over   n 2^{(n-1)(3-2D)} }
\left({1\over a}\right)^{n-1}  \left({\theta\over\delta}\right)^{D(n-1)}
\left( {P\theta\over\Lambda} \right)^{2a(\omega(\epsilon,n)-n)}.
\labl{cnfas}
\end{equation}
We emphasize again that the double logarithmic approximation
determines only the leading behaviour, Eq.(\ref{hmomas}), and the
normalization can be at best considered as the qualitative indication
of the nonleading contributions.

A second example of nonleading effects is the difference between the
factorial and cumulant moments. Consider for simplicity the second
moments. It follows easily from Eqs.(\ref{cmomvo}) and (\ref{fnd})
that the difference $f_2-c_2$ is nonleading \footnote{For large $\delta$
($\theta/\delta$ fixed, $P\rightarrow\infty$) it is of order $a$ and for
smaller $\delta$
($\epsilon$ fixed, $P\rightarrow\infty$), it is exponentially small relatively
 to the moments themselves.}.
Therefore, in our approximation, both factorial and cumulant moments
have {\em the same} asymptotic behaviour and any distinction between
the two can be discussed quantitatively only if the complete
next-to-leading correction is known.
\subsection{Fractal behaviour of moments and Intermittency}

There are two angular regions where our solution reveals different
behaviour.

At small opening angle $\Lambda/P < \delta \ll \theta$ there are two large
logarithms in the problem which can be chosen as $\ln{(\vartheta/\delta)}$
and $\ln{(P \vartheta/\Lambda)}$. Both of them are summed in
Eq.(\ref{hmomas}). At fixed $\epsilon$ and $P\rightarrow\infty$
we predict the new type of scaling in this variable.
However the effective exponent of the power law
%The prefactors are such that
% In the linear approximation ${\cal P}_n=1$ and we have
\begin{equation}
C^{(n)}(\theta,\delta) \sim \left({\theta\over\delta}\right)^{\phi_n}.
\end{equation}
depends on $\delta$ in a rather complicated way
\begin{equation}
\phi_n=D(n-1)-2a(P\theta)(n-\omega(\epsilon))/\epsilon,
\labl{cmomno}
\end{equation}
since $\epsilon$ depends on $\delta$. The QCD cascade is not selfsimilar
for small opening angles.

On the other hand for larger opening angles with the ordering
$1\ll \ln{(\theta/\delta)} \ll \ln{(P\theta/\Lambda)}$ we
expand $\omega(\epsilon)$ and ${\cal P}_n(\epsilon)$ in powers
of $\epsilon\sim$\al$(P\theta)\ln{(\theta/\delta)}$ and keeping the lowest
nontrivial terms gives
\begin{equation}
\overline{h} ^{(n)}(\delta,\theta,P)
 = {\cal D}_n
\left({P\theta\over \Lambda}\right)^{2na(P\theta)}
\left({\theta\over\delta}\right)^{-(n-{1\over n})
a(P\theta)}.
\labl{hmomnu}
\end{equation}
In the same approximation the cumulant moments read
\begin{eqnarray}
c^{(n)}(\theta,\delta)&\simeq & 4^{2-D}\pi {\cal D}_n
\left({P\theta\over\Lambda}\right)^{2na(P\theta)}
\left({\theta\over \delta}\right)^{-D-(n-{1\over n})a(P\theta)}
\labl{cmomnu}\\
C^{(n)} (\theta,\delta)&\simeq&{F_n\over n 2^{(n-1)(3-2D)} a^{n-1}}
\left({\theta\over \delta}\right)^{D(n-1)-(n-{1\over n}) a(P\theta)}
\labl{cmomnn}
\end{eqnarray}
where again $D=1,2$ corresponds to the one- and two-dimensional
cells in the angular phase space respectively.
We see that the normalized moments $C$ have a power dependence on the
opening angle $\delta$. This result establishes the self-similarity
of the QCD cascade when analysed in the angular bins of the size $\delta$.
Although the exponents in Eq.(\ref{cmomnn}) depend on the polar angle
$\theta$, they are {\em independent} of $\delta$. Hence in this region
normalized cumulant moments are simple powers of the opening angle.
The well developed QCD cascade is strictly intermittent.

%which show the power behaviour in the angular ratio
%$\theta/\delta$.
Corresponding expressions for fixed $\alpha_s$ are obtained
by letting $(P\theta/\Lambda)^{2a(P\theta)}$ $ \rightarrow
(P\theta/Q_0)^a$ and $a(P\theta)\to a$ as before. The result
for the normalised moment (\ref{cmomnn})
%remains unchanged
translates automatically to the constant \al case
as it does not depend on the scale $Q_0$: it is infrared
safe and free of narrow divergences.
 %Similarly, expanding the scaling function $\omega(\epsilon)$ for
%small $\epsilon$  (cf. Eq.(\ref{olin})) gives the power laws
%({\em see} Eqs.(\ref{cmomnu},\ref{cmomnn})). This limit
%also applies to the case of constant $\alpha_s$ for
%$\theta/\delta$ = const, $P\to \infty(\epsilon\to 0)$.
%
%
%As for the previous angular observables the behaviour of
%moments is qualitatively different for the range of smaller
%and for larger angles $\delta$. The dependence on the full
%range of the appropriate scaling variable $\epsilon$ in the
%high energy limit was derived above.
%The moments for larger angles (small $\epsilon$)
%will be obtained here by integration of the correlation
%function $\Gamma^{(n)}$ over the respective angular regions;
%which yields immediately the proper normalisation.

We have derived here the power behaviour as the small
$\epsilon$ limit of the general result (\ref{cnfas}). Alternatively
these moments can be easily obtained by integration of the full
correlation function as is demonstrated in Appendix K for n=2.
We have also checked that the cumulant moment $h^{(2)}$
integrates to the global moment $F^{(2)}=4/3$ (see Appendix H).

\subsection{Discussion}

Only in the case of fixed $\alpha_s$ the power law for moments
is found in the full angular region at high energies
($\epsilon$ fixed, $P\to \infty)$. The
parton cascade is selfsimilar down to very small angles.
%$\delta gsim? Q_0/P$, the limit of the \cite{DLA}.

Results for running $\alpha_s$ are shown in Fig.12 for the
sidewise cone (D=2) and the ring (D=1) according to
Eq.(\ref{cnas}). For sufficiently large $\delta$ (small
$\epsilon$) one observes the effective power behaviour
in the opening angle,
whereas at smaller angles (larger $\epsilon$) the moments
reach a maximum value and bend downwards.
This was explained in the previous Section.
 In Fig.12c we show
the energy dependence of the moments for D=2. At fixed $\epsilon$
the dependence on energy $P$ and angle $\theta$ enters through
the anomalous dimension $a(P\theta)$. Note that these
calculations use the asymptotic form of $\omega(\epsilon)$ which
vanishes for $\epsilon\to 1$. A full inclusion of finite energy
corrections would lead to an earlier drop and a vanishing of $\ln
c^{(n)}$ for $\delta \sim Q_0/P$ or $\epsilon \sim
\epsilon_{max}=\ln(P\theta/Q_0)/\ln(P\theta/\Lambda)$.
Numerical solution of the DLA equation for $\rho(\theta_{12})$
suggests that for $P=45$ GeV (Fig.7) such corrections become
important for $\epsilon >     0.6$.

%The power behaviour at small $\epsilon$ is valid only
%approximately for running $\alpha_s$ and the exponent
%$\phi_n$ in (\ref{cmomno}) depends on the kinematic
%variables $P$ and $\theta$ through the anomalous
%dimension $a(P\theta)$.
As was already mentioned, the exponent of the power dependence,
at small $\epsilon$, {\em does not} depend on the opening
angle $\delta$. Even for running \al normalized moments
show straightforward power behaviour in $\delta$ albeit with
the $P\theta$ dependent anomalous dimensions.
This result shows that the QCD
cascade has the intermittency property considered by Bialas
   and Peschanski \cite{BP}, although not directly in the limit $\delta\to 0$
but for the opening angle much larger than the soft cut-off
$\theta(\Lambda/P\theta)^{\overline{\epsilon}}$
\footnote{$\overline{\epsilon} (\simeq .2$, say)  denotes the
upper bound of the linear regime of $\omega(\epsilon)$.}
 that is for the well developed
cascade. This cut-off
tends to $0$ at high energy. Therefore only at infinite energies
the cascade is intermittent down to to the very small opening angles.
%only for sufficiently large angle (small $\epsilon$) and high
For larger $\epsilon$ the formulae (\ref{cnas}) or
(\ref{cmomvo}) should be used. The power behaviour at large
opening angles $\delta \sim \theta$ (small$\epsilon$) had been
postulated before in a phenomenological model \cite{SASA}.

The above power behaviour can be considered as resulting
from the fractal structure of the cascade, i.e. the
selfsimilarity of the branching process \cite{VENFR}, \cite{GUST} again
only valid in the region where the running of $\alpha_s$ can
be neglected. One can associate a fractal dimension to the
scaling property of the multiplicity fluctuations and  --
using the above definition (\ref{dfdef}) and neglecting the
difference between factorial and cumulant moments -- we find
\begin{equation}
D_n={n+1 \over n} a(P\theta).
\labl{dfrac}
\end{equation}
This implies that the fractal dimension for all moments are
determined by the anomalous dimension $a(P\theta)$
which governs the scale dependence of the average
multiplicity, in particular $D_n \approx a(P\theta)$ for
large $n$ \footnote{A similar result for the fractal dimension
has been given before \cite{GUST}, but refers to the leading
exponent of $P\theta/\Lambda$ of the unnormalized moments
in Eq.(\ref{cmomnu}), unlike our definition (\ref{dfdef}),
see also \cite{OW2}. }.
%It is interesting to note that the
The fractal
dimension is independent of the geometrical  dimension $D$ of the
problem,
as a consequence
the deviation from
$D$ is larger for $D=2$ than for $D=1$.
\section{Comparison with Monte Carlo Calculations}
         \rem1{pow7v4}
\rem1{pow7v4}

In this section we compare our analytic results from the
DLA with results derived from Monte Carlo methods which
take into account fully the kinematic constraints.
We use here for illustration the HERWIG-MC program
\cite{HERWIG} for the process $e^+e^-\to u\bar u$
with subsequent QCD evolution.
This MC program is based on the coherent branching
algorithm and takes into account the most important leading and
subleading effects\footnote{For a recent discussion,
see also \cite{DMO}. We take the parameters
$\Lambda =0.15\hbox{ GeV}, m_q=m_g=0.32\hbox{ GeV}$.
In our computations at the parton level
we don't include the non-perturbative splitting of gluons
into $q \bar q$ pairs at the end of the cascade.}.
%****************ende der fussnote************************
This allows for checks of the validity of our analytic results in DLA.
Furthermore we can estimate the
effect of cluster formation and hadronization on these results.

\subsection{Correlations in the relative angle $\t12$}

First we consider the correlation function $\hat r (\varepsilon)$
normalized by the full
multiplicity in the forward cone.
This quantity approaches at high energies a limiting distribution
in $\varepsilon$, see Eq.(\ref{rh12ra}),
\begin{equation}
-\frac{\ln \hat r(\varepsilon)}{2\sqrt{\ln\frac{P\Theta}{\Lambda}}}
\simeq 2\beta (1-\frac{1}{2}\omega (\ve,2))
\labl{rhadlim}
\end{equation}
where we dropped the contribution from the $\varepsilon$-independent
prefactor.
As can be seen from Fig.~7.1 the results from
the parton MC nicely confirm the prediction of scaling in $\varepsilon$
in the energy interval considered and in the range of
$\varepsilon\leq 0.6$. For larger $\varepsilon$ one observes a
violation of scaling as expected from the influence of the
cutoff $Q_0$ (see also the insert in the figure).
The full curve represents the asymptotic limit of the
correlation function (\ref{rhadlim}). Here and in the following
we adjust the absolute normalization
of the curve to the data. Only the large leading terms in the exponent
can be considered as reliably calculated in our present approach, the
normalization is  an effect
of nonleading order and therefore may obtain large corrections.
The dashed curve represents the prediction
from the quark jet (\ref{rhqg}). If we used
our calculated normalization
the dashed curve would be shifted downwards by 0.1 unit.

Next we consider the same correlation in $\vartheta_{12}$, but with
the differential normalization of Eq.(\ref{r12})\footnote
{In the MC the normalizing function
$d^{(2)}_{prod}(\ve)$ (see after Eq.(\ref{the12})) has been
calculated by ``event mixing'', i.e.\ $\t12$ is the
angle between two particles in different events each one determined
with respect to the sphericity axis.}.
%****************ende der fussnote******************************
We would again expect a scaling in $\varepsilon$ for this
quantity, however, see
Fig. 7.2, the data do not approach the scaling limit,
Eq.(\ref{rascal}). The finite energy corrections derived
for quark jets from Eq.(\ref{r12at}), see Fig. 5.3a, shift
the peak of the distribution towards smaller $\varepsilon$,
as observed in the MC, but the energy dependence is
predicted to be much weaker. The scaling violations in the MC are found most
pronounced for large $\varepsilon$, i.e. small relative
angles $\t12$. Apparently the disagreement comes from
the normalizing function $d^{(2)}_{prod}
(\vartheta_{12})$ in Eq.(\ref{r12}). This function is
built from the 1-particle angular distribution $\rho^{(1)}
(\vartheta)$ which for small angles
becomes very sensitive to the orientation of
the jet axis. This will be discussed further below in the
last subsection together with a possibly improved definition
of this quantity.

\subsection{Multiplicity moments}
In Fig. 7.3 we show the normalized cumulant and factorial
moments in dependence
of $\varepsilon$ (by variation of $\delta$ in (\ref{epsi}))
for two primary energies and both dimensions $D=1,2$\footnote{
In the MC we obtained the 2D data by selecting tracks from
quadrangular areas with
$\Delta \theta=\Delta \phi = 2 \delta$ inside the ring.}.
The
theoretical curves for  $C^{(2)}$ are computed from Eq.(\ref{cmomno}).
The description of the MC data improves with
increasing energy. Again the normalization of the curves is
adjusted. From our
calculation in Sect. 6 we obtain for the $D=1$ moment
at $\epsilon=0$, i.e. for the full forward
cone of half opening $2\theta$, the result
$C^{(2)} (0) =1/(3a(P\theta)c_a)$ which yields 0.62 and 0.78 at 45 and
900 GeV resp. for gluon jets (times 9/4 for quark jets).
This is larger than the exact DLA result $C^{(2)}=
F^{(2)}-1=1/3$, see Eq.(\ref{f2sol}). Our larger values at
$\varepsilon=0$ are due to the approximation
$\delta\ll\theta$  in the calculations for our typically small
angular intervals.
The strong decrease of data at small $\ve$ is therefore not reproduced
by our calculation at the lower energies (see also Fig. 6.2).
Furthermore it is known
that the DLA overestimates the value of the global
$F_2$ and that correct inclusion of
energy-momentum conservation leads to a reduction of its
value \cite{MALW}.

In the r.h.s. of Fig. 7.3 we show the factorial
moments $(F^{(2)}=C^{(2)}+1)$. At high energies $C^{(2)}\sim
F^{(2)}$ in the exponential accuracy of DLA. We have plotted
our analytic results on $C^{(2)}$ moments also with the data for the $F^{(2)}$
moments. We observe that they fit actually better to
the $F$-moments which indicates a peculiar cooperation of
non-asymptotic effects.

At large $\varepsilon$ the size of the $D=1$
moments is underestimated. This may be related to our requirement
 $\vartheta_{12} <
\delta$ in our integration over the ring.

The moments with $D=2$ deviate strongly from the predicted
shape at large $\epsilon$. The origin of this behaviour is
presumably the influence of the
$Q_0$-cutoff at finite energies
(see also the impact of an exact calculation for $r
(\vartheta_{12})$ at large $\ve$ in Fig. 5.2), also a sensitivity to
the choice of the jet axis (see below) could play a role - differently
to the symmetric 1D moments. The
normalization calculated for $\ve=0$, $C^{(2)}
(0)=4/(3a(P\theta)c_a)$ in Eq. (\ref{c2nr})
would yield moments considerably larger
than the MC-data.

Next we test the universal behaviour of moments at high
energies predicted from Eq.(\ref{cmomno}). In order to
exhibit this universality we consider the quantity
\begin{equation}
\hat C^{(n)} ={\ln\lbrack(\delta/\vartheta)^{D(n-1)}
C^{(n)}\rbrack
\over    n\sqrt{\ln({P\vartheta\over\Lambda})}}
\labl{chat}
\end{equation}
and in complete analogy the quantity $\hat F^{(n)}$. This quantity
approaches for $\ve$ fixed, $P\to \infty$ the scaling limit
\begin{eqnarray} -\hat C^{(n)}&\simeq& 2\beta(1-\omega(\varepsilon,n)/n)
\labl{chatasy}\\
&\approx& 2\beta(1-\sqrt{1-\varepsilon})\labl{chatnasy}
\end{eqnarray}
where the last expression is independent of $n$ and follows
in leading $n$ approximation for $\omega$.
In Fig. 7.4 we plot $-\hat C^{(2)}$ at different energies vs.
$\varepsilon$ and the approach to the scaling limit Eq.(\ref{chatasy})
is nicely demonstrated. The 2D moments have
about the same dependence on $\varepsilon$ at all energies
although with an energy dependent normalization, but the
dependence on $\varepsilon$ is steeper than expected from
universality Eq.(\ref{chatasy}). In Fig. 7.5 we
show the MC results on $\hat F^{(2)}$. The 1D
moments show a rather nice scaling behaviour in
$\varepsilon$ with the expected $\varepsilon$-dependence of
Eq.(\ref{chatasy}) which sets in already at low energies.

Finally we consider the $n$-dependence of moments.
Results of analytic calculations
were shown already in Fig. 6.3. In the quantity $\hat C^{(n)}$ the
leading $n$ dependence of $\omega(\ve,n)\sim n f(\ve)+O(1/n)$
is removed. The residual $n$-dependence is shown by the curves
in Fig. 7.6. They are
reproduced reasonably well by the MC data.

Remarkably, all angular correlations calculated here by the MC
(except $r(\vartheta_{12})$)
approach a scaling limit in $\varepsilon$ as predicted. Even
very different quantities like $\hat r (\ve )$ and  the $D=1$ moments
$C^{(2)},F^{(2)}$ - they are calculated for particles in different regions of
phase space, also $\hat r$ is a differential, $C,F$ are integral quantities -
approach the same asymptotic
behaviour predicted by Eqs.(\ref{rhadlim},\ref{chatasy}),
disregarding the normalization.
An experimental verification of this prediction
would be an interesting confirmation of the universalities occuring
in the QCD parton cascade.

\subsection{Hadronization}
So far we have compared our analytic calculations with the
MC results at the parton level to check the quality of
our approximations. Next we investigate how the process of
hadron formation modifies the results obtained at the parton
level. For this purpose we apply the cluster model as
incorporated in the HERWIG MC \cite{HERWIG}. In fig. 7.7 we
compare MC results at the parton and hadron level. There is
good agreement in particular for small $\epsilon$ (large angles) but
some discrepancy towards the small angle cutoff.
The agreement further improves with increasing energy.

At sufficiently large angles the normalized quantities
considered in this paper are infrared safe, i.e. become
independent of the cutoff $Q_0$. It appears that in this
kinematic regime the hadronization effects are negligable
also, i.e. the model realizes local parton hadron duality.
This is an interesting new criterion for the applicability
of this phenomenological rule.

\subsection{Limitations of the DLA}
One important problem is the choice of the jet axis which
should correspond to the primary parton direction in the DLA.
This we have chosen to be the sphericity axis
(in some cases we also tried
the thrust axis without appreciable difference).
Some quantities, like the angular distribution at
small angles with respect to the axis, are quite sensitive
to the precise direction of the axis in an event,
whereas other, integral quantities,
like the multiplicity in the forward cone
with half angle $\Theta$ around the jet axis,
are less sensitive to this choice.
This problem is illustrated in Fig.~7.8 which shows the
distribution of partons $\rho^{(1)}(\eta)$
in pseudorapidity $\eta=-\ln(\vartheta/2)$ with
respect to the sphericity axis. Also shown is the DLA result -
using either the exact form (\ref{rhoang}) or the
high energy approximation (\ref{asyan1}) for $\rho^{(1)}(\vartheta)$ -
which scales in $x=P\th/\Lambda$ or $\ln x = \ln (2P/\Lambda)-\eta$.
We see that this scaling property is strongly violated by the MC data,
in particular for large $\eta\approx\eta_{max}$ or small
angles $\vartheta$. Clearly, the jet axis defined in a
multiparticle event does not coincide with the direction of
the primary parton. Therefore the distribution
$\rho^{(1)}(\vartheta)$ is not well defined at small angles.

This also explains the bad scaling properties of
the correlation function $r(\vartheta_{12})$,
Eq.(\ref{r12}), observed in Fig. 7.2. Namely, the normalizing function
$d^{(2)}_{prod} (\vartheta_{12})$ is proportional to
$\rho^{(1)} (\vartheta_{12})\bar n (\vartheta_{12})$, see
Eq. (\ref{dprodpa}), and therefore no good scaling properties
can be expected for small $\vartheta_{12}$. On the other
hand, the quantities $\hat r (\epsilon)$ and the $D=1$ moments
depend on the jet axis only through the large angles
$\Theta$ or $\vartheta$, furthermore there is a
rotational symmetry, therefore these observables are stable against
small fluctuations of the jet axis.

There is an interesting possibility to improve on this
\cite{DMO}. Instead of defining an observable with respect
to a global jet axis one calculates this observable with
respect to the directions of all particles in an event in turn, each one
weighted by the respective particle's energy. For the angular distribution,
for example, one defines
\begin{equation}
\rho_E^{(1)}(\th)=\int d^3k_1 d^3k_2\frac{|k_1|}{W}
\rho^{(2)}(k_1,k_2)\delta (\th -\Theta_{k_1k_2})
\labl{rhoe}
\end{equation}
This quantity is also shown in Fig.~7.8. As can be seen,
the scale breaking is largely removed. In this paper
we restrict ourselves to observables using the sphericity
axis. It will be interesting to study the possible improvements
using the modified definitions like (\ref{rhoe})
for our observables $r(\vartheta_{12})$ and also the $D=2$
moments to investigate the sensitivity to the jet axis.
A further improvement would be the calculation of quantities like (\ref{rhoe})
within DLA (see ref. \cite{DMO} for the azimuthal angle correlations).
This would eliminate the global jet axis from both the theoretical
and experimental analysis.

Another serious effect of the DLA is the neglect of the recoil.
The energy-recoil can be taken into account in the so-called
MLLA \cite{QCD}. Its effect for moments has been studied in \cite{DD}
and is found to be typically of the order of 10\%.
There is not yet any scheme available to include
the angular recoil. Its significance can be recognized from studying
$r(\t12)$ in the full angular region
$0\leq\t12\leq 2\Theta$. Note that in the DLA applied in
this paper there is a peak near $\t12\approx 0$ but the correlation is minimal
($r=1$) for $\Theta\leq\t12\leq 2\Theta$ because of angular ordering.
On the other hand the MC calculation yields a
second peak near $\t12\simeq 2\Theta$ (not shown) of the same strength
which is due to the
recoiling primary parton. Therefore the inclusion of the angular recoil
in the analytic calculations seems to be the most important problem for the
more
quantitative analysis of angular correlations.
\section{Summary and concluding remarks}
          \rem1{pow8v4}

Beginning from a single principle, i.e. the
master equation for the generating functional of the QCD cascade,
we have derived
a variety of predictions for angular correlations between partons.
In particular we have
calculated the distribution of the relative angle $\rho(\theta_{12})$, the
fully differential two-body angular correlations $\Gamma(\Omega_1,\Omega_2)$
and subsequently also the general $n-$parton correlations.
Multiplicity moments of general order
in a restricted angular phase space were also computed.
All calculations were done for constant and running \al, and
the correspondence between these results was always demonstrated.
The double logarithmic approximation applied here provides the correct
high energy limit. Some nonleading effects, namely the ones which are
included in the master equation were also investigated.

This work summarizes and extends our earlier studies
\cite{OW1}-\cite{WOSIEKC}.
Some of the results on multiplicity moments were obtained also by other
groups \cite{DMO,DD,BMP} as mentioned in the introduction.

An interesting aspect of the angular correlations considered here
is their sensitivity to the soft gluon interference as the angular
ordering prescription enters all calculations in an essential way.
Indeed, in a recent study of polar angle correlations \cite{SYED},
using a quantity very similar to our $\rho(\theta_{12})$, a large
effect  from the angular ordering has been noted - unlike in the
case of the azimuthal angle or momentum correlations.

{\it Universal $\epsilon$ - scaling.}
We have studied in detail the emergence of the
power dependence on the scale characteristic in the given process.
Whenever the observable in question depends on the
single scale only, then, in the high energy limit, it is a simple
power of that scale. This is the case for the total multiplicity
and factorial moments in a forward cone, and for the angular distribution,
e.g. $\bar n\sim Q_1^{a(Q_1)}$. For fixed \al (constant anomalous
dimension $a$) one obtains a power law in $Q_1$ whereas in case of
running \al the power law is violated by the scale dependence of $a(Q_1)$.

For more complicated reactions, which in general depend on two or more
 independent scales, no straightforward power behaviour exists. We have
found instead a scaling behaviour of a new type which applies universally to a
large number of at first sight not related angular observables.
In case of two scales $Q_1, Q_2$, the distribution
of the relative angle and the second cumulant moment
in the sideway cone, for example,
show the universal scaling behaviour
\begin{equation}
\rho \sim \exp{\left(\beta \sqrt{\ln Q_1} f(\epsilon) \right)}, \labl{scal}
\end{equation}
where $f(\epsilon)$ is a calculable
universal function of the ratio of the logarithm of the two scales
$\epsilon={\ln{Q_2}\over \ln{Q_1} }$. Up to some simple
kinematic factors, the scaling function $f(\epsilon)$ is the same for
both quantities and analogous functions in $\epsilon$ occur for
the higher correlations.
The general result, Eq.(\ref{scal}), implies that the simple power
dependence on only one scale can be recovered in the "diagonal" scaling
limit for fixed $\epsilon$,
e.g. when $Q_1, Q_2 \rightarrow \infty, Q_1=Q_2^{\epsilon}$. Indeed
then Eq.(\ref{scal}) reduces to
\begin{equation}
\rho\sim Q_1^{a(Q_1)f(\epsilon)}
\labl{scalp}
\end{equation}
with the anomalous dimensions depending on $\epsilon$. The same
phenomenon was found for the single particle momentum distribution,
Eq.(\ref{rosim}),
which also depends essentially on two scales.

The ``$\epsilon$-scaling'', Eq.(\ref{scal}), has interesting
experimental consequences.
The observable quantity $f(\epsilon)
\sim \ln \rho /\sqrt{\ln Q_1}$ should depend on the three relevant variables
(for example $\theta,\delta,P$) only through the single variable
$\epsilon$, i.e. two variables are redundant.
Preliminary results from LEP \cite{MANDL} have verified recently
this scaling prediction for the polar angle correlation \cite{OW2}.
%showing its independence of the cone opening angle in the range
The $\epsilon$-scaling holds approximately for the opening angles
$\Theta$ in the range
$30^o < \Theta < 60^o$.

{\it Intermittency and fractal structure.}
Results of this work have a direct application to the intermittency
studies which have recently attracted a lot of attention.
This property of the parton cascade is obtained from the
general formula (\ref{scal}) in the limit of small $\epsilon$,
i.e. sufficiently large opening angles, in the
linear approximation of $f(\epsilon)$.
The suitably normalized observables then behave like $r\sim Q_2^{ca(Q_1)}$,
i.e. there is a power law in one scale $Q_2$ (say $\theta/\delta$) if the
other scale $Q_1$ (say $P\theta$) is kept fixed. So the
perturbative cascade is intermittent in the appropriate angular variables
and intermittency exponents
are available analytically, c.f. Eq.(\ref{cmomno}).
For fixed \al the power behaviour for the angular
correlations is actually
approached in the full angular region for high energies.

This power dependence can be related to the selfsimilar structure
of the partonic branching process. One can assign a fractal dimension $D_n$
to the normalized multiplicity moments and this is found to be
given in terms of the anomalous dimension of total multiplicity
$a=\gamma_0=\sqrt{6\alpha_s/\pi}$,
see Eq. (\ref{dfrac}).

{\it Monte Carlo tests.}
We have also confronted the analytical predictions with Monte Carlo
simulations which don't have many of the limitations of the
double logarithmic approximation. The results are rather satisfactory.
One indeed sees the approximate $\epsilon$-scaling
(\ref{scal}) and the shape of the
scaling function is approached with increasing energy.
The factorial moments are closer to the asymptotic limit than the
cumulant moments.
In the small $\epsilon$ region
the correlation functions and moments are independent
of the $p_T$ cutoff $Q_0$. In this region we find the hadronization effects
negligable in the Monte Carlo (HERWIG) calculations.
On the other hand deviations occur for particles with small angular
separation.

To conclude, there exists a rich spectrum of analytic predictions from
the perturbative QCD for multiparton correlations. The double
logarithmic approximation provides the leading order predictions
applicable quantitatively
at very high energies ($\sim 1 $ TeV) but some interesting experimental tests
of the theory in its relatively unexplored regime
are suggested already at present energies.

\section*{Acknowledgements}
J. W. thanks the Theory Group at the Max-Planck-Institite for the
support and for the warm atmosphere enjoyed during this work.
\section*{Appendix A}
 \rem1{powapav2}
     %This is the file powapgv2.tex as of 21.IV.94.J
In this Appendix we shall derive asymptotic form of the solution
 Eq.(\ref{hnrsim}).
Consider first the case of constant \al. This will allow to rederive
the results obtained also directly from the
correlation function, cf. Eq.(\ref{hm2}), and will provide 
some orientation about the relevant range of variables contributing to the
integrals in question. Inserting moments, Eq.(\ref{muca}), into
Eq.(\ref{hnrsim}) gives for the leading term
\begin{eqnarray}
\overline{h}_n(\delta,\theta,P)&=&\int_{\gamma}{ds\over 2\pi i}
{\sqrt{s^2+4a^2}+s\over 2\sqrt{s^2+4a^2}} \\ \nonumber
&& \int_{Q_0/\delta}^P {dK\over K}
\exp{\left[{1\over 2}\ln{P\over K}(\sqrt{s^2+4a^2}-s)\right]} \\ \nonumber
&& \int_{\delta}^{\theta}
{d\Psi\over\Psi}
\exp{\left[{1\over 2}\ln{\theta\over\Psi}(\sqrt{s^2+4a^2}+s)\right]}
d_n(K,\delta), \labl{ds1}
\end{eqnarray}
with the source term $d_n(K,\delta)={\cal  D}_n (K\delta/Q_0)^{na}$.
After changing the variable in the inner integration to $u=\ln{(\theta/\Psi)}$,
we can extend the upper limit to the infinity. If the contour $\gamma$
is properly chosen this does not affect the leading behaviour of the result.
One obtains
\begin{eqnarray}
\lefteqn{\overline{h}_n(\delta,\theta,P)={\cal D}_n \int_{\gamma}{ds\over 2\pi i}
{1\over \sqrt{s^2+4a^2}} } \\ \nonumber
&& \int_{Q_0/\delta}^P {dK\over K}
\exp{\left[{1\over 2}\ln{P\over K}(\sqrt{s^2+4a^2}-s)\right]}
\left({K\delta\over Q_0}\right)^{na}, \labl{ds2}
\end{eqnarray}
Similar procedure for the $K$ integration gives the pole in $s$
\begin{eqnarray}
\overline{h}_n(\delta,\theta,P)&=&{\cal D}_n \left({P\delta\over Q_0}\right)^{na}
\int_{\gamma}{ds\over 2\pi i} {1\over \sqrt{s^2+4a^2}}
{2\over (\sqrt{s^2+4s^2}-2s)-na}  \\  \nonumber
&& \exp{\left[{1\over 2}\ln{\theta\over\delta}(\sqrt{s^2+4a^2}+s)\right]},
 \labl{ds3}
\end{eqnarray}
This pole saturates the contour integral giving finally
\begin{equation}
\overline{h}_n(\delta,\theta,P)={\cal D}_n \left({\delta\over\kappa}\right)^{na}
\left({\theta\over\delta}\right)^{\frac{a}{n}},
 \labl{ds4}
\end{equation}
\rem1{(III.II.F.2)}
where $\kappa=Q_0/P$. Normalization ${\cal D}_n$ follows from Eq.(\ref{frul}).
We obtain
\begin{equation}
{\cal D}_n={a F_n \over \pi 2^{n+1} n},
\end{equation}
with $F_n\equiv \lim_{x\rightarrow\infty} f_n(x)/n_{tot}(x)^n$. With
$F_2=4/3$, cf. (\ref{f2sol}), Eq.(\ref{ds4}) reproduces
Eq.(\ref{hm2}) for $n=2$.
 
\noindent{\em Runnung} \al. Similarly to the previous case the saddle
point $s^{*}$ in Eq.(\ref{hnrhyp}) scales as the coupling constant hence
$s^{*}\sim 1/\sqrt{\ln{P}}$.
Therefore it is sufficient to use the appropriate asymptotic forms of
the confluent hypergeometric functions. They can be derived from the
well known integral representations of the solutions of the
conflunt hypergeometric equation \cite{erdy}. We obtain 
\rem1{(III.II.6.1-3)}
\begin{eqnarray}
\lim_{a\rightarrow +\infty,z\rightarrow +\infty,b=const} F(a,b,z)&=&
{1\over\sqrt{2\pi}}({z\Delta\over 4})^{1-b\over 2} (z^2+z\Delta)^{-{1\over 4}}
 \\  \nonumber
\exp{\left[{1\over 2}(\sqrt{z^2+z\Delta}+z)\right]}&&
\exp{\left[{\Delta\over 2}\ln{(\sqrt{z\over\Delta}+\sqrt{{z\over\Delta}+1}) }
\right]  }.   \labl{agfas}
\end{eqnarray}
\begin{eqnarray}
\lim_{a\rightarrow\infty,z\rightarrow\infty,b=const} \Phi(a,b,z)&=&
z^{{1\over 4}-{b\over 2}} (z+\Delta)^{-{1\over 4}} e^{\Delta\over 4}\\
  \nonumber
\exp{\left[-{1\over 2}(\sqrt{z(z+\Delta)}-z)\right]} & &
\exp{\left[-{\Delta\over 2}\ln{\sqrt{z}+\sqrt{z+\Delta}\over 2}\right]}.
\labl{uas}
\end{eqnarray}
where $\Delta=4a-2b$.
This implies the following asymptotics for the ${\cal A}$ and ${\cal F}$
functions required in Eq.(\ref{hnrhyp}). \rem1{(III.II.F.0)}
\begin{eqnarray}
{\cal F}(s,x)\simeq&\sqrt{s\over 2\pi \beta^2 p_x}& e^{{x\over 2}g_{-}(s,x)}
\exp{\left[{2\beta^2\over s}\ln{({\sqrt{x}\over 2\beta}g_{+}(s,x)})\right]},
\labl{fas}\\
{\cal A}(s,\lambda)\simeq &\sqrt{2\pi \beta^2\over s} 
{1+p_{\lambda}\over 2\sqrt {p_{\lambda}}}
& e^{-{\lambda\over 2}g_{-}(s,\lambda)}
\exp{\left[-{2\beta^2\over s}\ln{({\sqrt{\lambda}\over 2\beta }
g_{+}(s,\lambda)})\right]}, \labl{aas} \\
g_{\pm}(s,x)&=&\sqrt{s^2+4\beta^2/x} \pm s, \labl{afas}
\end{eqnarray}
where $p_{\tau}\equiv\sqrt{1+4{\beta^2\over s^2\tau}}.$
Now  Eq.(\ref{hnrhyp}) can be rewritten as
\begin{equation}
\overline{h}_n(\delta,\theta,P)=\int_{\gamma}   \exp{(s\ln{\theta\over\delta})}
{\cal F}(s,\ln{P\theta\over\Lambda}) \overline{u}_n(s)\exp{w_n(s,\ln{P\delta\over\Lambda})}
{ds\over 2\pi i},
\labl{ds5}
\end{equation}
where
\begin{eqnarray}
\overline{u}_n(s) \exp{w_n(s,\ln{P\delta\over\Lambda})}=&
\int_{\ln{Q_0/\Lambda}}^{\ln{P\delta/\Lambda}} d\sigma  & u_n(s,\sigma)
\exp{v_n(s,\sigma)},
\labl{ds66} \\
v_n(s,\sigma)=&-{\sigma\over 2} g_{-}(s,\sigma)-&
{2\beta^2\over s}\ln{({\sqrt{\sigma}\over 2\beta} g_{+}(s,\sigma))}
+2n\beta\sqrt{\sigma},   \nonumber  \\
u_n(s,\sigma)=&\sqrt{2\pi \beta^2\over s} &{1+p_{\sigma}\over 2\sqrt{p_{\sigma}}} 
{\cal D}_n,
\labl{dsigma}
\end{eqnarray}
and the prefactor resulting from the inhomogenous term $d_n(\sigma)=$
${\cal D}_n \exp{(2n\beta
\sqrt{\sigma)}}$ follows from Eqs.(\ref{frul},\ref{fqas}) and (\ref{ffflll}). 
\rem1{(III.II.H.1)}
\begin{equation}
{\cal D}_n={ a(\sigma) F_n\over 2\pi n} \left( {f\over 2\sqrt{a(\sigma)}}\right)^n .
\end{equation}
Saddle point condition for the $\sigma$ integration
$\partial_{\sigma} v_n(\sigma,s)=0$ gives \rem1{(III.II.F.3)}
\begin{eqnarray}
g_{-}(s,\sigma)={2n\beta\over\sqrt{\sigma}},
\end{eqnarray}
which has the solution
\begin{equation}
\sigma^{\star}={\beta^2(1-n^2)^2\over s^2 n^2}. \labl{sadsig}
\end{equation}
At the saddle
\begin{equation}
v_n(s,\sigma^{\star})={\beta^2\over s}(1-n^2)+
{2\beta^2\over s}\ln{n}\equiv w_n(s). \labl{vnstar}
\end{equation}
which is independent of the $P\delta$ and corresponds to the insensitivity
of the leading contribution to the integral (\ref{ds66}) to the upper limit
of integration. The prefactor reads
\begin{equation}
\overline{u}_n(s)=\sqrt{2\pi \over \ddot{v}_n(\sigma^{\star})} u_n(\sigma^{\star})= 
 {\beta^2 F_n\over  s^2 n} \sqrt{2(n^2-1)}   \left( {f\over 2}\right)^{1-{n\over 2}} 
a(\sigma^{\star})^{1-{n\over 2}}. 
\end{equation}
We now follow similar procedure for the $s$ integral. Writing
\begin{equation}
h_n(\delta,\theta,P)\simeq\int_{\gamma} U_n(s)\exp{W_n(s,L,l)} {ds\over 2\pi i},
\end{equation}
we have from (\ref{fas},\ref{aas},\ref{ds5},\ref{vnstar})
 $L=\ln{(P\theta/\Lambda)},
l=\ln{(P\delta/\Lambda)}$ \rem1{(III.II.F.4)}
\begin{eqnarray}
W_n(s,L,l)&=&
 s(L-l)+\\{1\over 2} L g_{-}(s,L)& +&
{2\beta^2\over s} \ln{[{n\sqrt{L}\over 2\beta} g_{+}(s,L)]}
-{\beta^2\over s}(n^2-1), \labl{Wn}
\end{eqnarray}
and
\begin{equation}
U_n(s,L)=\sqrt{s\over 2\pi\beta^2 p_L} \overline{u}_n(s).
\end{equation}
 Saddle point condition, $\partial_s W_n(s,L,l)=0$, gives after some
algebra Eq.(\ref{alg}) with $z=s\sqrt{L}/\beta$ \rem1{(III.II.F.4)}.
In addition Eq.(\ref{Wn})
can be simplified at the saddle point reducing finally to
\begin{equation}
W_n(s^{\star},L,l)=2\beta L \omega(\epsilon,n),
\end{equation}
 and the overall prefactor
\begin{equation}
 \Pi_n(s^*)={1\over \sqrt{2\pi W^{''}_n(s^{\star})}} U_n(s^{\star})
\equiv {\cal D}_n {\cal P}_n,
\end{equation}
which after some algebra gives Eq.(\ref{presol}) for ${\cal P}_n$ .
 It is easy to 
check that in the constant \al limit our Eq.(\ref{finsol}) reduces to the
result (\ref{ds4}) derived by the exact integration of the resolvent
 representation (\ref{ds1}).

\section*{Appendix B}
 \rem1{powapbv1}
     In this Appendix we shall derive the constant \al  limit of
the energy moments, Eq.(\ref{momsol}). This is done by leting
$\beta,\lambda\rightarrow\infty $ while keeping the ratio
$\beta^2/\lambda\equiv a $ constant. In this process
the first parameter and the independent variable of the
hypergeometric confluent functions become large. Appropriate
asymptotic forms of these functions are derived in Appendix A.
Using Eqs.(\ref{agfas}) and (\ref{uas}) one obtains for the ingredients
of the first part of Eq.(\ref{momsol})
\begin{eqnarray}
\lefteqn{
\Phi(\alpha-1,1,s\lambda) \simeq (s\lambda)^{-1/4}  (s\lambda+4\beta^2/s-2)^{-1/4} 
\exp{ ( \beta^2/s-1/2) } 
             } \nonumber  \\  & &  
  \exp{\left[ -{1\over 2}\left(\sqrt{s\lambda(s\lambda+4\beta^2/s-2)}-
                   s\lambda\right) \right]} \nonumber \\ &&
    \exp{\left[-\left(  {2\beta^2\over s}-1\right) 
  \ln{
            \sqrt{s\lambda}+\sqrt{s\lambda+4\beta^2/s-2}\over 2
        } \right]
                  },  \nonumber
\end{eqnarray}

\begin{eqnarray}
 \lefteqn{ 
     F(\alpha,2,s(X+\lambda))  \simeq  
 {  1   \over  \beta(X+\lambda)\sqrt{2\pi s}  }
      \left( 1+{4\beta^2\over s^2(X+\lambda)}\right)^{-1/4}  
  \exp{\left({\beta^2\over s}\ln{ s^2(X+\lambda)\over 4\beta^2  }\right) } 
              }  \nonumber  \\  &  &
    \exp{    \left[
          {s(X+\lambda)\over 2}
           \left(
                    \sqrt{1+    {4\beta^2\over s^2(X+\lambda)}   }+1
          \right)
               \right]
           }   
     \exp{  \left[       {2\beta^2\over s}
                             \ln{  
                                    \left( 
                                    1+\sqrt{   1+{4\beta^2\over s^2(X+\lambda)}   }
                                    \right)
                                   }
                    \right]   
                }.   \nonumber
\end{eqnarray}
All symbols are as in Eq.(\ref{momsol}) except that $n$ was replaced by the 
continuous
variable $s$.
Constant \al limit is performed at fixed values of the kinematical variables.
In particular, $X/\lambda << 1$ and $s \lambda >> 1$ in this limit.
Expanding above expressions and keeping track of the $O(\lambda)$ and
$O(1)$ terms in the exponents we obtain
\begin{eqnarray}
\lefteqn{
              \Phi(\alpha-1,1,s\lambda)\simeq  
                        {        s+\sqrt{s^2+4a^2}     \over      2\sqrt{s  \sqrt{s^2+4a^2}   }          }
                       \exp{(\beta^2/s-1/ 2)} 
             }  \\  & &
\exp{\left(-{\lambda\over 2}(\sqrt{s^2+4a^2}-s) \right)}  
\exp{\left( s\over 2\sqrt{s^2+4a^2}\right) }
\exp{\left( -{\beta^2\over s}\ln{(s\lambda)}\right) }   \nonumber  \\ &&
\exp{\left(-{2\beta^2\over s}\ln{s+\sqrt{s^2+4a^2}\over 2 s}\right) }
\exp{\left(2a^2\over \sqrt{s^2+4a^2}(s+\sqrt{s^2+4a^2})\right)},   \nonumber
\end{eqnarray}
\begin{eqnarray}
\lefteqn{
F(\alpha,2,s(X+\lambda))  \simeq 
{1\over \beta(X+\lambda) \sqrt{2\pi \sqrt{s^2+4a^2}} } 
\exp{\left(-{a^2 X\over \sqrt{s^2+4a^2}}\right)} 
              } \\  & &
\exp{ \left( {X+\lambda\over 2}(\sqrt{s^2+4a^2}+s) \right) }  
%  \exp{\left( {\beta^2\over s}\ln{s^2(X+\lambda)\over \beta^2}\right)  }  \\    \nonumber   & & 
  \exp{\left( {\beta^2\over s}\ln{s^2\lambda\over \beta^2}\right)  } 
\exp{\left({a^2 X\over s} \right)}    \nonumber  \\   & & 
\exp{\left( {2\beta^2\over s}\ln{s+\sqrt{s^2+4a^2}\over 2s}\right ) }
\exp{\left( -{4a^4 X\over s \sqrt{s^2+4a^2} (s+\sqrt{s^2+4a^2})} \right)}.  \nonumber
\end{eqnarray}
 Where we have already replaced the $O(1)$ ratio $\beta^2/\lambda$ by $a$.
Finally, inserting these expressions into Eq.(\ref{momsol})  
 and using  Stirling formula for the  gamma function,  we see that the
  divergent,  leading terms canel and the first nonleading corrections
give the finite result which reproduces the first part of  
Eq.(\ref{muca}). Second part is obtained in the analogous way from the
second term of Eq.(\ref{momsol}).

 \section*{Appendix C}
  \rem1{powapcv1}
     %This is the file powapfv1.tex as of 9.III.94
We shall discuss here some details of the derivation of the simplified
resolvent representation, Eq.(\ref{hnrsim}). First we will derive
Eq.(\ref{hnres}). To this end recall the resolvent solution
for the two dimensional inclusive density, Eq.(\ref{enfres}). Performing
the steps outlined there one obtains with the running \al
\begin{equation}
\rho(k,\theta,P)=a^2(k,\theta)+\int_{k}^{P} \frac{dK}{K}
\int_{Q_0/k}^{\theta}
\frac{d \Psi}{\Psi} R\left({P\over K},{\vartheta\over\Psi}
, {K\Psi\over\Lambda} \right) a^2(K,\Psi).
 \labl{rhoenres}
\end{equation}
Formula for the momentum distribution follows from Eq.(\ref{rhoenres})
upon the angular integration. Changing orders of the $\theta$ and $\Psi$
integrations, recalling the definition (following Eq.(\ref{req2}))
of $\overline{R}$ and using the substitution rules (\ref{subst}) gives
Eq.(\ref{hnres}). In the reduced variables
$x=\ln{\theta\over \delta}, t=\ln{P\delta\over Q_0}, z=\ln{\Psi\over\delta},
\tau=\ln{K\delta\over Q_0}, \lambda=\ln{Q_0\over\Lambda}$ it reads
\begin{equation}
h_n(x,t,\lambda)=d_n(t+\lambda)+\int_{0}^{x} dz
\int_{0}^{t} d\tau
 \overline{R}(x-z,t-\tau,
z+\tau+\lambda) \partial_{\tau} d_n(\tau+\lambda).
 \labl{enresred}
\end{equation}
Integrating by parts and noting that $d$ is independent of $z$ we obtain
\begin{equation}
h_n(x,t,\lambda)=d_n(t+\lambda)+
\int_{0}^{t} d\tau
 \tilde{R}(x,t-\tau,\tau+\lambda) d_n(\tau+\lambda).
 \labl{enresredtwo}
\end{equation}
where we have introduced
\begin{equation}
\tilde{R}(x,t-\tau,\tau+\lambda)=\int_{0}^{x}
\partial_t \overline{R}(x-z,t-\tau,z+\tau+\lambda).
\end{equation}
Because $\overline{R}$ has the Laplace representation, Eq.(\ref{hnrhyp}),
which is dominated by $s\sim O(1/\ln{P\over Q_0})$, one can set to the
leading order $\partial_1 \overline{R}(x_1,x_2,\lambda)=\partial_2
\overline{R}(x_1,x_2,\lambda)$. Therefore finally
\begin{equation}
\tilde{R}(x,t-\tau,\tau+\lambda)=\int_{0}^{x} dz \partial_x
\overline{R}(x-z,t-\tau,z+\tau+\lambda)=\overline{R}(x,t-\tau,\tau+\lambda),
\end{equation}
since the contribution from the upper limit $z=x$ is negligible \footnote{We
were also consistently neglecting terms proportional to $\partial_{\lambda}
\overline{R}(x_1,x_2,\lambda)$ since they are nonleading comparing to the
derivatives with respect to the first two arguments.}.
This concludes the proof of Eq.(\ref{hnrsim}) which can be also simply
summarized as the dominance of the lower limit in the $\Psi$ integration
in Eq.(\ref{hnres}).

 \section*{Appendix D}
  \rem1{powapdv2}
    %This is the file powap0v3.tex as sent by W on 21 Jul 1994.
\rem1{powap0v3}
In this Appendix we shall derive the asymptotic behaviour, Eq.(\ref{fqas}),
of the total factorial moments, and the recursive relation (\ref{recur})
for constant \al. First, we simplify the resolvent solution 
for $f_n(X)$. Changing the integration
variables $Z$ and $\zeta$  in Eq.(\ref{fnres}) to $t=Z+\Xi $ and $v=Z+\zeta $
and interchanging the order of $t$ and $v$ integrals one can perform the
$t$ integral to obtain
\begin{equation}
f_n(X)=b_n(X)+a\int_0^X \sinh{(a(X-v)) } b_n(v) dv, \labl{shres}
\end{equation}
with $X=Y+\Xi=\ln{(P\Theta/Q_0)}$.
We now prove Eq.(\ref{fqas}) by induction. For $n=2$,
\begin{equation}
b_2(X)=<n>^2\simeq
B_2 \exp{(2aX)}/2^2.
\end{equation}
Assume now that
\begin{equation}
b_n(X)=B_n \left( {1\over 2} e^{aX} \right)^n.  \labl{bas}
\end{equation}
Inserting (\ref{bas}) into (\ref{shres}) one easily verifies (\ref{fqas}) with
\begin{equation}
F_n={n^2\over n^2-1} B_n. \labl{nsquared}
\end{equation}
Since $b_n(X)$ is built recursively from the products of moments of
{\em lower} order (for example $b_3(X)=<n>^3+3<n>(f_2(X)-b_2(X))$) with the
same combined anomalous dimension, the form (\ref{bas}) follows. This
completes the inductive proof of Eq.(\ref{fqas}).

In order to derive now the recurrence relation (\ref{recur}) we rewrite
the master equation, Eq.(\ref{mez}), for the generating function
of the global moments, i.e. we set $u(K)\equiv u$. In the high energy
limit the nested terms are negligible which amounts to puting $u=1$
in the exponent of Eq.(\ref{mez}). With these simplifications Eq.(\ref{mez})
reads
\begin{equation}
Z_P(u)=\exp{(\int d^3 K {\cal M}_P(K) (Z_K(u) - 1) ) }.
\end{equation}
Differentiating $n$ times one obtains
\begin{equation}
f_n(X)=\sum_{k=1}^{n-1} {n-1\choose k} f_k(X) g_{n-k}(X) + g_n(X),
\end{equation}
where we have introduced the cumulant moments
\begin{equation}
g_m(X)\equiv \int d^3K {\cal M}_P(K) Z^{(m)}_K(u)_{|u=1}= f_m(X)-b_m(X).
\end{equation}
Using the last equality and Eqs(\ref{fqas},\ref{bas}) together with
Eq.(\ref{nsquared}) to eliminate $g_m$ in favour of $f_m$ gives
the recursion Eq.(\ref{recur}).

 \section*{Appendix E}
  \rem1{powapev1}
  %this is the file powapav2.tex as sent by W on 21 Jul 1994.
\rem1{powapav2}
In this Appendix we shall derive the series representation Eq.(\ref{rhobess})
for the distribution of the relative angle $\theta_{12}$  for constant
\al . The contributions from the product and nested terms will be calculated
separately. \newline
A. {\em Product term.} Inserting Eqs.(\ref{resI}) and (\ref{rhoprod})
into Eq.(\ref{sol12}) and expanding in powers of $a$ one obtains upon
integration over $\Psi$ ($\kappa=Q_0/P$)
\begin{eqnarray}
\gamma_{prod}&\equiv&\rho^{(2)}_{prod}(\theta_{12})-
d_{prod}^{(2)}(\theta_{12})
\\  \nonumber
   &= & {2a^2\over\theta_{12}}\sum_{r=0}^{\infty}{a^{2r+1}\over r!(r+1)!}
\ln^{r+1}{\left(\Theta\over\theta_{12}\right)} f_r\left({\theta\over
\kappa}\right),
\end{eqnarray}
with
\begin{equation}
f_r(y)=\sum_{m=0}^{\infty} {4^m-1\over (2m+1)!} \int_0^y (y-z)^r z^{2m+1} dz.
\end{equation}
Integration over $z$ is proportional to the Euler beta function \newline
$B(\mu,\nu)=\int_0^1 t^{\mu -1} (1-t)^{\nu-1}$. Changing the orders of
the $r$ and $m$ summations we obtain
\begin{equation}
\gamma_{prod}={2a\over \theta_{12}} \sum_{m=0}^{\infty}(4^m-1) \left\{
\overline{y}^{m+1/2}
I_{2m+1}(z) - {(a\ln{\vartheta/\kappa})^{2m+1}\over (2m+1)!} \right\}.
\end{equation}
Where $\overline{y}=y^2/4$ and $y$ and $z$ are defined below Eq.(\ref{rhobess}).
The second term in the curly bracket gives after summation the inhomogenous
term $d_{prod}$, therefore the inclusive density itself is given by the first
term only. \remark{(II.15.4.3)} \newline
B. {\em Nested term}. Calculation of this contribution is entirely analogous
to the previous one. After expanding $d_{nest}$, Eq.(\ref{rhonest})
and the resolvent (\ref{resI}) one is left with elementary integrals.
Final result for the nested part reads \remark{(II.16.2.2)}
\begin{equation}
\rho^{(2)}_{nest}(\theta_{12})={2a\over \theta_{12}}\left( \sum_{m=0}^{\infty}
\overline{y}^{m+1/2} I_{2m+1}(z) - \sinh{(a\ln(\theta_{12}/\kappa))} \right).
\end{equation}
Adding both contributions gives the result (\ref{rhobess}).

\section*{Appendix F}
       \rem1{powapfv2}
          In this Appendix we integrate the formula, Eq.(\ref{rhobess}) to obtain the
total factorial moment $f_2$ as given by Eq.(\ref{f2sol}). Integration
of the second term in (\ref{rhobess}) is trivial hence we write
\begin{equation}
\int_{\kappa}^{\Theta} d \theta_{12} \rho^{(2)}(\theta_{12},P,\Theta)=
J\left(a\ln\left({\Theta\over\kappa}\right) \right) - 2
\left(\cosh\left(\ln{\Theta\over\kappa}\right)-1\right).  \labl{Jint}
\end{equation}
with $J(aX), X=\ln{(\Theta/\kappa)}$ defined as the integral of the first part of Eq.(\ref{rhobess}).
Expanding modified Bessel functions one obtains
\begin{eqnarray}
J(aX)&=&\sum_{m=0}^{\infty}\sum_{k=0}^{\infty} {2^{2m+1} a^{2m+2k+2}
\over k! (k+2m+1)! } \int_0^X t^{2m+k+1} (X-t)^k dt \\ \nonumber
 &=&\sum_{m=0}^{\infty} 2^{2m+1} \sum_{k=0}^{\infty}
{(aX)^{2m+2k+2}\over (2m+2k+2)!},
\end{eqnarray}
which can be resummed
\begin{eqnarray}
J(Y)&=&\sum_{l=0}^{\infty} {Y^{2l+2}\over (2l+2)!} \sum_{m=0}^{l}
2^{2m+1} \\ \nonumber
 &=& {2\over 3}\left(\cosh{(2Y)}-\cosh{(Y)} \right).  \labl{Jfin}
\end{eqnarray}
Inserting (\ref{Jfin}) into (\ref{Jint}) gives
\begin{equation}
\int_{\kappa}^{\Theta} \rho^{(2)}(\theta_{12}) d\theta_{12} = {4\over 3}
\left( \cosh(aX) - 1\right)^2,
\end{equation}
in agreement with Eq.(\ref{f2sol}).  \remark{(II.16.4)}

\section*{Appendix G}
       \rem1{powapgv4}
         %This is the file powapcv1.tex as of 16.III.94.J.
We derive here the asymptotic form, Eq.(\ref{selfsim}), of the
correlation function (\ref{rhobess}) at high energies
$\kappa=Q_0/P \to 0$.
In this limit
\begin{equation}
y=2\sqrt{{\ln{(\theta_{12}/\kappa)}\over
\ln{(\Theta/\theta_{12})} } }
\to \infty,\;\;\;
z=2a\sqrt{\ln{(\theta_{12}/\kappa)} \ln{(\Theta/\theta_{12})}}
\to \infty,
\end{equation}
 with
\begin{equation}
 z/y=a\ln{(\Theta/\theta_{12})}\equiv \sigma^2/2=const. \label{zlim}
\end{equation}
Using the sum rule,
\begin{equation}
\sum_{k=0}^{\infty} y^k I_{\nu+k}(z) = z^{-\nu}e^{yz/2}\int_0^z
e^{-y\tau^2/(2z)}I_{\nu-1}(\tau) \tau^{-\nu} d\tau,\;\;\; \nu>0,
  \label{kam}
\end{equation}
one can rewrite the sum in Eq.(\ref{rhobess}) as
\begin{equation}
\rho^{(2)}(\theta_{12},P,\Theta)= {a \over \sigma^2 \theta_{12}}
\left(
  e^{ z y/2} \int_0^z e^{-\tau^2/\sigma^2} I_0(\tau) \tau d\tau
+ e^{-z y/2} \int_0^z e^{ \tau^2/\sigma^2} I_0(\tau) \tau d\tau
\right)
%+O\left( \left(\theta_{12}/\kappa\right)^a \right)
, \label{w1}
\end{equation}
with the first  term $ O( (\theta_{12}/\kappa)^{2a})$
 dominating the high energy behaviour. We have neglected the terms
$ O( (\theta_{12}/\kappa)^{a})$.
Since the integrand in the leading term
is well behaved at large $\tau$, we can extract the large $P$
limit  by setting
the upper bound to infinity. This gives
\begin{equation}
\rho^{(2)}(\theta_{12},P,\Theta)= {a\over\theta_{12}}
\left( {\theta_{12} \over \kappa} \right)^{2a} \int_0^\infty e^{-u^2}
I_0(\sigma u) u du,
\end{equation}
which can be integrated exactly to give finally
\begin{equation}
\rho^{(2)}(\theta_{12},P,\Theta)={a\over 2\theta_{12}}
\left( {\theta_{12} \over \kappa} \right)^{2a} e^{ \sigma^2/4}.
\end{equation}
This is identical with Eq.(\ref{selfsim}) after taking into account
 Eq.(\ref{zlim}).

\section*{Appendix H}
        \rem1{powaphv1}
         %This is the file powapdv2.tex as of 8.VIII.94.J.
In this Appendix we shall perform several consistency tests of
our results for the two-parton correlations.

First, we 
show that the form, Eq.(\ref{cora1}), of the
fully differential connected angular correlation is consistent with the
Eq.(\ref{sol12}) for the correlation in the relative angle. It suffices
to show this compatibility for the connected parts only, since the relation
between the direct terms was already established (cf. Eqs (\ref{dirterm})
and (\ref{rhoprod}) ). To this end we define
\begin{equation}
\Gamma^{(2)}(\theta_{12}) = \int d\Omega_1 d\Omega_2 \delta(\theta_{12}-
\Theta_{k_1 k_2}) \Gamma^{(2)}(\Omega_1,\Omega_2),
\end{equation}
with $\Gamma^{(2)}(\Omega_1,\Omega_2)$ given by Eq.(\ref{cora1}).
Due to the symmetry between partons if is sufficient to analyse only
the first term ($\Gamma_1$, say) in Eq.(\ref{cora1}). Choosing the polar
coordinates of the $d\Omega_2$ integration relative to the direction of
the momentum of the first parton one obtains
\begin{equation}
\Gamma^{(2)}(\theta_{12})=2\int d\Omega_1 \int \Theta_{k_1 k_2}
d \Theta_{k_1 k_2} d\phi_{12} \Gamma_1^{(2)}(\Omega_1,\Omega_{12})
\delta(\theta_{12}-\Theta_{k_1 k_2}),
\end{equation}
which gives after performing integrations over $\Theta_{Kk_1}$
(in (\ref{cora1})), $\Theta_{n_1 n_2}$ and $\phi_{12}$
\begin{eqnarray}
\Gamma^{(2)}(\theta_{12})&=&{2a^3\over \theta_{12}}
\int_{Q_0/\theta_{12}}^{\Theta} {dK\over K}
\left( \cosh{(a\ln(K\theta_{12}/Q_0))}-1\right) \\ \nonumber
&&\sinh{(a\ln(K\theta_{12}/Q_0))}
  \int_{\theta_{12}}^{\Theta}
 I({P\over K},{\Theta_{Pk_1}\over \theta_{12}})
 {d\Theta_{Pk_1}\over\Theta_{Pk_1}}.
\end{eqnarray}
This is identical with the resolvent solution (\ref{sol12}) after
changing the integration variable $\Theta_{Pk_1}$ to
$\Psi=\Theta\theta_{12}/\Theta_{Pk_1}$.

As a second consistency check we
 integrate the cumulant correlation
$\Gamma^{(2)}(\Omega_1, \Omega_2)$ up to the
multiplicity moment $f_2$ for the full forward cone
which is known \cite{KUV} \cite{QCD} to be $f_2 / \bar
n^2=4/3$, cf. Eq.(\ref{recur}), with either $\rho^{(2)}
(\theta_{12})$ or $h^{(2)} (\Omega, \delta)$ and $c^{(2)}
(\theta,\delta)$ as intermediate results.
 
In the first case we derive the cumulant correlation in the
relative angle by integrating Eq.(\ref{Gas}). One obtains
\begin{eqnarray}
\Gamma^{(2)} (\theta_{12}) & = &
\int d \Omega_1 d \Omega_2 \delta(\theta_{12}-\Theta_{k_1,k_2})\Gamma^{(2)}
(\Omega_1,\Omega_2) \nonumber \\
& = & {1\over 8} ({a\over 2\pi})^2 \int^\Theta_{\theta_{12}}
{2\pi d \theta_1\over \theta_1}{1\over \theta_{12}} ({\theta_{12}\over \kappa})
^{2a} ({\theta_1\over \theta_{12}})^{a\over 2}+(1\leftrightarrow 2)
\nonumber \\
& = & {a\over 2 \theta_{12}} ({\theta_{12}\over \kappa})^{2a}
\lbrack({\Theta\over \theta_{12}})^{a\over 2}-1\rbrack.
\labl{gam12}
\end{eqnarray}
 
Here we used again the pole approximation with the same
results for both terms. With the results
Eqs.(\ref{gam12},\ref{selfsim}) for $\rho^{(2)}(\theta_{12})$
and $d^{(2)}_{prod}
(\theta_{12})$ we recover the relation (\ref{roga}) $\Gamma^{(2)}
(\theta_{12})=\rho^{(2)}(\theta_{12})-d^{(2)}_{prod}(\theta_{12})$.
 
We can now calculate the multiplicity moment
$f_2(\delta,\Theta)$ in a cone of half opening $\delta$
averaged over the forward cone $\Theta$ and the
corresponding normalisation $\bar n^2(\delta,\Theta)$
\begin{equation}
f_2 (\delta,\Theta) = \int^\delta_\kappa d\theta_{12} \rho^{(2)}
(\theta_{12}) \simeq {1\over 3} ({\delta\over \kappa})^{2a}
({\Theta\over \delta})^{a\over 2},
\labl{f2del}
\end{equation}
\begin{equation}
\bar n^2(\delta,\Theta)=\int^\delta_\kappa d\theta_{12} d^{(2)}
_{prod} (\theta_{12})\simeq {1\over 4} ({\delta\over\kappa})^{2a},
\labl{n2del}
\end{equation}
where the approximation is for $\delta \gg \kappa$.
For the normalised moment we then obtain $F_2(\delta,\Theta)
\simeq {4\over 3} ({\Theta\over \delta})^{a\over 2}$
. This
reproduces $F_2={4\over 3}$ in the full forward cone
$(\delta\to\Theta)$.
%or for the full forward cone  $(\delta\to\Theta)$ $F_2={4\over 3}$
%the well known result of LLA \cite{} and DLA \cite{}
So we find here that the density
$\rho^{(2)}(\theta_{12})$ and not the cumulant correlation
$\Gamma^{(2)}(\theta_{12})$ follows a power law, although the
difference between both quantities disappears for
$\theta_{12}\ll\Theta$. 
%which is always %possible at high energies.
 
In the second case we note that the cumulant moments
$h^{(2)}(\Omega,\delta)$ and $c^{(2)}(\Omega,\delta)$ derived from
$\Gamma^{(2)}(\Omega_1, \Omega_2)$ are power behaved, cf.
Eqs.(\ref{hm2},\ref{cm2}), and not the factorial moments. To
check the consistency with the previous results we calculate
in analogy to Eqs.(\ref{f2del},\ref{n2del}) the cumulant
moment in cone $\delta$ averaged over the full cone
$\Theta$.
 
\begin{equation}
c^{(2)} (\delta,\Theta)=\int^\delta_\kappa d \Omega
h^{(2)} (\Omega,\theta)+\int^\Theta_\delta d \Omega
h^{(2)} (\Omega,\delta),
\labl{c2del}
\end{equation}
The first term appears because of the requirement $\delta
\leq \theta$. Using our result (\ref{hm2}) for
$h^{(2)}(\Omega,\delta)$ we obtain for $\delta \gg\kappa$
\begin{equation}
c^{(2)} (\delta,\Theta)\simeq {1\over 4}
({\delta\over \kappa})^{2a} \lbrack{4\over 3} ({\Theta\over
\delta})^{a\over 2}-1\rbrack.
\labl{c2delt}
\end{equation}
With the normalisation (\ref{n2del}) and the relations
(\ref{roga}) we find again the consistency with the previous
result on $f^{(2)}(\delta,\Theta)$ in (\ref{f2del}), and
therefore the correct integration to the full cone with
$F_2={4\over 3}$.

\section*{Appendix I}
        \rem1{powapiv1}
         \rem1{powapiv1}
Here we discuss the derivation of Eq.(\ref{gamnom}) for $\Gamma^{(n)}_P$
for general order $n$.
First we have to calculate $\Delta^{(n)}$ from (\ref{deli}).
In case the inhomogeneous term $d^{(n)}$ is built up from
the products of $\rho^{(1)}(i)$
we apply again the pole approximation to
the angular integral which yields n terms, after partial
integration in leading order of
$1/\sqrt{\ln(K\theta_1^m/\Lambda)}$ for $i=1$
\begin{equation}
\Delta^{(n)}_{P,1} = {\beta f^n\over (4\pi)^n}
%\beta\int^P_{Q_0/\theta_1^m} {dK\over K}
\int^P_{K_{min}} {dK\over K}
{(\xxem)^{{1\over 4}}\over\xxeL}
{e^{\left\{2\beta
\left(\sqrt{\xxem} + \sum^n_{i=2} \sqrt{\xxiL}\right)\right\}}
\over\theta_1^2\prod^n_{i=2}
\left\{\theta^2_{1i}\left(\xxiL\right)^{(1/4)}\right\}}.
\labl{delp1}
\end{equation}
In the exponent and in the denominator
we expand
\begin{equation}
\sqrt{\xxiL}\approx\sqrt{\xxem} (1+{1\over2}
\delta^{1i}_{1m}+\dots)
\labl{dexpand}
\end{equation}
%with $\epsilon^{1i}_{1m}
% = \ln (\theta_{1i}/\theta^m_1)/\ln(K\theta^m_1/\Lambda)$.
and neglect the $\delta$-terms because of (\ref{delijlm}).
%Similarly in the
%denominator we write $\ln(K\theta_{1
%i}/\Lambda) = \ln (K\theta_i/\Lambda) (1-\epsilon^i_{1i})$
%and neglect $\epsilon^i_{1i}$. This is justified in the region
%where the power law applies $(\epsilon\ll1)$,
%otherwise it affects the prefactor but not the exponential behavior. After
Then, after partial integration, we obtain in leading order of
$1/\sqrt{P\theta_1^m/\Lambda}$ and with the corresponding terms $i>1$ added
\begin{equation}
\Delta_P^{(n)}(\{\Omega\})=
   \left({f\over 4\pi}\right)^n {1\over n}
   \sum_{i=1}^n {a(P\bar\theta_i)^{{n\over 2}} \over \theta_i^2}
   \biggl({P\theta_i^m\over \Lambda}\biggr)^{2na(P\theta_i^m)}
   F_{i}^n(\{\tij\}),
\labl{delnr}
\end{equation}
where $F_i^n = 1/(\theta^2_{i1}\dots\theta^2_{i-1,i}
\theta^2_{i+1,i}\dots\theta^2_{in})$.

We move on to
the case with $d^{(n)}$ as products of $\Gamma^{(m)}$.
Consider first $d^{(3)} = \rho^{(1)}(1)\Gamma^{(2)}(2,3)$.
In the derivation of $\Delta^{(3)}_{P,1}$ in pole approximation
we replace
$\omega(\epsilon_{12}^1,2)\approx \omega(-\delta_{23}^{12},2) \approx 2$
in $\Gamma^{(2)}(2,3)$ of Eq.(\ref{gam2as}), the same replacement applies
in leading approximation of the angular integral for
 $\Delta^{(3)}_{P,2}$  and $\Delta^{(3)}_{P,3}$ so that again the form
in Eq.(\ref{delnr}) is obtained, but with different $F_i^n$.
%Eq. (\ref{gam2as}) that $\Gamma^{(2)}(2,3)$ in leading order of
%$\epsilon~(\omega(\epsilon,2)\approx 2)$ has the same structure as
%$\rho^{(1)}(2)\rho^{(1)}(3)$, except for the angular
%factors. Whereas for the above case
%$d^{(3)}\sim 1/(\theta_1^2\theta^2_2\theta^2_3)$
The complete inhomogeneous term from (\ref{dprod}) has the structure
(without couplings and exponential factors)
\begin{equation}
d^{(3)}\sim
{1\over\theta^2_1\theta_2^2\theta_3^2}+
{1\over\theta^2_1} {1\over 2}
\left({1\over\theta_2^2\theta^2_{23}} +
{1\over\theta_3^2\theta^2_{23}}\right) + \rm{cycl.}
\labl{d3ang}
\end{equation}
Again, the angular and momentum integrals yield a result as in
(\ref{delnr}), except for
different functions $F^n_i$, see Eq.(\ref{Ffcts}).
In the same way the result
(\ref{delnr}) follows for an
arbitrary number $n$ of particles and $F^n_i$ is again homogeneous
of degree $p=-2(n-1)$.
With the inhomogeneous term
$\Delta^{(n)}_{P,i}$ from Eq. (\ref{delnr})
which is of the form (\ref{gendir})
we can then solve the integral equation (\ref{gami}) for
$\Gamma^{(n)}_{P,i}$ using the results of section 4 which yields
finally Eq.(\ref{gamnom}).

\section*{Appendix J}
         \rem1{powapjv1}
         %This is the file powaphv1.tex as of 9.III.94
In this Appendix we shall discuss assumptions leading to
Eq.(\ref{dbyt}). For $n=2$ Eq.(\ref{Delta2D}) reads
\begin{equation}
d^{(2)}_P(\delta)=\int d^3K d\Omega_2 {\cal M}_P(K) \rho^{(1)}_K(\Omega_{KC})
\rho^{(1)}_K(\Omega_2).
\end{equation}
According to the definition, Eq.(\ref{hdef}), we have chosen the cone axis
$\vec{C}$  to coincide with the direction of the first parton $\vec{k}_1$.
 Integration over $\Omega_2$ gives
\begin{equation}
d^{(2)}_P(\delta)=\int d^3 K {\cal M}_P(K) \rho^{(1)}_K(\Omega_{KC})
\overline{n}(K\Theta_{KP}).
\end{equation}
Remaining integral is dominated by the singularities in $\Theta_{KC}$ due to
the angular ordering $(\Theta_{KP} > \Theta_{Kk_1}=\Theta_{KC})$. Using the
pole dominance $(\Theta_{KC} \sim 0, \Theta_{KP}\simeq\Theta_{PC}=\delta)$
allows to integrate over $d\Omega_K=d\Omega_{KC}$ and
we obtain Eq.(\ref{dbyt}) since $t^{(2)}(K\delta)=\overline{n}^2(K\delta)$.
 
The case of $n=3$ is more instructive since it illustrates other complications
due to the multiparton kinematics. We shall discuss it in more detail.
The subsequent generalization to arbitrary $n$ will be straightforward.
First,
the product term $d^{(3)}_{prod}$ is given by Eq.(\ref{dprod}), and consequently
one obtains Eq.(\ref{rbyt}) with
\begin{equation}
t^{(3)}_K(\Psi)=\overline{n}^3(K\Psi)+3\overline{n}(K\Psi) g_2(K\Psi),
\end{equation}
where
\begin{equation}
g_2(K\Psi)=\int_{(K,\Psi)} \Gamma^{(2)}_K(\Omega_1,\Omega_2)
 d\Omega_1 d\Omega_2. \label{gcum}
\end{equation}
 
We move now to Eq.(\ref{dbyt}) for $d^{(3)}$. Contribution from the
product of the single densities to $d_{prod}^{(3)}$ can be treated
in the full analogy to the $n=2$ case. One obtains Eq.(\ref{dbyt}) with
the $\overline{n}^3$ piece of $t^{(3)}$. Similarly calculation of the contribution
from the $\rho^{(1)}(1)\Gamma^{(2)}(2,3)$ term is simple. One obtains
\begin{equation}
\int d^3 K {\cal M}_P(K) \rho^{(1)}_K(\Omega_{KC}) g_2(K\Theta_{KP}),
\end{equation}
which due to the dominance of the $\Theta_{KC} \sim 0$ region gives
Eq.(\ref{dbyt}) with a third of the whole mixed contribution 
$3 \overline{n} g_2$ to $t^{(3)}$.
 
The last two contributions are equal up to the relabeling $2\leftrightarrow 3$.
Consider the $\rho^{(1)}(2)\Gamma^{(2)}(1,3)$ term in $d_{prod}$. Its
contribution to $d^{(3)}$ is
\begin{equation}
\int d^3 K {\cal M}_P(K)\int_{\gamma} d\Omega_2 d\Omega_3 \rho^{(1)}_K(2)
\Gamma^{(2)}_K(1,3).
\end{equation}
Integration over $\Omega_2$ gives
\begin{equation}
\int d^3 K {\cal M}_P(K) n(K\Theta_{KP}) \int_{\gamma} d\Omega_3
\Gamma_K^{(2)}(\Omega_{KC},\Omega_{Kk_3}).
\end{equation}
There are two regions which dominate the remainimg integrals: A) $\vec{K}||
\vec{k}_1(\vec{C})$, this corresponds to the sequence $P\rightarrow
K\rightarrow 1(C)\rightarrow 3 (\Theta_{PK}>\Theta_{KC}>\Theta_{Ck_3})$ and B)
$\vec{K}||\vec{k}_3$ corresponding to the process $P\rightarrow K\rightarrow
3\rightarrow1(C)$ with the ordering $\Theta_{PK}>\Theta_{Kk_3}>\Theta_{Ck_3}$.
In each of them the connected correlation function depends on different set
of variables. However in both cases $d\Omega_K d\Omega_3$ integration
is equivalent to the complete integration over the phase space of the final
partons, which reproduces the total cumulant $g_2$. To show this let us recall
the generic expression for $g_2$, Eq.(\ref{gcum}), if the underlying process
$(S)$ has a two-step structure $K\rightarrow k_1 \rightarrow k_2$. Then
$\Gamma_K^{(2)}(\Omega_1,\Omega_2)=\Gamma^{(S)}_K(\Omega_{Kk_1},\Omega_{12}),
\Theta_{Kk_1}>\Theta_{12})$ and Eq.(\ref{gcum}) becomes
\begin{equation}
g_2^{(S)}(K\Psi)=\int_{\kappa}^{\Psi} d\Omega_{Kk_1}
 \int_{\kappa}^{\Theta_{Kk_1}}
d\Omega_{12} \Gamma_K^{(S)}(\Omega_{Kk_1},\Omega_{12}), \labl{g2ph}
\end{equation}
where $\kappa=Q_0/K$ and limits of the integrations apply to the corresponding
 polar angles.
 
Now, in the case A) we obtain the following contribution to Eq.(\ref{dbyt})
\begin{eqnarray}
(A)&=&\int_{Q_0/\delta}^P {dK\over K} {a^2(K\Theta_{KP})\over 2\pi
 \Theta_{KP}^2}
{d\Omega_K\over \Theta_{KC}^2} {d\Omega_3\over \Theta_{Ck_3}^2}
\overline{n}(K\Theta_{KP})\\ \nonumber
& & \left[ \Theta_{KC}^2\Theta_{Ck_3}^2
\Gamma^A(\Theta_{KC},\Theta_{Ck_3}) \right] \\ \nonumber
&=&{1\over 2\pi}\int_{Q_0/\delta}^P {dK\over K} a^2(K\delta)
\overline{n}(K\delta)
\int_{\kappa}^{\delta}{d\Theta_{KC}
\over \Theta_{KC}} 2\pi \int_{\kappa}^{\Theta_{KC}} {d\Theta_{Ck_3}\over
\Theta_{C3} } 2\pi \\  \nonumber
&&  \left[\Theta_{KC}^2 \Theta_{Ck_3}^2 \Gamma^A(\Theta_{KC},
\Theta_{Ck_3}) \right] \\  \nonumber
&=& {1\over 2\pi\delta^2}\int_{Q_0/\delta}^P {dK\over K} 
\overline{n}(K\delta) {1\over 2}
 g_2(K\delta),
\end{eqnarray}
which gives the half of the $n g_2$ contribution to $t^{(3)}$ in the virtue
of Eq.(\ref{g2ph}).
 
In the case B) corresponding contribution to $d^{(3)}$ reads
\begin{equation}
(B)={1\over 2\pi}\int_{Q_0/\delta}^P {dK\over K} a^2(K\delta) 
\overline{n}(K\delta)
\int d\Omega_{K} d\Omega_3\Gamma^B(\Theta_{Kk_3},\Theta_{Ck_3})
\end{equation}
which can be rewritten using the pole approximation for the inner vertex
$1(C)\rightarrow 3 : \Theta_{Ck_3}\sim 0, \Theta_{Kk_3}\simeq \Theta_{KC}$
\begin{equation}
(B)=
{1\over 2\pi\delta^2}\int_{Q_0/\delta}^P {dK\over K} a^2(K\delta) 
\overline{n}(K\delta)
\int_{\kappa}^{\delta}d\Theta_{KC}
 2\pi \int_{\kappa}^{\Theta_{KC}} d\Theta_{C3}
 2\pi \Gamma^B(\Theta_{KC},
\Theta_{Ck_3})
\end{equation}
Comparison with Eq.(\ref{gcum}) shows that this provides the second half
of the $\overline{n} g_2$ term to $t^{(3)}$. Adding the contribution from the
$\rho^{(1)}(3)\Gamma^{(2)}(1,2)$ gives finally complete mixed term
$3 \overline{n} g_2$. This establishes Eq.(\ref{dbyt}) and hence the derivative rule
(\ref{srul}).

\section*{Appendix K}
         \rem1{powapkv1}
         Here we derive the second multiplicity moment by the direct
integration of the cumulant correlation function.
We first calculate the differential cumulant moment
$h(\Omega, \delta)$ with one particle kept at angle $\Omega$
and the others within a cone of half opening $\delta$. We
start again with $n=2$, first for fixed $\alpha_s$, using
Eq.(\ref{Gas})
\begin{eqnarray}
h^{(2)} (\Omega_1,\delta) & = & \int_{\gamma(\theta,\delta)}
d\Omega_2\quad \Gamma^{(2)} (\Omega_1,\Omega_2) \nonumber \\
& = &{1\over 8} ({a\over 2\pi})^2 \int^\delta_\kappa
{\theta_{12} d \theta_{12} d\phi_{12}\over \theta^2_{12}}
({\theta^{a\over 2}_1\over \theta^2_1}
{\theta^{3\over 2}_{12}\over \kappa^{2a}} +
{\theta^{a\over 2}_2 \theta^{{3\over 2}a}_{12}\over
\theta^2_2 \kappa^{2a}}),
\labl{hmom2}
\end{eqnarray}
with $\Omega_{12}=(\delta_{12}, \phi_{12})$. In the second
term we use
$\theta^2_2=\theta^2_1+\theta^2_{12}+\theta_1\theta_{12}
\cos \phi_{12}$ and apply the pole dominance approximation
for the $\theta_{12}$ integral, i.e. we set $\theta_{12}=0$
in the nonsingular factors. This returns $\theta_2=\theta_1$
and makes the second integral equal to the first one, so we
obtain
\begin{equation}
h^{(2)} (\Omega,\delta)={a\over 12\pi} {1\over \theta^2} ({\theta
\over \delta})^{-{3\over 2}a}({\theta\over \kappa})^{2a}.
\labl{hm2}
\end{equation}
The cumulant moment, calculated from the definition
(\ref{cnd}) has a complicated boundary for the $\Omega_1,
\Omega_2$ integration. With the approximation
\begin{equation}
c^{(2)} \simeq \int_{\gamma(\theta,\delta} h^{(2)}
(\Omega,\delta) d \Omega \simeq \pi\delta^2
h^{(2)}(\Omega,\delta),
\labl{c2app}
\end{equation}
we obtain
\begin{equation}
c^{(2)} \simeq{a\over 12} ({\delta^2\over \theta^2})^
{1+{3\over 4}a} ({\theta\over \kappa})^{2a}.
\labl{cm2}
\end{equation}
The average multiplicity in the cone
$\gamma(\theta,\delta)$ can be derived within the same
approximation from Eq.(\ref{asyan2}) as
\begin{equation}
\bar n (\theta,\delta)={a\over 4} ({\delta\over \theta})^2
({\theta\over \kappa})^a,
\labl{nbard}
\end{equation}
and we obtain for the normalised moment from
Eq.(\ref{momn})
\begin{equation}
C^{(2)} (\theta,\delta)={4\over 3a} ({\theta^2\over \delta^2})
^{1-{3\over 4}a}.
\labl{c2nor}
\end{equation}
In the case of running $\alpha_s$ one obtains in
complete analogy by integrating Eqs. (\ref{gon2s},
\ref{asyan1})
\begin{equation}
h^{(2)} (\Omega,\delta)={f^2\over 12\pi\theta^2}
({P\theta\over \Lambda})^{4a(P\theta)}
({\theta^2\over \delta^2})^{-{3\over 4}a(P\theta)},
\labl{h2r}
\end{equation}
\begin{equation}
c^{(2)} (\theta,\delta) \simeq \pi \delta^2 h^{(2)} (\Omega,\delta),
\labl{c2vr}
\end{equation}
\begin{equation}
C^{(2)}(\theta,\delta)\simeq {4\over 3a(P\theta)}
({\theta^2\over \delta^2})^{+1-{3\over 4}a(P\theta)}.
\labl{c2nr}
\end{equation}
 
The consistency of these results with the case of fixed
$\alpha_s$ is easily established by letting $\beta,
\lambda\to \infty,~ \beta/\sqrt{\lambda}=a$, (see Eq.(\ref{alimit})).
%~$f=2 \beta
%K_0(2\beta\sqrt\lambda) / \sqrt \pi \to \sqrt a~
%e^{-2\beta\sqrt\lambda}$ and $a(P\theta)\to a$.
 
We have checked the quality of our approximations in case of
fixed $\alpha_s$. In Fig.21 we see that the pole
approximation in Eq.(\ref{hmom2}) which yields the power law
Eq.(\ref{hm2}) is rather close to the exact numerical result
in the full range of $\delta$. The approximation
Eq.(\ref{cm2}) for the moment $c^{(2)}$ approaches the
correct power for sufficiently small $\delta (\ln{
\theta / \delta} \geq 1)$, but rises stronger towards larger
$\delta \approx \theta$ where the integral over $\Omega_1$
approaches the singularity of $\Gamma^{(2)}$ at
$\theta_1=0$. This demonstrates the importance of the
$h^{(n)}$ moments for the test of power behaviour.
 
%It  may be surprising at first, that a power law is found for
%the density $\rho^{(2)}(\theta_{12})$ but for the local factorial
%moment which is calculated as integral over the density
%$\rho^{(2)}$ in Eq.(\ref{fnd}). 
It  may be surprising at first sight, that a power law is found for 
the cumulant moments and not for the factorial moments 
which are the integrals of the density $\rho^{(n)}$ in Eq.(\ref{fnd}),
while on the other hand $\rho^{(2)}(\theta_{12})$ is power behaved.
In the Appendix D we perform
some consistency checks proving that indeed this is the case.
In particular we
 integrate successively the cumulant correlation
$\Gamma^{(2)}(\Omega_1, \Omega_2)$ up to the
multiplicity moment $f_2$
with either $\rho^{(2)}
(\theta_{12})$ or $h^{(2)} (\Omega, \delta)$ and $c^{(2)}
(\theta,\delta)$ as intermediate steps.

%\section*{References}
        \rem1{powref8}
%        \input{powref8}
 %powref5  1.8.94   adds last ref. to powref4
%powref6 22.9.94
%powref7 15.12.94
%powref8 18.12.94

\newpage
%\section*{Figures}
%        \rem1{powfig8}
%         \input{powfig8}
\clearpage
\rem1{figures}
         \pagestyle{empty}
\begin{figure}[htb]
\vspace{9pt}
%\framebox[55mm]{\rule[-21mm]{0mm}{43mm}}
\epsfxsize=7cm \epsfbox{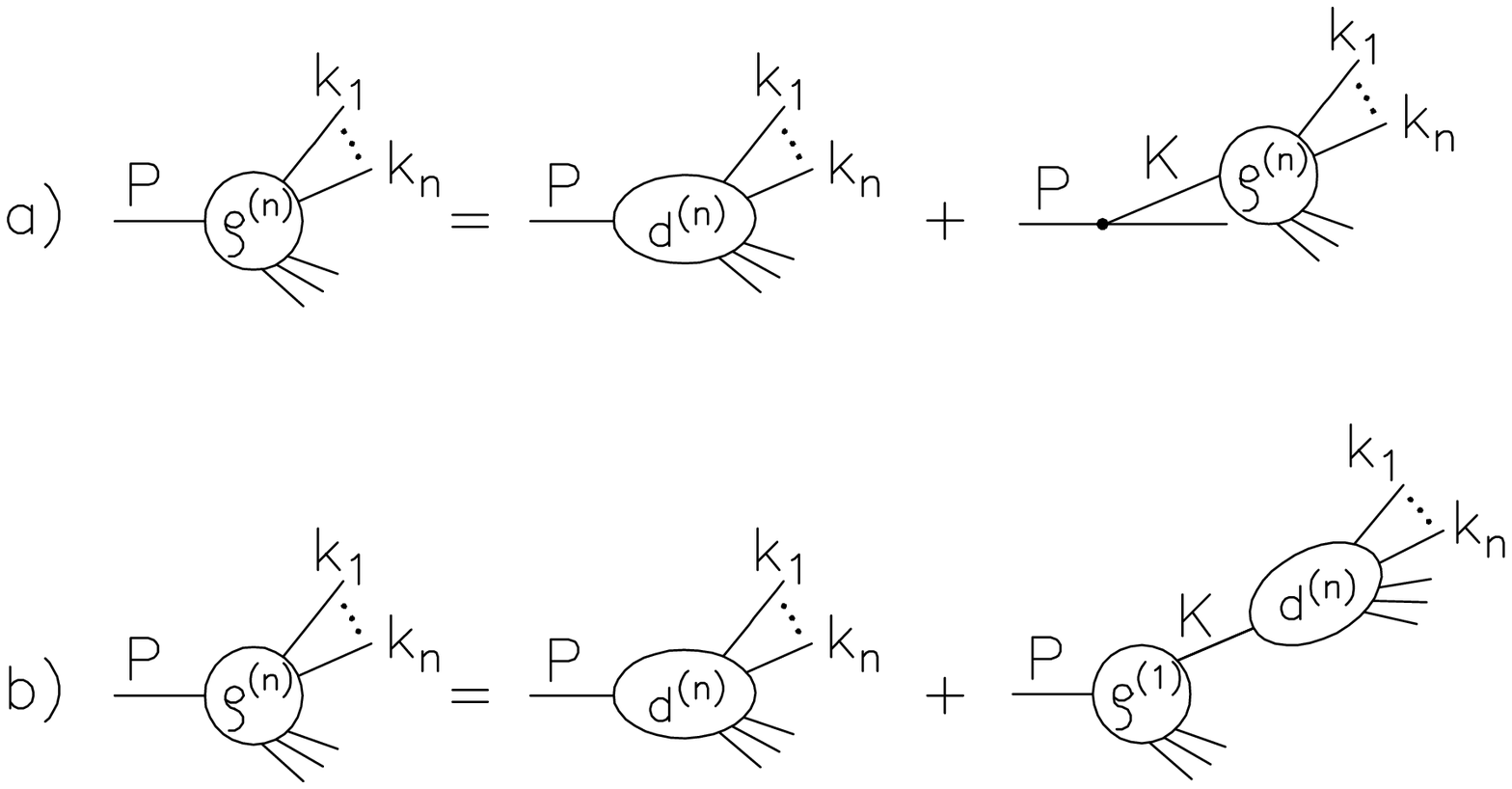}
\caption{Diagrammatic representation of a) the integral 
equation (\protect\ref{men})
for partonic densities  and b) of its solution, Eq.(\protect\ref{res}).
 }
\label{fig:f1}
\end{figure}

\begin{figure}[htb]
\vspace{9pt}
%\framebox[55mm]{\rule[-21mm]{0mm}{43mm}}
\epsfxsize=8cm \epsfbox{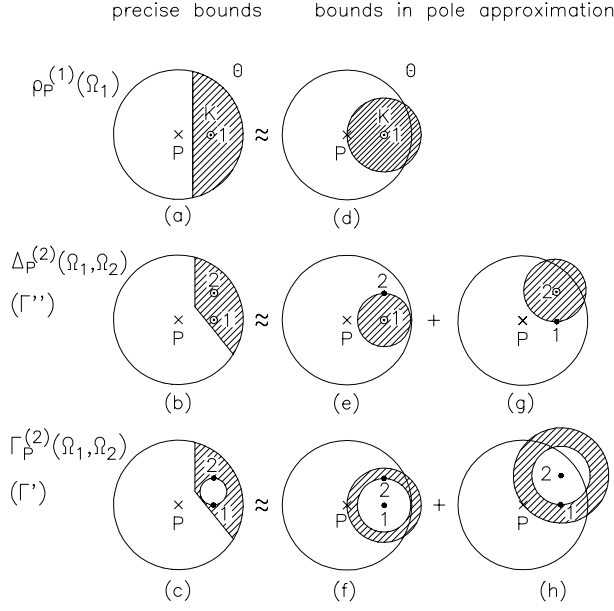}
\caption{
Kinematics of a QCD jet in the transverse plane. Allowed phase space of the
intermediate parent $K$ is shown as the shaded regions.
The large circles indicate the cone
of half opening $\Theta$ around the primary
parton of momentum $P$; left column: (a) precise bounds
for $\rho^{(1)}_P$ in Eq.(\protect\ref{ro1}), (b) for
$\rho^{(2)}_P$ in Eq.(\protect\ref{mero}) or $\Delta^{(2)}_P$
in Eq.(\protect\ref{F4}), (c) for $\Gamma^{(2)}_P$ in Eq.(\protect\ref{F3});
middle and right column: integration areas in the pole
approximation.
  }
\label{fig:f2}
\end{figure}

\begin{figure}[htb]
\vspace{9pt}
%\framebox[55mm]{\rule[-21mm]{0mm}{43mm}}
\epsfxsize=5cm \epsfbox{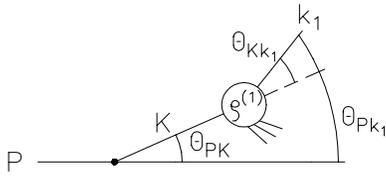}
\caption{Kinematics of the inclusive production of a single parton.
  }
\label{fig:f3}
\end{figure}

\begin{figure}[htb]
\vspace{9pt}
%\framebox[55mm]{\rule[-21mm]{0mm}{43mm}}
\epsfxsize=5cm \epsfbox{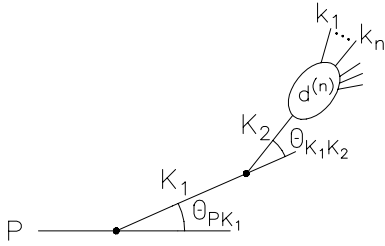}
\caption{Second iteration of the integral equation 
(\protect\ref{men}).
The vertex $K_1\rightarrow K_2$ can be interpreted as the first
term of the power expansion of the resolvent $R_{K_1}(K_2,\sigma)$,
c.f. Eq.(\protect\ref{res}).
    }
\label{fig:f4}
\end{figure}

\begin{figure}[htb]
\vspace{9pt}
%\framebox[55mm]{\rule[-21mm]{0mm}{43mm}}
\epsfxsize=13cm \epsfbox{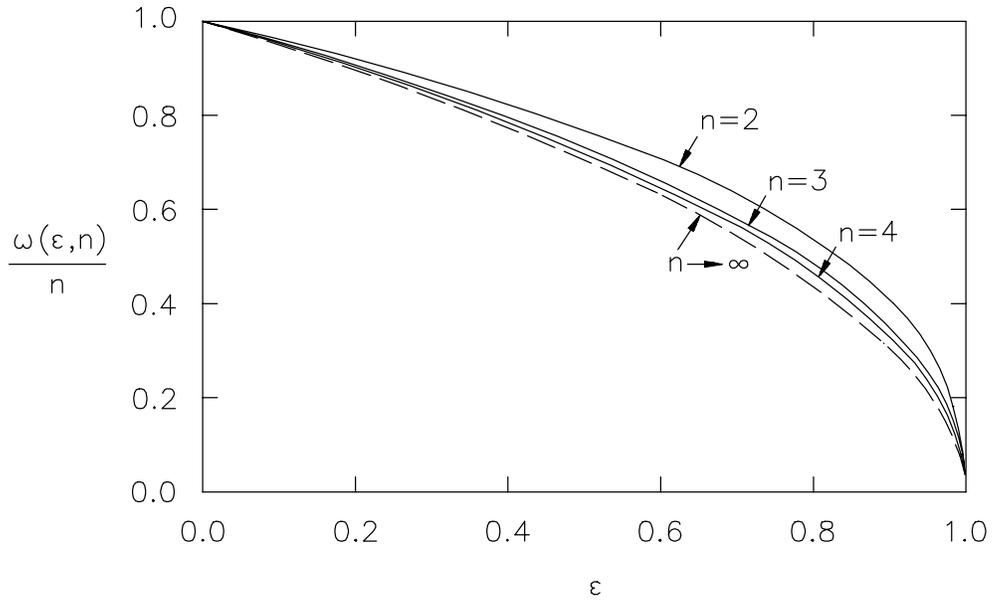}
\caption{The functions $\omega(\epsilon,n)/n$ and their
limit $(1-\epsilon)^{1/2}$ for $n\rightarrow\infty$. }
\label{fig:f5}
\end{figure}

\begin{figure}[htb]
\vspace{9pt}
%\framebox[55mm]{\rule[-21mm]{0mm}{43mm}}
\epsfxsize=5.5cm \epsfbox{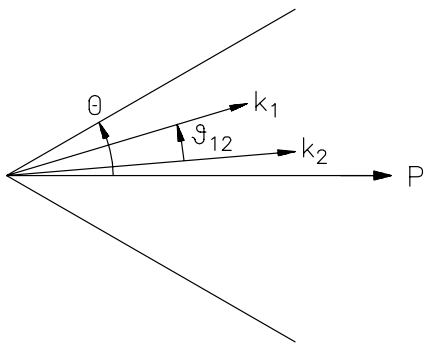}
\caption{Definition of the relative angle $\theta_{12}$ for
particles in a forward cone around a primary parton of
momentum $P$.
    }
\label{fig:f6}
\end{figure}

\begin{figure}[htb]
\vspace{9pt}
%\framebox[55mm]{\rule[-21mm]{0mm}{43mm}}
\epsfxsize=11cm \epsfbox{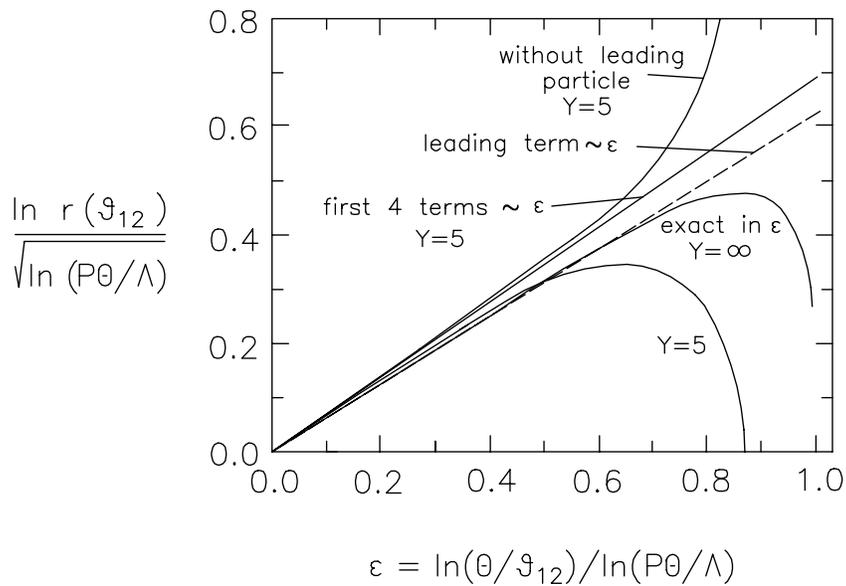}
\caption{Normalized 2-parton angular correlation $r(\theta_{12})$ in the
forward cone of half opening angle $\Theta=1$ for asymptotic and finite energies
$(Y=\ln(P/Q_0)=5, Q_0/\Lambda=2)$.
    }
\label{fig:f7}
\end{figure}

\begin{figure}[htb]
\vspace{9pt}
%\framebox[55mm]{\rule[-21mm]{0mm}{43mm}}
\epsfxsize=10cm \epsfbox{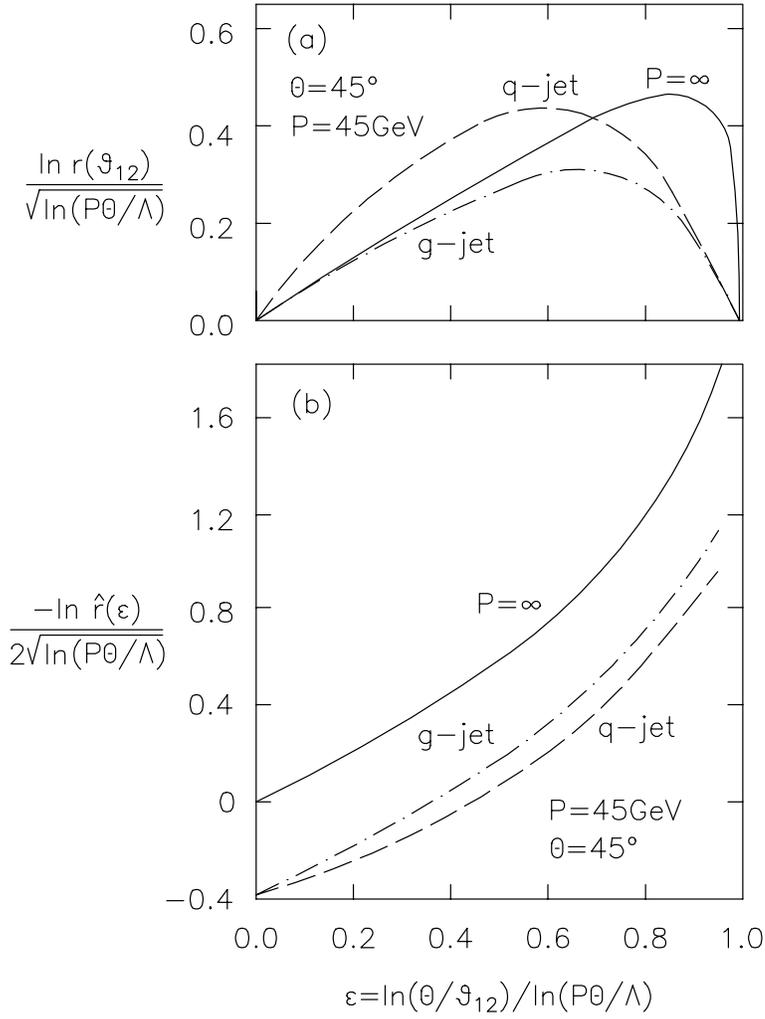}
\caption{Normalized 2-parton correlation functions at finite energies
for quark and gluon jets and in the infinite energy limit. (a) Differential
normalization (\protect\ref{r12}) and (b) global normalization (\protect\ref{r12hat}).
    }
\label{fig:f8}
\end{figure}

\begin{figure}[htb]
\vspace{9pt}
%\framebox[55mm]{\rule[-21mm]{0mm}{43mm}}
\epsfxsize=6.5cm \epsfbox{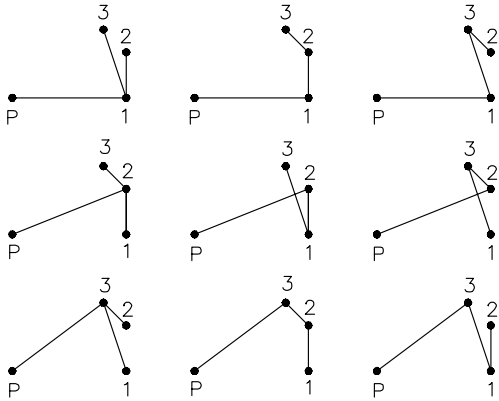}
\caption{Singularity structure of the terms building
up the connected correlation function
$\Gamma^{(n)}$ (for n=3). A line between particles $i$ and $j$
denotes a singularity $\theta^{-2}_{ij}$.
    }
\label{fig:f9}
\end{figure}

\begin{figure}[htb]
\vspace{9pt}
%\framebox[55mm]{\rule[-21mm]{0mm}{43mm}}
\epsfxsize=10cm \epsfbox{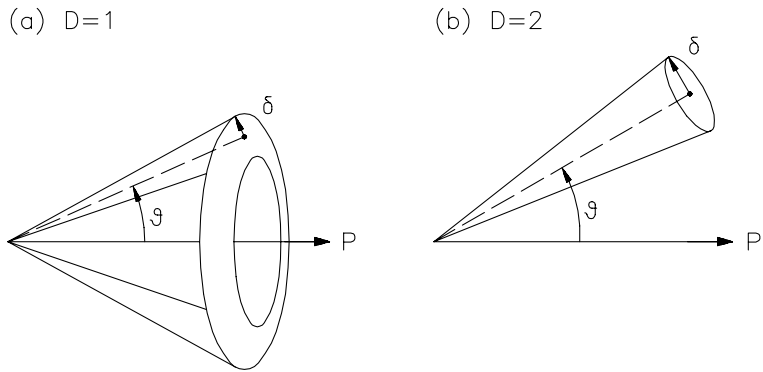}
\caption{Definition of angles in a sidewise ring and a sidewise cone 
refered to as 1-dim and 2-dim configurations respectively.
    }
\label{fig:f10}
\end{figure}

\begin{figure}[htb]
\vspace{9pt}
%\framebox[55mm]{\rule[-21mm]{0mm}{43mm}}
\epsfxsize=6.5cm \epsfbox{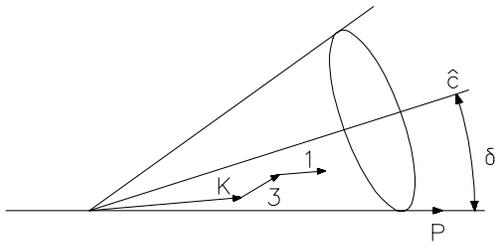}
\caption{Maximal sidewise cone used to define $h(\delta)$
in Eq.(\protect\ref{boun}). Kinematics corresponds to the subprocess (B)
discussed in the Appendix J, i.e. $P\rightarrow K\rightarrow
3 \rightarrow 1(c)$.
    }
\label{fig:f11}
\end{figure}

\clearpage
 
\begin{figure}[htb]
\vspace{9pt}
%\framebox[55mm]{\rule[-21mm]{0mm}{43mm}}
\epsfxsize=12cm \epsfbox{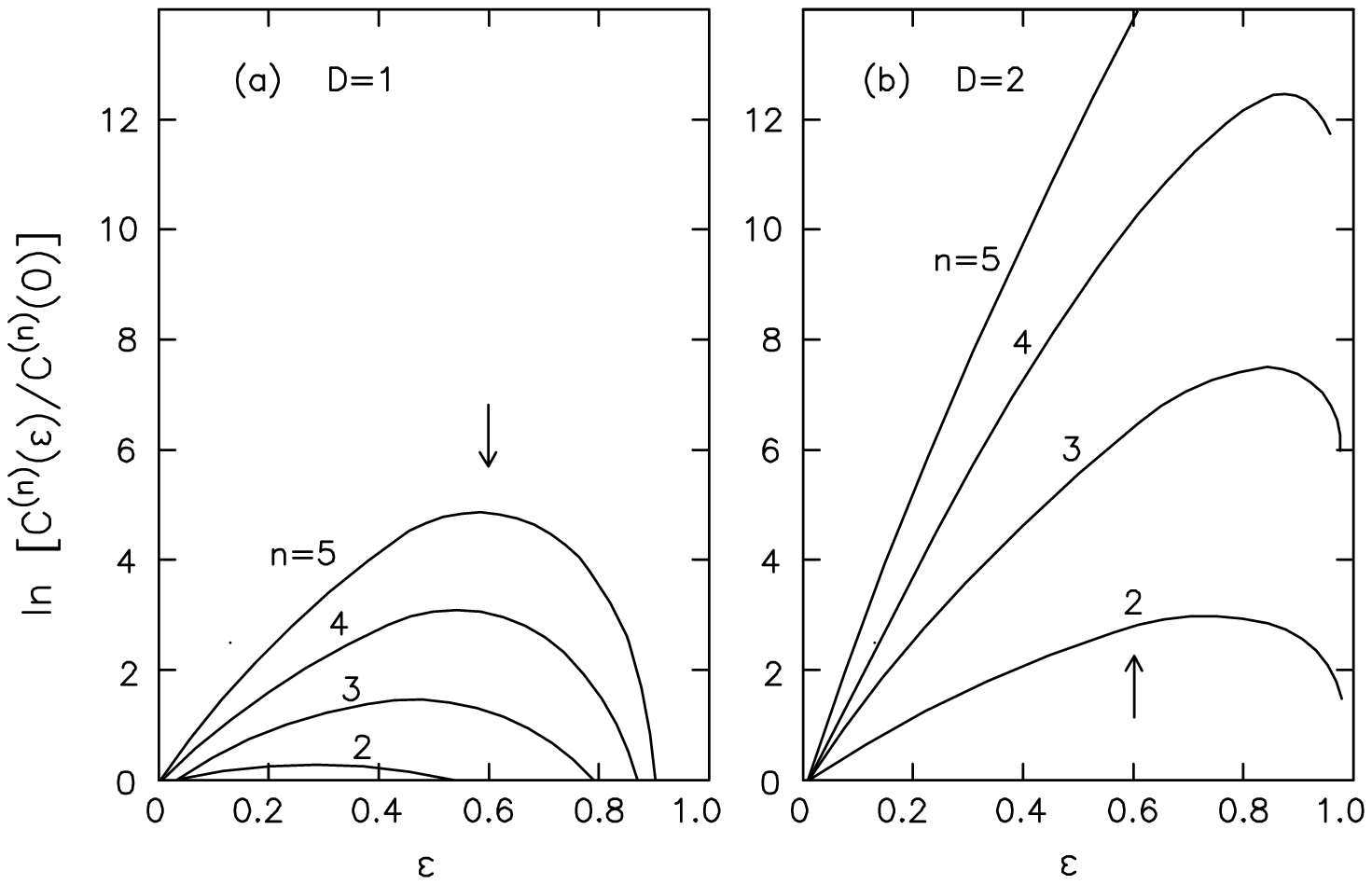}
\vspace*{1cm}
\hspace*{2cm} \epsfxsize=7cm  \epsfbox{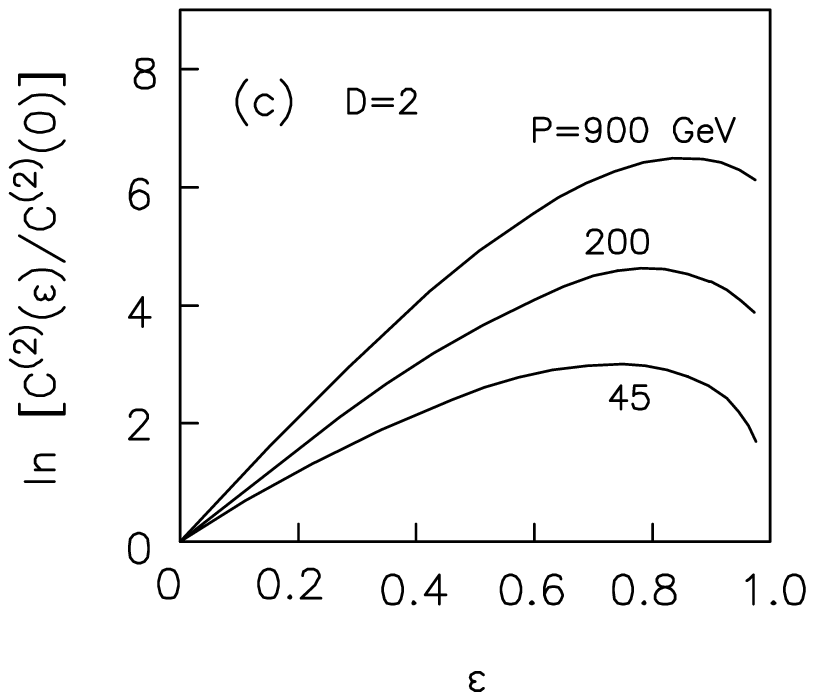}
\caption{Normalised moments for (a) the sidewise ring (D=1) and (b) the sidewise
   cone (D=2) from Eq.(\protect\ref{cmomno}) using the high energy approximation  
 (evaluated for momentum $P=45\;\; GeV,~\theta = 30^o$).
 For $\epsilon$ smaller than indicated by the arrows the asymptotic
 expressions apply.
   (c) Energy dependence of the normalized moments.
    }
\label{fig:f13}
\end{figure}

\begin{figure}[htb]
\vspace{9pt}
%\framebox[55mm]{\rule[-21mm]{0mm}{43mm}}
\epsfxsize=10.5cm \epsfbox{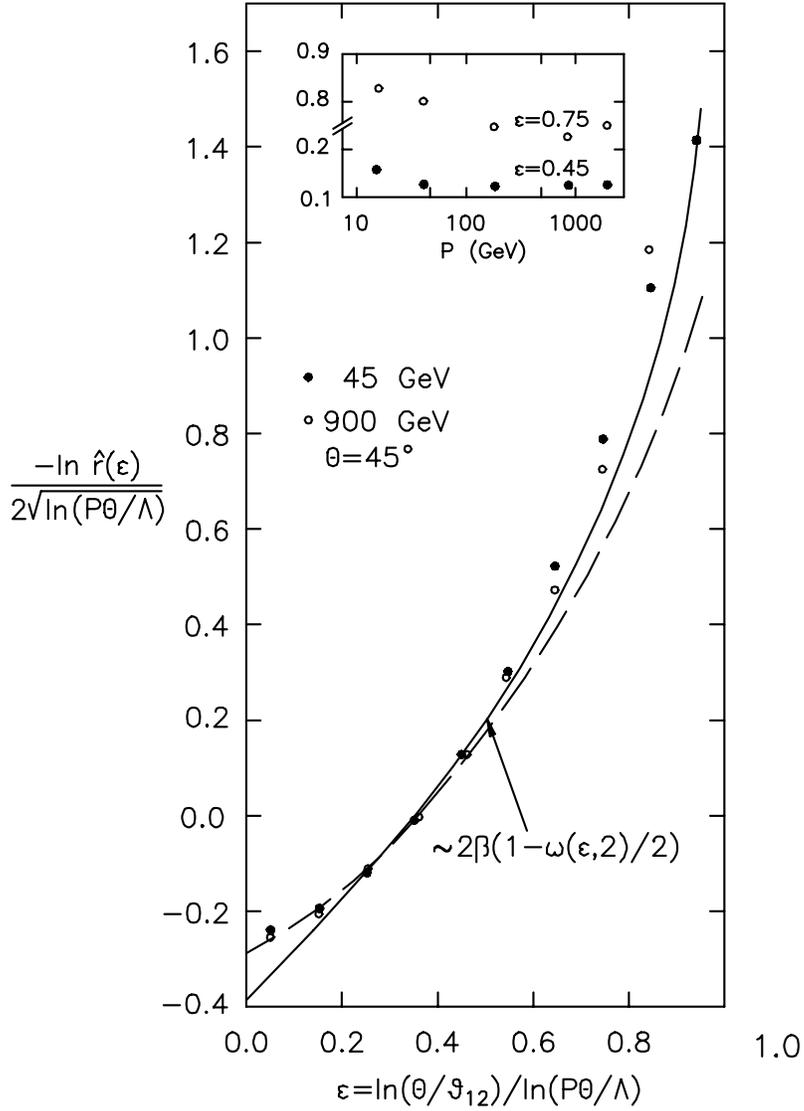}
\caption{Rescaled 2-particle correlation $\hat r (\varepsilon)$ 
vs.\ scaling
variable $\varepsilon$ for different primary energies $P$ from the parton
MC. The full curve represents
the high energy limit %(\protect\ref{rhadlim})
of the DLA, the dashed curve the prediction for quark jets at $P=45$ GeV;
the normalization of the curves is adjusted. The insert shows the energy
dependence of the same quantity at fixed $\epsilon$.
    }
\label{fig:f14}
\end{figure}

\newpage

\begin{figure}[htb]
\vspace{9pt}
%\framebox[55mm]{\rule[-21mm]{0mm}{43mm}}
\epsfxsize=11cm \epsfbox{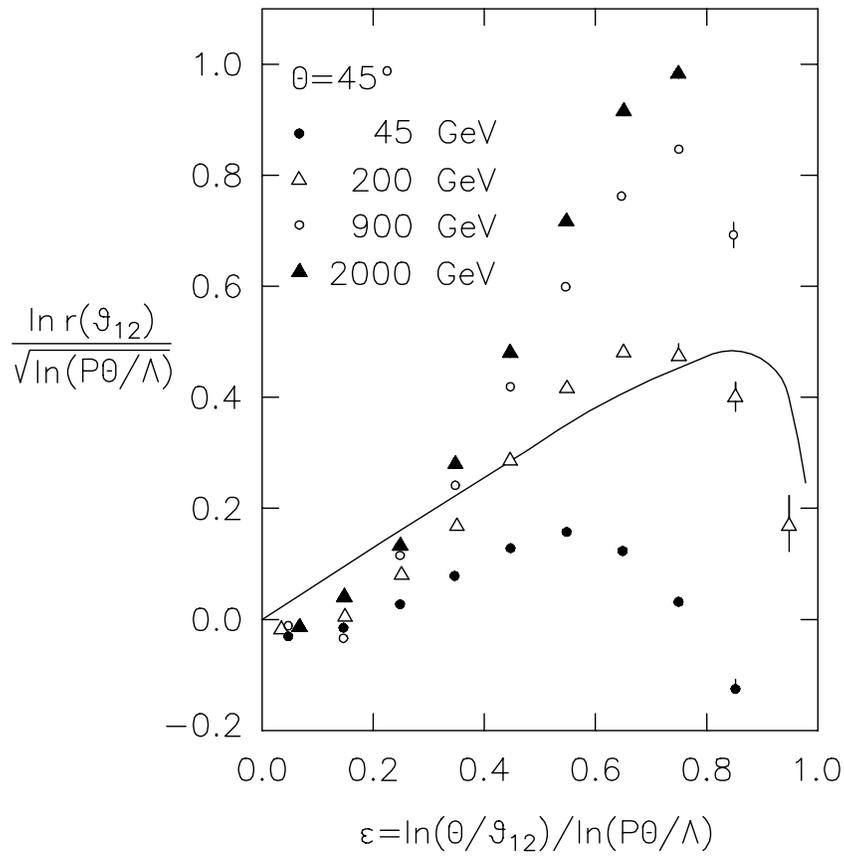}
\caption{Rescaled two particle correlation as in Fig.13
but with differential normalization as in Eq.(\protect\ref{r12}).
The MC data don't approach the scaling limit represented by the curve.
    }
\label{fig:f15}
\end{figure}

\newpage

\begin{figure}[htb]
\vspace{9pt}
%\framebox[55mm]{\rule[-21mm]{0mm}{43mm}}
\epsfxsize=12cm \epsfbox{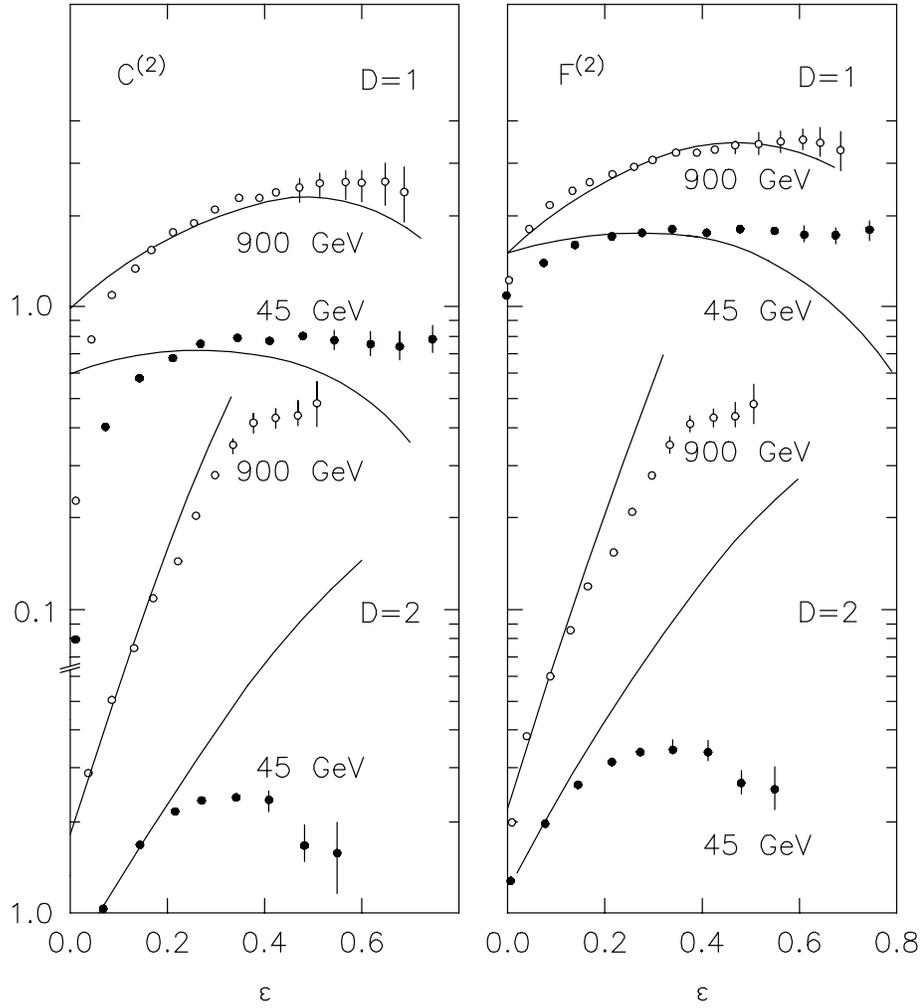}
\caption{Cumulant and factorial moments $C^{(2)}$ and $F^{(2)}$ for different
dimensions $D$ and primary momenta $P$ from the parton MC. The curves
represent our results for $C^{(2)}$ whereby the normalization is adjusted
to fit the MC data.
    }
\label{fig:f16}
\end{figure}

\begin{figure}[htb]
\vspace{9pt}
%\framebox[55mm]{\rule[-21mm]{0mm}{43mm}}
\epsfxsize=14cm \epsfbox{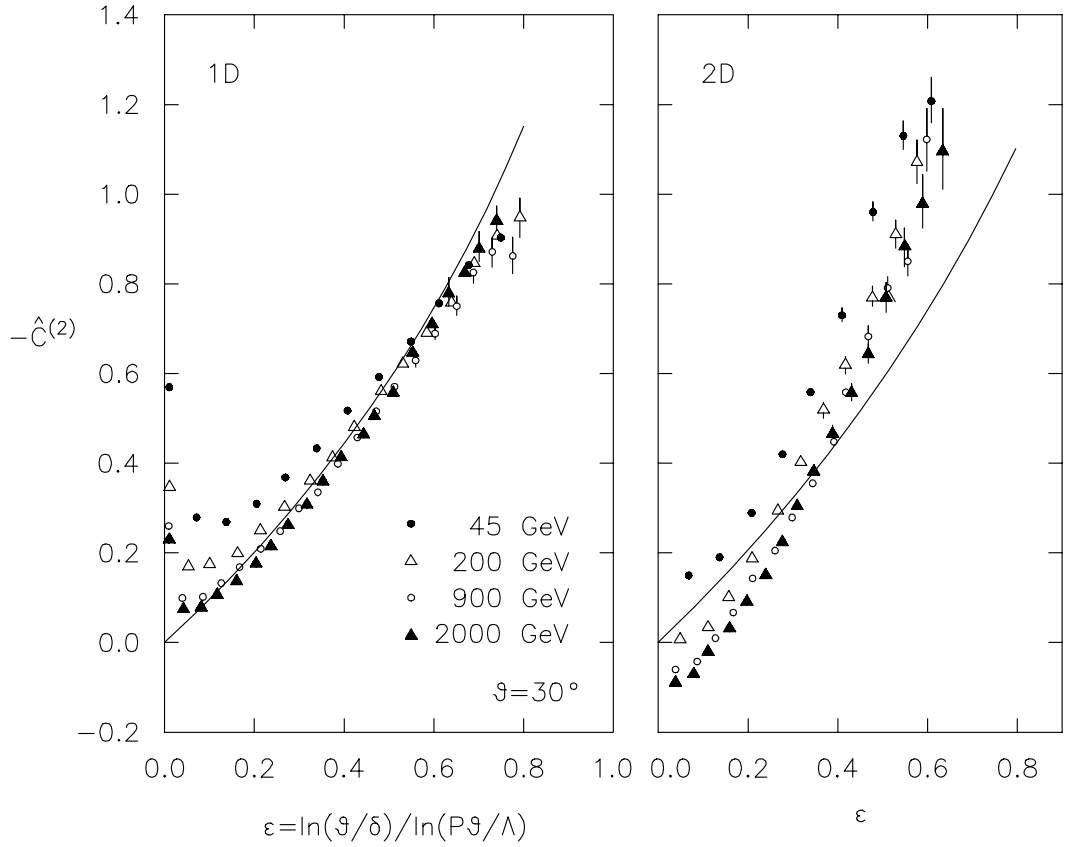}
\caption{Rescaled cumulant moments $\hat C_2$ as defined in
Eq.(\protect\ref{chat}) for the sidewise ring (1D) and the
sidewise cone (2D) for different jet momenta $P$ as
function of $\ve$ in comparison with the high energy DLA
prediction $2\beta (1-\omega(\ve, n)/n)$, normalized
to zero at $\ve=0$.
    }
\label{fig:f17}
\end{figure}

\begin{figure}[htb]
\vspace{9pt}
%\framebox[55mm]{\rule[-21mm]{0mm}{43mm}}
\epsfxsize=14cm \epsfbox{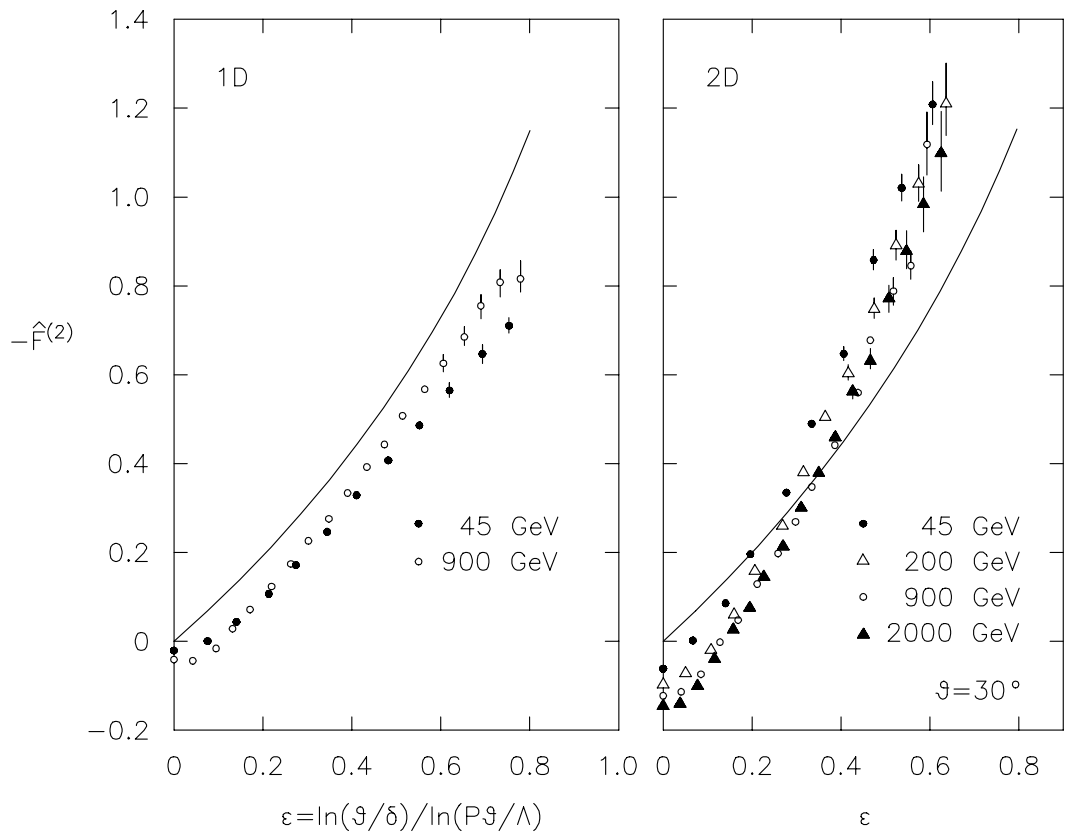}
\caption{Rescaled factorial moments $\hat F_2$ as in Fig.16.
%\protect\ref{fig:f17}
    }
\label{fig:f18}
\end{figure}

\begin{figure}[htb]
\vspace{9pt}
%\framebox[55mm]{\rule[-21mm]{0mm}{43mm}}
\epsfxsize=14cm \epsfbox{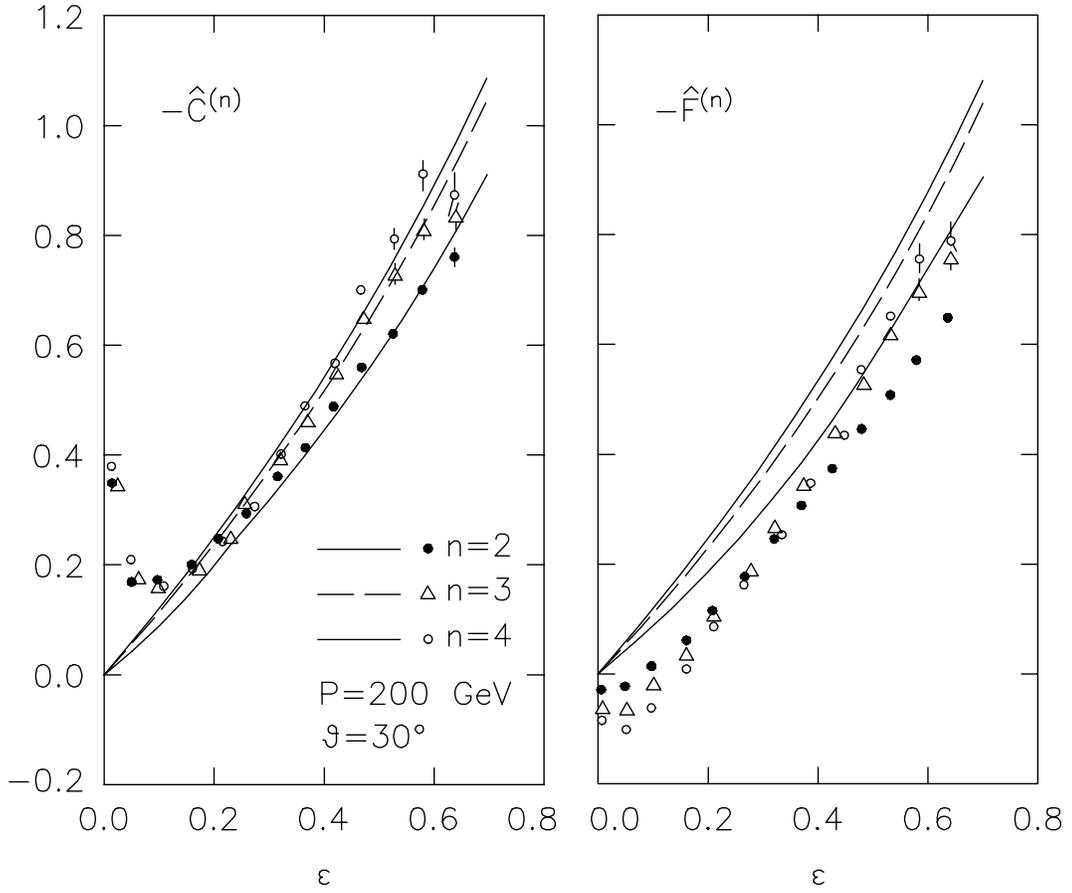}
\caption{Dependence on the order $n$ of the $D=1$ rescaled factorial moments.
The curves represent the DLA predictions, normalized to
$\hat C^{(n)}(0)=\hat F^{(n)}(0)=0$.
    }
\label{fig:f19}
\end{figure}

\begin{figure}[htb]
\vspace{9pt}
%\framebox[55mm]{\rule[-21mm]{0mm}{43mm}}
\epsfxsize=9cm \epsfbox{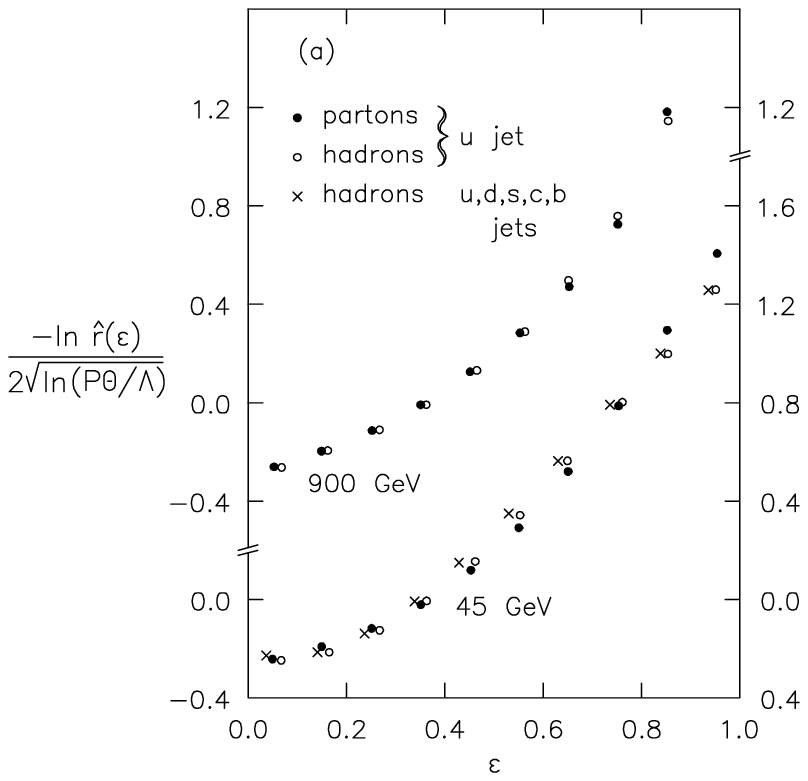}
 \hspace*{.5cm}  
\epsfxsize=9cm \epsfbox{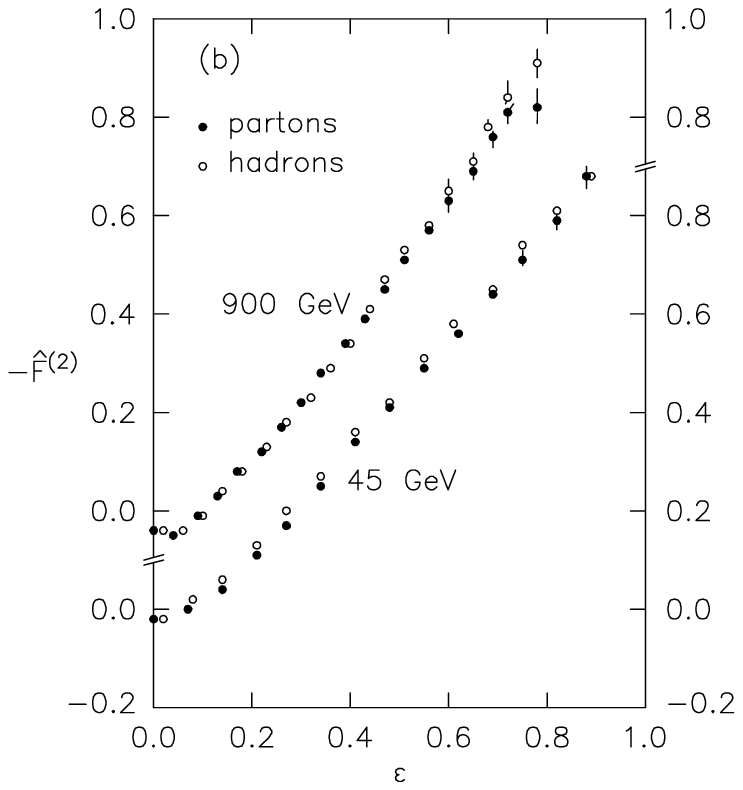}
\caption{Comparison of results on correlations and moments obtained at the
parton level and after hadronization for u-quark jets. Results
for a superposition of all jet flavours produced in $e^+e^-\to$ hadrons
don't differ appreciably.
    }
\label{fig:f20}
\end{figure}

\begin{figure}[htb]
\vspace{9pt}
%\framebox[55mm]{\rule[-21mm]{0mm}{43mm}}
\epsfxsize=12cm \epsfbox{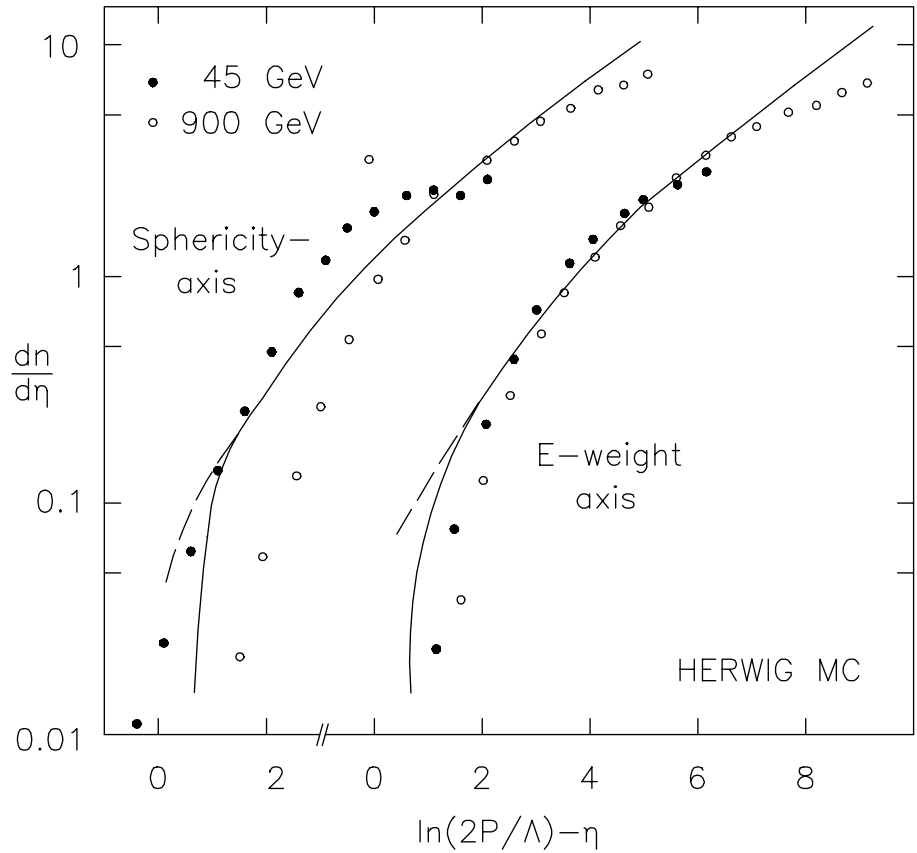}
\caption{Angular distribution $dn/d\eta$ for different jet
momenta $P$ with respect to the sphericity axis and
with respect to the particle's directions weighted by their energies.
The full (dashed) curve represents the exact (approximate) DLA results where
the normalization is adjusted.
    }
\label{fig:21}
\end{figure}

\begin{figure}[htb]
\vspace{9pt}
%\framebox[55mm]{\rule[-21mm]{0mm}{43mm}}
\epsfxsize=12cm \epsfbox{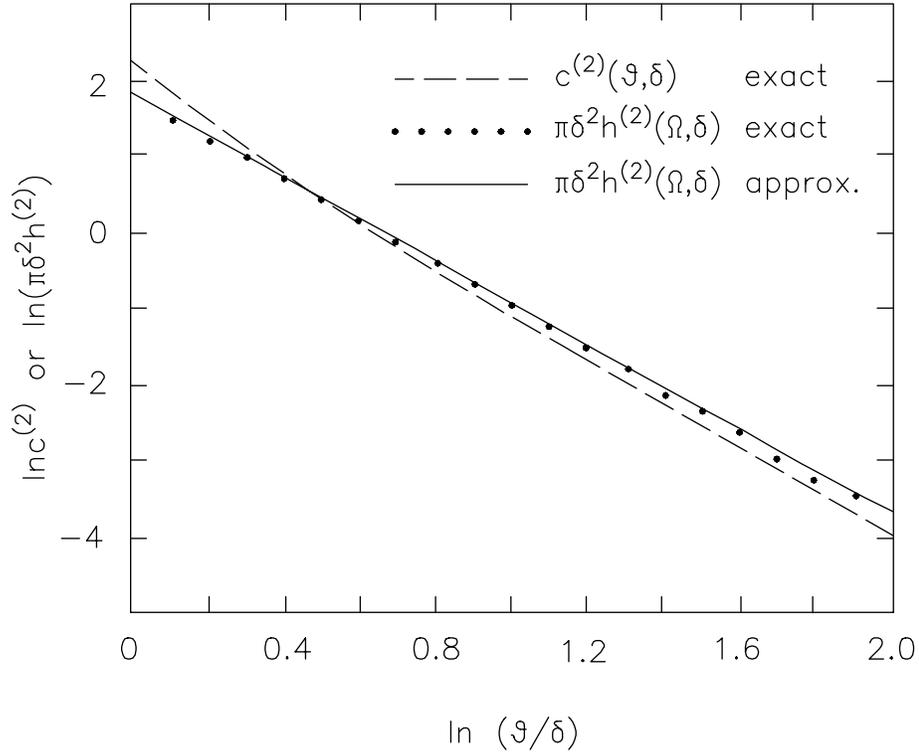}
\caption{Numerical check of various approximations in moment
calculations: the straight line represents the function
$\pi \delta^2 h (\Omega, \delta)$
using pole approximation Eq.(\protect\ref{hm2}) and the dots the result of exact
numerical integration
in Eq.(\protect\ref{hmom2}).
The dashed curve represents the exact
calculation of the moment $c^{(2)}(\theta, \delta)$ in
Eq.(\protect\ref{cnd}), for $Y =\ln P/Q_o=5$ and $\Theta = 1$.
    }
\label{fig:f12}
\end{figure}

 \end{document}